\documentclass{aa}
\usepackage{natbib}
\usepackage{graphicx}
\usepackage{array}
\usepackage{tabularx}
\usepackage{txfonts}
\usepackage[x11names]{xcolor}

\usepackage{orcidlink}
\usepackage{amsmath}
\usepackage[normalem]{ulem}

\usepackage{hyperref}
\hypersetup{
    colorlinks,
    allcolors=blue
}

\newcommand{\boldvec}[1]{\vec{\mbox{\boldmath{$#1$}}}}
\newcommand{\NHH}{$N_{\rm{H_2}}$}

\newcommand{\jonLEt}[1]{\textcolor{black}{#1}}

\begin{document}

   \title{Statistical analysis of the relative orientations between filaments and magnetic fields using \textit{Herschel} and \textit{Planck} data in star-forming regions}


   \author{J. Oers
          \inst{1}{\orcidlink{0009-0001-5143-0188}}
          \and
          I. Ristorcelli\inst{1}{\orcidlink{0000-0002-1469-6323}}
          \and
          K. Ferrière\inst{1}{\orcidlink{0000-0002-4395-2840}}
          \and
          M. Juvela\inst{2}{\orcidlink{0000-0002-5809-4834}}
          \and
          L. Montier\inst{1}
          \and
          D. Alina\inst{3}{\orcidlink{0000-0001-5403-356X}}
          \and
          J. Montillaud\inst{4}
          }

   \institute{
    IRAP, Université de Toulouse, CNRS, 9 avenue du colonel Roche, BP 44346, 31400 Toulouse Cedex 4, France
    \and
    Department of Physics, PO Box 64, 00014, University of Helsinki, Finland
    \and
    Department of Physics, School of Sciences and Humanities, Nazarbayev University, Kabanbay batyr ave, 53, Nur-Sultan 010000, Kazakhstan
    \and
    Université Marie et Louis Pasteur, CNRS, Institut UTINAM, UMR 6213, F-25000 Besançon, France \\ \\
    \email{jonathanoers@gmail.com}
    }

   \date{Received ???/ ???}

 
  \abstract
   {Filamentary structures are ubiquitous in the interstellar medium (ISM). This is particularly true in molecular clouds, with most clumps and cores forming inside the densest regions of filaments. Observations and simulations both suggest that magnetic fields play a key role in the formation and evolution of filaments and in the process of star formation, yet their exact role is still poorly understood. In this context, the study of the relative orientations between filaments and magnetic fields has become a go-to method to obtain new insight.}
   {We aim to statistically examine the relative orientations between filaments and magnetic fields in various star-forming regions with different physical properties and Galactic environments.}
   {We used a dedicated method, {\tt FilDReaMS}, that relies on a template that has the shape of a rectangular bar with a variable width to detect and extract filaments at multiple scales. We applied {\tt FilDReaMS} to the 116 fields of the \textit{Herschel} "Galactic Cold Cores" (GCC) key project (18$^{\prime\prime}$-36$^{\prime\prime}$ resolution), which measured dust emission in diverse star-forming regions. We then compared the filament orientations to the orientation of the plane-of-sky (PoS) magnetic field (\textbf{\textit{B}$_{\rm{PoS}}$}), inferred from \textit{Planck} observations of the dust polarized thermal emission (7$^{\prime}$ resolution) using histograms of relative orientations (HROs). Additionally, we used a catalog of dense cold cores detected in the GCC fields to study the relative orientations of filaments hosting cores.}
   {We present the results of our statistical analysis of these relative orientations as functions of filament effective width, H$_2$ column density ($N_{\rm{H_2}}$), evolutionary stage, and Galactic environment. We find that low-$N_{\rm{H_2}}$ filaments tend to be roughly parallel to \textbf{\textit{B}$_{\rm{PoS}}$} at all scales, while narrow high-$N_{\rm{H_2}}$ filaments do not have any preferred orientations and wide high-$N_{\rm{H_2}}$ filaments tend to be roughly perpendicular. This change in preferred orientations occurs at transition column density $(N_{\rm{H_2}})_{\rm{t}}$ values typically in the range $[0.8$, $8]\,\times\,10^{21}\,\rm{cm^{-2}}$, a range that is consistent with results of previous \textit{Planck} studies. We also analyzed the HROs for filaments with embedded cores and find them to be consistent with HROs for high-$N_{\rm{H_2}}$ filaments, although the trend is less pronounced. However, several fields do not follow the general trends, with a variety of behaviors that can be due to factors such as projection effects, confusion along the line of sight (LoS), magnetic field tangling, or different magnetic field strengths. Our analysis of projection effects shows that, statistically, preferred orientations in the PoS are indicative of true preferred orientations in 3D. Our results suggest that higher polarization fractions, $p$, entail weaker projection effects, consistent with the presumed link between $p$ and the magnetic field inclination to the LoS.}
   {Our results confirm the existence of a coupling between magnetic fields at cloud scales and filaments at smaller scales while also highlighting the complexity of the ISM. They also call for further statistical analyses studying magnetic fields and other physical processes at smaller scales to better understand the variety of behaviors seen in the HROs.}


   \authorrunning{J. Oers et al.}
   \titlerunning{Statistical analysis of the relative orientations between filaments and magnetic fields.}
   \maketitle
%

\section{Introduction}
\label{sec:introduction}
The early stages of star formation involve a combination of physical processes, including turbulence, gravity, and magnetic fields, whose respective roles across spatial and temporal scales remain poorly constrained. In particular, while the impact of magnetic fields is increasingly supported by observations and simulations, advancing our understanding requires systematic multiscale observational studies of star-forming structures, from molecular clouds to filaments, clumps, and dense cores.

Filamentary structures are now recognized as being ubiquitous in a wide range of environments of the interstellar medium (ISM). Filaments were initially observed via dust extinction and later through molecular spectral lines (see, e.g., review by \citealp{Hacar2023}, and references therein), but their prevalence and morphology were later revealed in unprecedented detail by the \textit{Herschel} Space Observatory \citep[see review by][]{Andre2014}. In addition, these observations established that star formation is tightly linked to filamentary networks, with a significant fraction of prestellar cores located within the densest filaments \citep{Polychroni2013, Konyves2015, Montillaud2015}. Understanding the origin and evolution of filaments, which are closely linked to dense cores, is thus essential for unveiling the early stages of star formation.

Numerical simulations of magnetohydrodynamic (MHD) turbulence have consistently reproduced filamentary structures. They have also demonstrated that the relative orientations between filaments and magnetic fields are sensitive to the magnetization of the medium \citep{Soler2013, Chen2016, Soler2017, Hennebelle2019}. In weakly magnetized environments, density structures tend to line up parallel to the magnetic field, whereas in strongly magnetized regimes, a transition from parallel to perpendicular alignment is observed at a critical column density.

Observationally, an increasing number of studies have revealed preferential relative orientations between filaments and plane-of-sky (PoS) magnetic fields (\textbf{\textit{B}$_{\rm{PoS}}$}), whose orientation is inferred from either starlight polarization or dust polarized emission (e.g., review by \citealp{Pattle2023} and references therein). In particular, dust polarization measurements from the \textit{Planck} all-sky survey enabled statistical studies of these relative orientations over interstellar structures resolved by \textit{Planck} all around the Galaxy, as the authors found that elongated structures are predominantly parallel to the magnetic field in the diffuse (neutral) medium, while they are mostly perpendicular in dense molecular clouds \citep{PlanckXXXII2016,PlanckXXXV2016}, with a transition at a column density\footnote{In a purely molecular medium, $N_{\rm{H}} = 2\,N_{\rm{H_2}}$.} $(N_{\rm{H}})_{\rm t}$$\,\approx\,10^{21.7}\,{\rm{cm^{-2}}}$ \citep{PlanckXXXV2016}.

Based on \textit{Planck} data alone on filaments hosting clumps, \cite{Alina2019} found that
low-density contrast filaments tend to line up parallel to the magnetic field in low-density regions, while these filaments tend to be perpendicular to the magnetic field in high-density regions. However, no preferential relative orientation prevails for the high-density contrast filaments regardless of the density of their environment. \cite{Alina2019} also found that the transition in relative orientations occurs in the dense medium at a density around $(n_{\rm{H}})_{\rm t} \simeq 10^{3}\,{\rm{cm^{-3}}}$. These results are consistent with the previous analysis by \cite{Fissel2019}, in Vela-C, based on observations with the BLAST-pol balloon-borne experiment and molecular spectral lines tracing different gas densities. These results confirm that even if low-density regions are mostly dominated by turbulence and magnetic fields, leading to preferential parallel alignments, self-gravity must play a significant role for high-density structures, leading to filaments oriented perpendicular to the magnetic field.

Combining \textit{Planck} polarization data with high-resolution \textit{Herschel} observations can provide new insights to investigate the above trends across various environments and scales. \cite{Malinen2016}, \cite{Cox2016}, and \cite{Soler2019} reported similar trends in relative orientations by comparing \textbf{\textit{B}$_{\rm{PoS}}$} orientation inferred from \textit{Planck} (at $\simeq 10^{\prime}$ resolution) and the orientations of filaments traced by \textit{Herschel} ($36^{\prime\prime}$ resolution) in a sample of nearby molecular clouds. They found that filaments with lower column densities tend to be parallel to \textbf{\textit{B}$_{\rm{PoS}}$}, while those with higher column densities are generally perpendicular to \textbf{\textit{B}$_{\rm{PoS}}$}, with similar thresholds in column density marking this transition, which are consistent with previous \textit{Planck}-only results. These findings support the idea that magnetic fields at cloud scales are linked to the structure and alignment of filaments at smaller scales and motivate further investigation with a statistical analysis.

The \textit{Herschel} Galactic Cold Cores (GCC) program \citep{Juvela2010} provides a suitable dataset, covering 116 fields with broad ranges of physical conditions and evolutionary stages. A systematic analysis of the relative orientations between filaments and magnetic fields in this sample requires robust methods capable of tracing filamentary structures across wide dynamic ranges in scale and column density.
Several methods have been developed to extract elongated structures from two-dimensional maps \citep[e.g., see description in][]{Carriere2022a}. Previous studies of relative orientations have often used local methods, which compute either the gradient \citep[first-order derivatives; e.g.,][]{Soler2013,PlanckXXXV2016} or the Hessian matrix \citep[second-order derivatives; e.g.,][]{Polychroni2013,Schisano2014,PlanckXXXII2016} at each pixel of the considered map. 
Nonlocal methods focus on a given scale around each pixel, such as template matching \citep{Juvela2016}, 
Rolling Hough Transform (RHT; \cite{Clark2014} used in \cite{Malinen2016} and \cite{Alina2019} studies), and Filament Detection and Reconstruction at Multiple Scales ({\tt FilDReaMS}; \citealp{Carriere2022a, Carriere2022b}). 
A comprehensive comparison of these various techniques would be highly valuable, but only a limited number of studies have addressed this issue so far. \cite{Micelotta2021} conducted a comparative analysis of the gradient and RHT methods, based on simulation data. Their findings revealed both similarities and differences between the results of the two methods, with disparities attributed to the intrinsic characteristics of the filamentary structures identified by each method. 

For this study, we used the recent method called {\tt FilDReaMS}, developed by \cite{Carriere2022a}, to detect and extract filaments of different effective widths. The method was first applied to a sample of four GCC fields by \cite{Carriere2022b}. Their analysis revealed that narrow low-column-density filaments tend to be parallel to the magnetic field, while wider high-column-density filaments are predominantly perpendicular. A transition between these relative orientations was observed in the two closest fields, around $(N_{\rm{H_2}})_{\rm t} \simeq 1.1 \times 10^{21}\,{\rm{cm^{-2}}}$ and $(N_{\rm{H_2}})_{\rm t} \simeq 1.4 \times 10^{21}\,{\rm{cm^{-2}}}$, respectively, consistent with results from previous studies. Interestingly, one of the fields showed preferred parallel relative orientations for all filaments, regardless of their widths and column densities.

In this paper, we extend the analysis to the whole sample of 116 GCC fields with the purpose of performing both individual and statistical analyses of the relative orientations and exploring possible variations with diverse physical parameters. In Sect.~\ref{sec:data}, we present the data used in our study. In Sect.~\ref{sec:method}, we describe the {\tt FilDReaMS} method used to extract filaments, and we explain how histograms of relative orientations (HROs) are built and exploited for our analysis. In Sect.~\ref{sec:properties}, we discuss the samples used for the statistical analysis, which depend on the physical properties of the GCC fields, the extracted filaments, and the detected cores. In Sect.~\ref{sec:results}, we apply {\tt FilDReaMS} to the column density maps of the 116 GCC fields and present both an overview of the results obtained for the individual fields and the statistical results obtained for various filament samples. In Sect.~\ref{sec:projection_effects}, we explore projection effects using both an analytical analysis and a toy model. In Sect.~\ref{sec:discussion}, we provide a summary of our results and discuss their implications. Finally, Sect.~\ref{sec:conclusion} presents our conclusions and perspectives.


\section{Data}
\label{sec:data}

\subsection{\textit{Planck}}
\label{sec:data_Planck}

\begin{figure}
\centering
\includegraphics[width=\hsize]{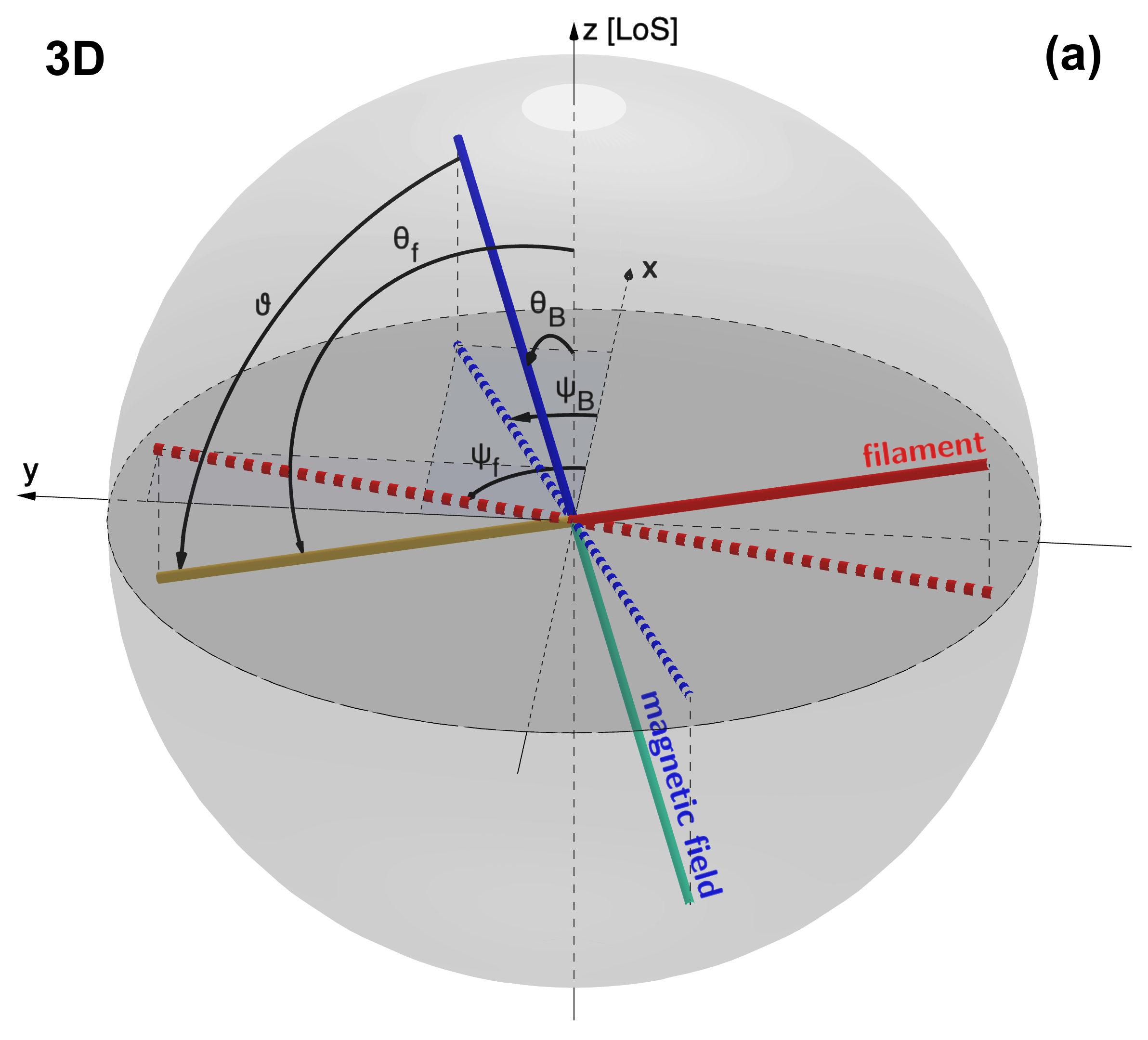}
\includegraphics[width=\hsize]{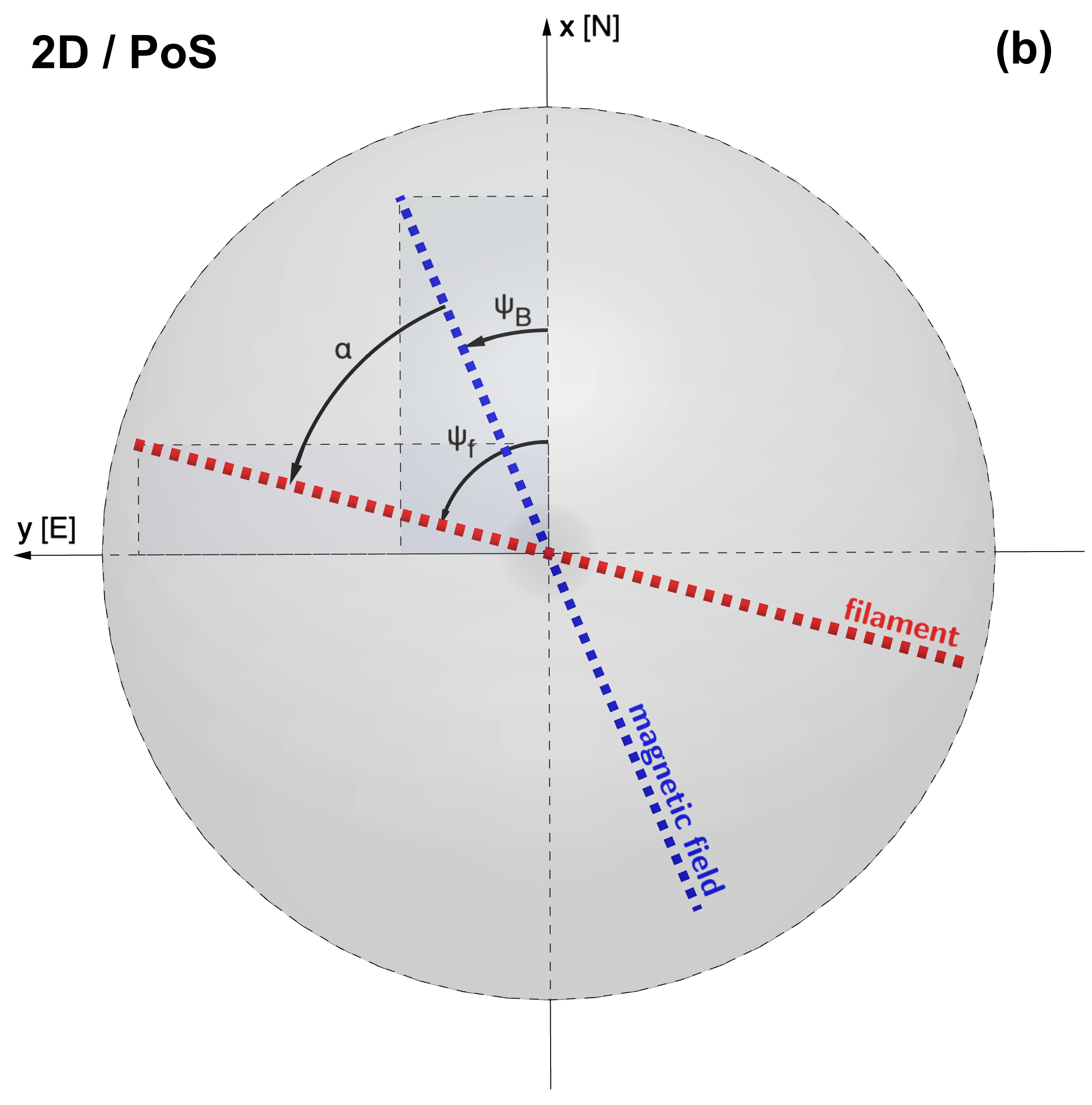}
\caption{Representation of the angular coordinates used in the paper, both in 3D (\textbf{a}) and in the 2D PoS (\textbf{b}). Note that $(\theta,\psi)$ are the standard spherical angular coordinates, with subscripts f and B referring to the local filament and local $\boldvec{B}$, respectively; $(x,y,z)$ are the cartesian coordinates. The $xy$-plane corresponds to the PoS, with $x$-axis pointing north, the $y$-axis pointing east, and the $z$-axis along the LoS toward the observer. The solid red and blue lines show the orientations of the local filament and local $\boldvec{B}$, respectively. The dashed red and blue lines represent the respective projections of the solid lines onto the PoS. The symbol $\vartheta$ is the angle between the local filament and the local $\boldvec{B}$ in 3D (given by Eq.~\eqref{eq:vartheta}). Finally, $\alpha\,=\,\psi_{\rm{f}} - \psi_{B}$ is the relative orientation angle in the PoS (Eq.~\eqref{eq:alpha}).}
\label{fig:data_angles}
\end{figure}

\textit{Planck}-HFI measured polarized emission at 353 GHz, which is dominated by dust thermal emission in the Galaxy \citep{PlanckXIX2015, PlanckXX2015}. The full mission data are available as part of the \textit{Planck} Product Release 2018. \textit{Planck} observations provide maps of the Stokes \textit{I}, \textit{Q}, and \textit{U} parameters, which correspond to the total intensity and the linear polarization components. We derived the PoS magnetic field (\textbf{\textit{B}$_{\rm{PoS}}$}) orientation angle, $\psi_{B}$, using
\begin{equation}\label{eq:Borientation}
    \psi_{B} = \frac{1}{2} \ \rm{arctan}\left(\frac{U}{Q}\right) \pm 90^{\circ} \ ,
\end{equation}
where $\arctan$ is the two-argument arctangent function defined from $-180^\circ$ to $+180^\circ$. Since polarization data only give access to the magnetic field orientation, and not its direction, there is a 180$^{\circ}$ ambiguity on $\psi_{B}$. Hence, we can define $\psi_{B}$ in the range [$-90^{\circ}$, $+90^{\circ}$] by choosing the $+$ or $-$ sign in the last term of Eq.~\eqref{eq:Borientation} accordingly. Here, we adopt the IAU convention, where the polarization angle increases from Galactic north to east (see Fig.~\ref{fig:data_angles}b), opposite to the {\tt Healpix} convention used by the \textit{Planck} community. As such, we flipped the sign of \textit{U} taken from the \textit{Planck} Legacy Archive.\footnote{\textit{Planck} Legacy Archive: \href{https://www.cosmos.esa.int/web/planck/pla}{https://www.cosmos.esa.int/web/planck/pla}.}
Following \cite{Montier2015}, we computed the uncertainty in $\psi_{B}$, $\sigma_{\psi_{B}}$, which is equal to the uncertainty in the polarization angle, using
\begin{equation}\label{eq:Borientationuncertainty}
    \sigma_{\psi_{B}} = \frac{1}{2} \sqrt{\frac{Q^2 \sigma_{\rm{U}}^2 + U^2 \sigma_{\rm{Q}}^2 - 2QU \sigma_{\rm{QU}}}{(Q^2 + U^2)^2}} \ \cdot
\end{equation}

We extracted from the all-sky \textit{Planck} data the Stokes parameters and their uncertainties at 353 GHz as 2$^{\circ} \times 2^{\circ}$ maps centered on the \textit{Herschel} fields from the GCC program (presented in Sect. \ref{sec:data_Herschel}). We smoothed the \textit{Planck} maps from $4.7^{\prime}$ to $7^{\prime}$ to improve the signal-to-noise ratio (S/N).

From Stokes \textit{I}, \textit{Q}, and \textit{U}, we derived the polarized intensity and the polarization fraction, respectively, given by
\begin{equation}
    P \equiv \sqrt{Q^2 + U^2} \ \label{eq:polarizationintensity1}
\end{equation}
and
\begin{equation}
    p \equiv \frac{P}{I} \ \cdot \label{eq:polarizationfraction2}
\end{equation}
In this analysis, we debiased the polarization fraction, $p$, by using the modified asymptotic (MAS) estimator \citep{Plaszczynski2014,Montier2015b}, only valid for polarized-intensity S/N $\geq$ 2 \citep{Montier2015,Montier2015b}. Following posterior characterizations of polarization efficiencies \citep{PlanckIII2020,PlanckXI2020} on 2018 data from the \textit{Planck} Legacy Archive, we assumed a 1.5\% photometric uncertainty in the 353 GHz Stokes \textit{Q} and \textit{U} polarization maps.

\subsection{\textit{Herschel}}
\label{sec:data_Herschel}

The \textit{Herschel} Space Observatory \citep{Pilbratt2010}, hereby \textit{Herschel}, measured dust thermal emission in far-infrared and submillimeter wavelengths, from 70$\,\mu$m to 500$\,\mu$m, at angular resolutions of $5^{\prime\prime}\,$-$\,36^{\prime\prime}$, which made it possible to probe star-forming regions. \textit{Herschel} provided observations of filaments and cores within molecular clouds at higher angular resolution than \textit{Planck}.

The \textit{Herschel} key program GCC \citep{Juvela2010} is a follow-up program designed to map out a sample of regions around cold clumps previously detected in the \textit{Planck} all-sky survey. The regions were selected using the \textit{Planck} catalog of Galactic cold clumps by ensuring a full coverage of physical properties and Galactic environments representative of the cold clump population, complementary to other \textit{Herschel} Galactic surveys \citep{Juvela2012}.

This follow-up is composed of 116 fields with a typical size of 40$^{\prime}$, up to 1$^{\circ}$, observed in the 100 and 160$\,\mu$m bands with the PACS instrument \citep{Poglitsch2010} and in the 250, 350 and 500 $\mu$m bands with the SPIRE instrument \citep{Griffin2010}, with angular resolutions of 7.7, 12, 18, 25, and 36$^{\prime\prime}$, respectively. The observed fields and their properties are presented in \cite{Montillaud2015}. The authors used the source extraction algorithm {\tt getsources} \citep[version v1.130401;][]{Menshchikov2012} developed for \textit{Herschel} surveys, to build a catalog of 4466 cold and compact sources across the GCC fields, identifying their positions and temperatures, and providing estimates of their evolutionary stages from gravitationally unbound to prestellar and protostellar cores. \cite{Montillaud2015} divided the catalog into two sets of sources depending on their physical sizes: cores, which are small ($\lesssim 0.2\,{\rm pc}$) and individual sources, and clumps, which are large ($\gtrsim 0.2\,{\rm pc}$) sources of size that may be composed of several cores. For simplicity, we employ the general term of "core" for all sources throughout this paper.

In this study, we used the maps of H$_2$ column density, $N_{\rm{H_2}}$, presented in \cite{Montillaud2015} and derived from the dust spectral energy distribution fit using only the 250, 350, and 500 $\mu$m SPIRE maps. The authors assumed a dust opacity $\kappa = 0.1\,\rm{cm}^{2}\,/\,\rm{g}\,(\nu / 1000\,\rm{GHz})^{\,\beta}$, with a fixed emissivity spectral index $\beta =$ 2.0, suited for high-density environments \citep{Hildebrand1983,Beckwith1990}. The $N_{\rm{H_2}}$ maps have an angular resolution of 36$^{\prime\prime}$.

Distance estimates are essential for our study. \cite{Montillaud2015} provided a distance estimate for the molecular cloud targeted in each GCC field. Here, we assign the distance of the cloud to the distance of its filaments and cores, and we refer to it as the distance of the GCC field itself. We revisited 33 of the 116 GCC fields with unreliable or missing distance estimates. We estimated their new distances using \textit{Gaia} observations and the dust extinction method described by Marshall et al. (in prep.).

\begin{figure*}
\centering
\sidecaption
\includegraphics[width=12cm]{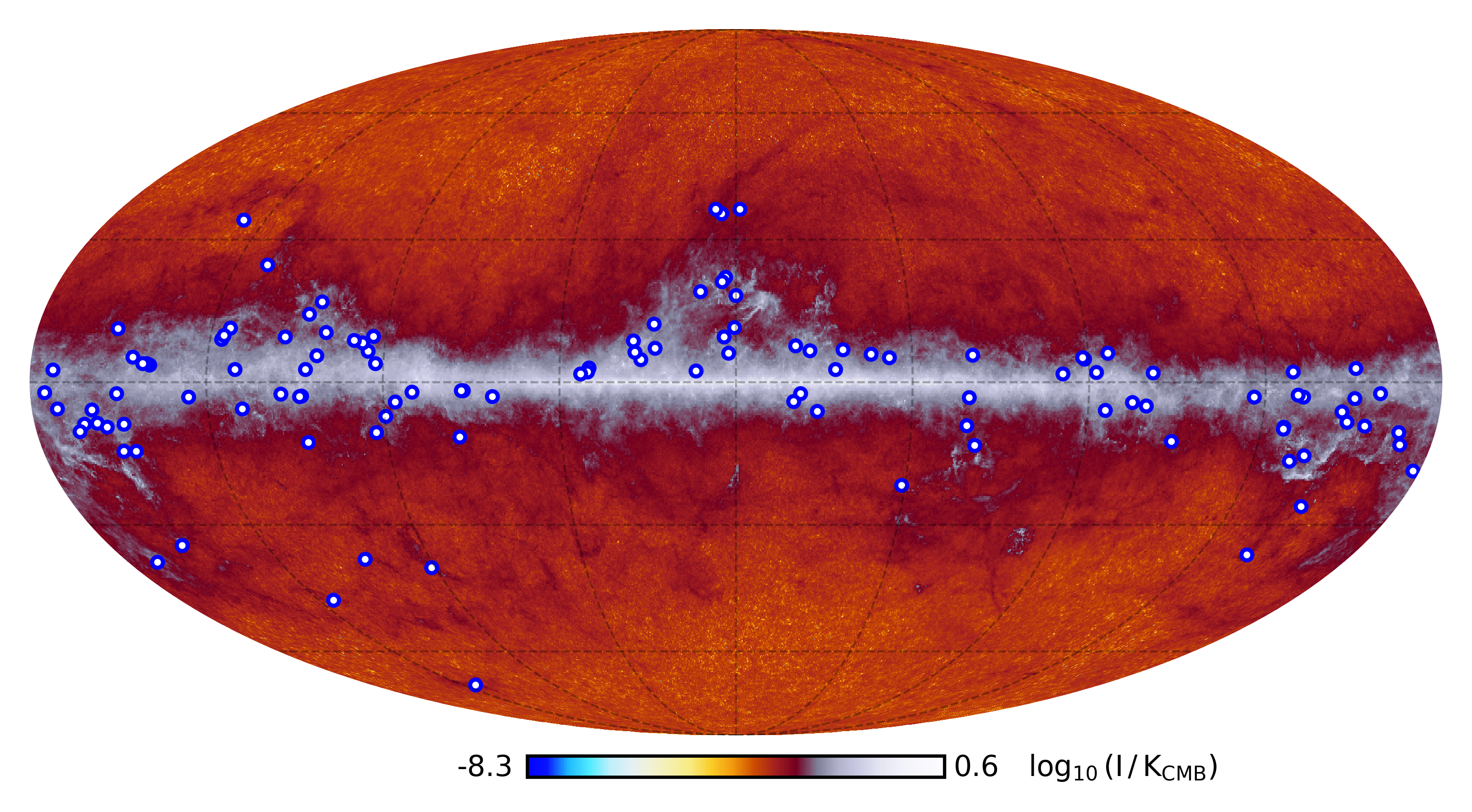}
\caption{Locations of the fields observed with \textit{Herschel} from the GCC program \citep{Juvela2010,Juvela2012}. The background map shows the \textit{Planck} intensity at 353 GHz.}
\label{fig:data_GCC_fields_locations}
\end{figure*}


\section{Method}
\label{sec:method}

\subsection{FilDReaMS}
\label{sec:method_fildreams}

In order to extract filaments and their physical properties in each $N_{\rm H_2}$ map, we used the {\tt FilDReaMS} method developed by \cite{Carriere2022a}.
{\tt FilDReaMS} is designed to detect filamentary structures, which we call filaments from hereon, in an image and provide information on their widths, their orientations, and the robustness of the detection.

More specifically, {\tt FilDReaMS} relies on a model template which has the shape of a rectangular bar (referred to as the model bar) defined by its width $W_{\rm{b}}$, its length $L_{\rm{b}}$, and hence its aspect ratio $r_{\rm{b}} = L_{\rm{b}}/W_{\rm{b}}$. In the following astrophysical application, we adopt $r_{\rm{b}}$ = 3 \citep{Panopoulou2014,Arzoumanian2019,Carriere2022a}, and we consider values of $W_{\rm{b}}$ spanning the range $[(W_{\rm{b}})_{\rm{min}},(W_{\rm{b}})_{\rm{max}}]$, with $(W_{\rm{b}})_{\rm{min}} = 5\,\rm{pixels}$ and $(W_{\rm{b}})_{\rm{max}}$ equal to the widest filament detected in the image, with an upper limit set to one-ninth the size of the \textit{Herschel} map. The orientation angle of the model bar, $\psi_{\rm{b}}$, is defined in the range $[-90^{\circ}, +90^{\circ}]$ and follows the IAU convention (see Sect.~\ref{sec:data_Planck} and Fig.~\ref{fig:data_angles}b).
For each considered value of $W_{\rm{b}}$, {\tt FilDReaMS} filters out structures wider than $W_{\rm{b}}$, then converts the resulting image into a binary map. Next, FilDReaMS successively considers all the active pixels of the binary map, which trace the remaining structures after filtering. At every active pixel $i$, {\tt FilDReaMS} retrieves the orientation angle of the model bar that best matches the map, $(\psi_{\rm{b}})_{i}$, and computes the corresponding significance, $S_{i}$, which compares the detected filament to an ideal case. If $S_{i} > 1$, {\tt FilDReaMS} confirms the detection of a filament with orientation angle $(\psi_{\rm{f}})_{i}$ = $(\psi_{\rm{b}})_{i}$ at pixel $i$. 
If a given pixel $i^{\prime}$ belongs to two or more filaments, the filament orientation angle assigned to pixel $i^{\prime}$ is the orientation angle of the most significant filament.
Once all the pixels have been processed, {\tt FilDReaMS} creates a filament mask by multiplying the above binary map by a new binary map formed by the model bars of all the detected filaments. This mask is then applied to the initial image to reconstruct the physical network of detected filaments of bar width $W_{\rm{b}}$, with their true shapes and their intensities. 
This process is repeated over the whole range of $W_{\rm{b}}$, leading to filamentary networks of different bar widths. 
Ultimately, {\tt FilDReaMS} provides pixel-by-pixel information on filament bar width, $W_{\rm{b}}$, and orientation angle, $\psi_{\rm{f}}$.
The uncertainties in the filament orientation angles are below 3$^{\circ}$, and they decrease with increasing $W_{\rm{b}}$.

\subsection{Histogram of relative orientations}
\label{sec:method_hro}

Following previous studies \citep{Soler2013, PlanckXXXII2016, PlanckXXXV2016, Soler2019, Alina2019, Carriere2022b}, we use HROs to quantify the statistics of relative orientation angles. \cite{Soler2013} first used this name to refer to their algorithm which uses column density gradients to detect structures in an image and extracts their orientations, which are then compared to the magnetic field orientation. The relative orientations between the column density structures and the magnetic field were characterized through histograms. In this paper, we use the acronym HRO with a different meaning, only to refer to the histograms characterizing the relative orientations. In most of the paper, we analyze the differences between the orientation angles of filaments extracted with {\tt FilDReaMS}, $\psi_{\rm{f}}$, and the orientation angle of \textbf{\textit{B}$_{\rm{PoS}}$} inferred from \textit{Planck} polarization data, $\psi_B$ (Eq.~\eqref{eq:Borientation}). In addition, we also consider the orientation angle of the major axis of each GCC core, $\psi_{\rm{c}}$, provided in the GCC catalog \citep{Montillaud2015}, and we analyze the differences between $\psi_{\rm{c}}$ and $\psi_{\rm{f}}$, or between $\psi_{\rm{c}}$ and $\psi_B$ (see Sect.~\ref{sec:results3_SourceHRO_s}). 

We now describe the relative orientations between filaments and \textbf{\textit{B}$_{\rm{PoS}}$}, but this can be generalized to any two objects with a 180$^{\circ}$ ambiguity. We define $\psi_{\rm{f}}$ and $\psi_B$ in the range $[-90^{\circ}, +90^{\circ}]$ and define the difference between $\psi_{\rm{f}}$ and $\psi_B$, $\psi_{\rm{f}} - \psi_{B}$ (also called $\alpha$ in Fig.~\ref{fig:data_angles}b and Eq.~\eqref{eq:alpha}), in the range $[-180^{\circ}, 180^{\circ}]$. Since the 180$^{\circ}$ ambiguity remains for ($\psi_{\rm{f}} - \psi_{B}$), we redefine ($\psi_{\rm{f}} - \psi_{B}$) in the range $[-90^{\circ}, 90^{\circ}]$ by adding $\pm180^{\circ}$. If the two objects are nearly parallel, $\lvert\psi_{\rm{f}} - \psi_{B}\rvert \simeq 0^{\circ}$. Conversely, if the two objects are nearly perpendicular, $\lvert\psi_{\rm{f}} - \psi_{B}\rvert \simeq 90^{\circ}$. We only compute the relative orientation angles where the uncertainty in the angle estimates is below $10^{\circ}$. We compute ($\psi_{\rm{f}} - \psi_{B}$) pixel by pixel over all pixels of all the maps of a sample, then build our histogram of $\lvert\psi_{\rm{f}} - \psi_{B}\rvert$ in 9 angle bins of 10$^{\circ}$ each, from 0$^{\circ}$ to 90$^{\circ}$.

We also construct 2D histograms (2D HROs) of ($\psi_{\rm{f}} - \psi_{B}$) as functions of $N_{\rm{H_2}}$ \citep{Malinen2016,Carriere2022b}. To this end, we divide the sample of pixels into continuous bins of $N_{\rm{H_2}}$ containing an equal number of pixels, and we perform the HRO analysis in each $N_{\rm{H_2}}$ bin. The number of angle or $N_{\rm{H_2}}$ bins is determined by requiring a sufficient number of bins to resolve high-$N_{\rm{H_2}}$ trends, while maintaining significant statistics. The HROs have 18 angle bins of 10$^{\circ}$ from $-90^{\circ}$ to 90$^{\circ}$. The 2D HROs are built in 18 $N_{\rm{H_2}}$ bins.

We estimate the uncertainties in our HROs by using the bootstrap technique \citep{Efron1979,Efron1993} consisting of random sampling with replacement, from the corresponding pixel samples. To properly estimate the uncertainties, we need to bootstrap on independent filament pixels. As a compromise between bootstrapping on independent pixels, computation time, and accuracy, we resampled the pixel population to 1 pixel every 36$^{\prime\prime}$ (down from 1 pixel every 12$^{\prime\prime}$). We verified that the resulting HRO of each resampled pixel used in the bootstrap is consistent with the HRO of the original pixel sample. The uncertainty in each HRO bin is calculated as the standard deviation of the HRO values obtained in the different bootstrap realizations, centered on the average of these values. Since the number of bootstrap realizations depends on the sample size and scatter, we determine the required number of realizations by verifying that the average and standard deviation of the bootstrapped HROs converge, and that the average of the bootstrapped HROs matches the HRO obtained directly from the data.

The \textit{Herschel} and \textit{Planck} maps that we combine have different angular resolutions (36$^{\prime\prime}$ and 7$^{\prime}$, respectively). Such a combination of different resolutions was already used in previous studies (see, e.g., \citealp{Malinen2016,Soler2019,Carriere2022b}).
With 1 pixel every 36$^{\prime\prime}$, our \textit{Herschel} pixels are now independent, while our \textit{Planck} pixels remain over-sampled.
However, this is in line with our purpose to consider the relative orientations between filaments at 36$^{\prime\prime}$ scale, observed with \textit{Herschel}, and the mean magnetic field of the surrounding medium at 7$^{\prime}$, observed with \textit{Planck}. From now on, we consider that pixels used throughout the analysis are independent.

\subsection{Preferred alignment criteria}
\label{sec:method_ratios}

To quantify the preferred alignment in the HROs, previous studies have used the so-called histogram shape parameter, first introduced by Soler et al. (2013) and later redefined by Soler et al. (2017) as $\xi = (A_{0} - A_{90}) / (A_{0} + A_{90})$, where $A_{0}$ and $A_{90}$ are the areas of the histogram for $|\psi_{\rm{f}} - \psi_{B}| \le 22.5^\circ$ and $|\psi_{\rm{f}} - \psi_{B}| \ge 67.5^\circ$, respectively.
An important limitation of this approach is that it only distinguishes between uniform ($\xi \approx 0$), preferentially parallel ($\xi > 0$), and preferentially perpendicular ($\xi < 0$) relative orientations, with no way of uncovering preferentially intermediate ($|\psi_{\rm{f}} - \psi_{B}| \sim 45^\circ$) or bimodal parallel-perpendicular distributions.

This limitation naturally led us to propose a new criterion based on the three following ratios,
\begin{equation}
\centering
    r_{\parallel} = \frac{N_{\parallel}}{N} \ ,
    \qquad
    r_{\rm{int}} = \frac{N_{\rm{int}}}{N} \ ,
    \qquad
    r_{\perp} = \frac{N_{\perp}}{N} \ ,
    \label{eq:perpara_ratios}
\end{equation}
where $N_{\|}$ is the number of pixels with $\lvert\psi_{\rm{f}} - \psi_{B}\rvert$ $\in$ $[0^{\circ}, 30^{\circ}]$ (mostly parallel orientations), $N_{\rm{int}}$ the number of pixels with $\lvert\psi_{\rm{f}} - \psi_{B}\rvert$ $\in$ $]30^{\circ}, 60^{\circ}[$ (intermediate orientations), $N_{\perp}$ the number of pixels with $\lvert\psi_{\rm{f}} - \psi_{B}\rvert$ $\in$ $[60^{\circ}, 90^{\circ}]$ (mostly perpendicular orientations), and $N = N_{\|} + N_{\rm{int}} + N_{\perp}$ is the total number of pixels. 
All three ratios range from 0 to 1, and they sum up to 1. 
They make it possible to distinguish between uniform ($r_{\parallel} \approx r_{\rm{int}} \approx r_{\perp}$), preferentially parallel ($r_{\parallel} > r_{\perp}, r_{\rm{int}}$), preferentially perpendicular ($r_{\perp} > r_{\parallel}, r_{\rm{int}}$), preferentially intermediate ($r_{\rm{int}} > r_{\parallel}, r_{\perp}$), and bimodal parallel-perpendicular ($r_{\parallel}, r_{\perp} > r_{\rm{int}}$) distributions. 
We also quantify the preferred alignment in the 2D HROs by computing the three ratios in each of the 18 $N_{\rm{H_2}}$ bins.
The uncertainties in the different ratios can be estimated by applying the bootstrap technique introduced above to HROs with three angle bins of 30$^{\circ}$ each.

The criterion based on the ratios defined in Eq.~\eqref{eq:perpara_ratios} suffers from a loss of information resulting from the binning process. To better utilize all the information contained in the data, we drew inspiration from the projected Rayleigh statistic, $Z_x$, used in \cite{Jow2018} and defined the three following improved ratios:
\begin{equation}
\begin{aligned}
    \mathfrak{r}_{\parallel} & = 2 \ \frac{\displaystyle \sum_{0^{\circ} \le |\alpha_i| \le 45^{\circ}} \cos^2 (2\alpha_i)}{\displaystyle \sum_{|\alpha_i|} \left( 1 + \cos^2 (2\alpha_i) \right)},
    \\
    \noalign{\bigskip}
    \mathfrak{r}_{\rm{int}} & = \frac{\displaystyle \sum_{0^{\circ} \le |\alpha_i| \le 90^{\circ}} \sin^2 (2\alpha_i)}{\displaystyle \sum_{|\alpha_i|} \left( 1 + \cos^2 (2\alpha_i) \right)},
    \\
    \noalign{\bigskip}
    \mathfrak{r}_{\perp} & = 2 \ \frac{\displaystyle \sum_{45^{\circ} \le |\alpha_i| \le 90^{\circ}} \cos^2 (2\alpha_i)}{\displaystyle \sum_{|\alpha_i|} \left( 1 + \cos^2 (2\alpha_i) \right)} \ ,
    \label{eq:perpara_ratios_bis}
\end{aligned}
\end{equation} 
with $\alpha_i = \psi_{{\rm f},i} - \psi_{B,i}$ defined in the range $[-90^{\circ}, 90^{\circ}]$. \\
These ratios have the sought-after properties: \\
- $\mathfrak{r}_{\parallel} + \mathfrak{r}_{\rm{int}} + \mathfrak{r}_{\perp} = 1$; \\
- $\mathfrak{r}_{\parallel} = \mathfrak{r}_{\rm{int}} = \mathfrak{r}_{\perp} = \frac{1}{3}$ for a uniform distribution; \\
- $\mathfrak{r}_{\parallel} > \mathfrak{r}_{\rm{int}}, \mathfrak{r}_{\perp}$ for a preferentially parallel distribution, with $\mathfrak{r}_{\parallel} = 1$ if all $\alpha_i = 0^{\circ}$; \\
- $\mathfrak{r}_{\rm{int}} > \mathfrak{r}_{\parallel}, \mathfrak{r}_{\perp}$ for a preferentially intermediate distribution, with $\mathfrak{r}_{\rm{int}} = 1$ if all $\alpha_i = \pm 45^{\circ}$; \\
- $\mathfrak{r}_{\perp} > \mathfrak{r}_{\parallel}, \mathfrak{r}_{\rm{int}}$ for a preferentially perpendicular distribution, with $\mathfrak{r}_{\perp} = 1$ if all $\alpha_i = \pm 90^{\circ}$; \\
- $\mathfrak{r}_{\parallel}, \mathfrak{r}_{\perp} > \mathfrak{r}_{\rm{int}}$ for a bimodal parallel-perpendicular distribution, with $\mathfrak{r}_{\parallel} = \mathfrak{r}_{\perp} = 0.5$ if the $\alpha_i$ are equally split between $\alpha_i = 0^{\circ}$ and $\alpha_i =\pm 90^{\circ}$.

Expressions for the uncertainties in $\mathfrak{r}_{\parallel}$, $\mathfrak{r}_{\rm{int}}$, and $\mathfrak{r}_{\perp}$ can be obtained from the general analytical formula
\begin{equation}
    \sigma^2 (\mathfrak{r}_{\star}) = \sum_i 
    \left( 
    \frac{\partial \mathfrak{r}_{\star}}{\partial x_i}
    \right)^2 \ \sigma^2 (x_i) \ ,
    \label{eq:perpara_ratios_bis_sigma2_formula}
\end{equation}
where subscript $\star$ stands for $\parallel$, $\rm{int}$, or $\perp$; $x_i = \cos^2 (2\alpha_i)$, and $\sigma (x_i)$ is the uncertainty in $x_i$.
The latter is the quadratic sum of the measurement error in $x_i$ and the statistical uncertainty in $x_i$, which can be approximated by the r.m.s. dispersion in the measured values of $x_i$, $\sigma_x$.
In practice, the statistical uncertainty gives the dominant contribution, and we may use $\sigma (x_i) = \sigma_x$.
Applying Eq.~\eqref{eq:perpara_ratios_bis_sigma2_formula} to Eq.~\eqref{eq:perpara_ratios_bis} successively yielded
\begin{equation}
\begin{aligned}
    \sigma^2 (\mathfrak{r}_{\parallel}) & = \frac{4}{(N + Y)^4} \ 
    \left[
    N_{<} \ (N + Y_{>})^2 + N_{>} \ Y_{<}^2
    \right] \ \sigma^2_x,
    \\
    \noalign{\bigskip}
    \sigma^2 (\mathfrak{r}_{\rm{int}}) & = \frac{4 \, N^3}{(N + Y)^4} \ \sigma^2_x,
    \\
    \noalign{\bigskip}
    \sigma^2 (\mathfrak{r}_{\perp}) & = \frac{4}{(N + Y)^4} \ 
    \left[
    N_{>} \ (N + Y_{<})^2 + N_{<} \ Y_{>}^2
    \right] \ \sigma^2_x \ ,
    \label{eq:perpara_ratios_bis_sigma2}
\end{aligned}
\end{equation}
where, as before, $N$ is the number of independent pixels, $Y = \sum_i x_i = \sum_i \cos^2 (2\alpha_i)$, and subscript $<$ [$>$] indicates that only pixels with $|\alpha_i| \le 45^{\circ}$ [$\ge 45^{\circ}$] are included.
Strictly speaking, Eq.~\eqref{eq:perpara_ratios_bis_sigma2_formula} and the ensuing Eq.~\eqref{eq:perpara_ratios_bis_sigma2} are valid only if $\mathfrak{r}_{\star}$ is approximately linear over the interval $[x_j - \sigma_x, x_j + \sigma_x]$ $\forall x_j$.
This is indeed the case, as the only nonlinearity arises from the term $x_j$ in the denominator of $\mathfrak{r}_{\star}$, and this term is much smaller than the denominator itself as soon as there are at least a dozen pixels.
We verified with random drawings that Eq.~\eqref{eq:perpara_ratios_bis_sigma2} is valid to good accuracy.

\subsection{Relative orientation analysis}
\label{sec:method_analysis}

Our study of HROs is divided into three parts. First, we analyze the 116 GCC fields individually, with their respective networks of filaments and HROs, and present an overview of the results obtained for this sample of maps (Sect.~\ref{sec:results_individual-fields}). Then, we consider all filaments extracted with {\tt FilDReaMS} over the 116 GCC fields and perform a combined analysis of relative orientations and their variations with a number of physical properties of the GCC fields and filaments (Sect.~\ref{sec:results2_FilamentHRO}). Finally, we focus on filaments with at least one embedded core from the catalog described in Sect.~\ref{sec:data_Herschel} \citep{Montillaud2015}, and perform the same combined analysis, additionally taking core properties into account (see Sect.~\ref{sec:results3_SourceHRO}).

When considering filament orientations, we start from the data cubes of filament $\psi_{\rm f}$ and $S$ extracted with {\tt FilDReaMS} for each $W_{\rm{b}}$, and for each pixel that belongs to at least one filament, we retain only the $W_{\rm{b}}$ and $\psi_{\rm{f}}$ values corresponding to the filament with the highest $S$, referring to them as $W_{\rm{b}}^{\star}$ and $\psi_{\rm{f}}^{\star}$, respectively. This procedure reduces the filament cubes to 2D maps of $W_{\rm{b}}^{\star}$ and $\psi_{\rm{f}}^{\star}$ for each considered pixel. We then compute the relative orientations between these filaments and \textbf{\textit{B}$_{\rm{PoS}}$}, given by $\psi_{\rm{f}}^{\star} - \psi_{B}$, and perform the HRO analysis described above.

In the final case, we focus on filaments with at least one embedded core from the GCC catalog \citep{Montillaud2015}. For each core, we considered the pixel corresponding to the center of the core, and we checked whether it lies within a filament. We then retained the orientation angle of the most significant filament associated with each core, $\psi_{\rm{f}}^{\star}$; the orientation of \textbf{\textit{B}$_{\rm{PoS}}$}, $\psi_B$; and the orientation angle of the core itself, $\psi_{\rm{c}}$. Hence, we built three different sets of HROs, for $\lvert\psi_{\rm{f}}^{\star} - \psi_{B}\rvert$, $\lvert\psi_{\rm{f}}^{\star} - \psi_{\rm{c}}\rvert$, and $\lvert\psi_{\rm{c}} - \psi_{B}\rvert$, respectively.


\section{Description of the samples}
\label{sec:properties}

We aim to better understand the physical agents that may influence the orientations of the considered structures and therefore their orientations relative to the magnetic field. Hence, we grouped the GCC fields, filaments, and cores according to physical parameters derived from \textit{Planck} and \textit{Herschel} data, and we performed our HRO analysis on each group, which we refer to as a sample. In the hypothesis that all the structures within a given \textit{Herschel} field interact with each other and may be dynamically linked, it is interesting to consider not only the properties of individual filaments or cores, but also those of the environment in which they form.

\subsection{\textit{Herschel}-GCC fields}
\label{sec:properties_field}

We built samples of GCC fields with similar physical properties and perform our HRO analysis on each sample to identify possible correlations between the regions' properties and preferred relative orientations. For each considered parameter, we examined the distribution of GCC fields and defined thresholds that separate different regimes or behaviors while ensuring that each sample contains enough fields to maintain statistical significance in the HRO analysis.

A key parameter to consider was distance, which enabled us to more accurately compare structures with similar physical scales. The field distances in our sample range from 110\,pc to 3.8\,kpc. We defined a distance threshold of 500\,pc, and we considered two samples of GCC fields with $d \leq 500\,{\rm{pc}}$ and $d > 500\,{\rm{pc}}$.

We also investigated the potential dependence on Galactic latitude. Fields at higher latitudes have weaker line of sight (LoS) confusion and background contamination compared to fields at lower latitudes, thus allowing for a more reliable analysis of individual regions. The field absolute latitudes, $\lvert b \rvert$, range from 1$^{\circ}$ to 70$^{\circ}$. We define two latitude thresholds and consider samples of fields with $\lvert b \rvert > 5^{\circ}$ and $\lvert b \rvert > 10^{\circ}$, to examine how the HROs evolve as the latitude threshold increases.

Another obvious parameter of interest is $N_{\rm{H_2}}$. We aim to determine whether there are differences in HROs for filaments in high-$N_{\rm{H_2}}$ or low-$N_{\rm{H_2}}$ environments. Here, we compute the average column density, ${\langle}N_{\rm{H_2}}{\rangle}$, of each GCC field. Based on the ${\langle}N_{\rm{H_2}}{\rangle}$ distribution, which ranges from 9$\,\times\,10^{19}\,{\rm{cm^{-2}}}$ to 8$\,\times\,10^{21}\,{\rm{cm^{-2}}}$, we define various thresholds to study different ${\langle}N_{\rm{H_2}}{\rangle}$ ranges. Two of the ${\langle}N_{\rm{H_2}}{\rangle}$ ranges are complementary, providing a full coverage of the sample, while the other ${\langle}N_{\rm{H_2}}{\rangle}$ ranges only cover the lowest and highest-${\langle}N_{\rm{H_2}}{\rangle}$ cases. Hence, we perform our HRO analysis on five samples of fields with: ${\langle}N_{\rm{H_2}}{\rangle} \leq 10^{21}\,{\rm{cm^{-2}}}$, ${\langle}N_{\rm{H_2}}{\rangle} \leq 1.5 \times 10^{21}\,{\rm{cm^{-2}}}$, ${\langle}N_{\rm{H_2}}{\rangle} > 1.5 \times 10^{21}\,{\rm{cm^{-2}}}$, ${\langle}N_{\rm{H_2}}{\rangle} > 2 \times 10^{21}\,{\rm{cm^{-2}}}$ and ${\langle}N_{\rm{H_2}}{\rangle} > 2.5 \times 10^{21}\,{\rm{cm^{-2}}}$.

Likewise, we investigated whether regions with higher polarization fractions show different behaviors in the HROs. To this end, we used the polarization fraction averaged over the GCC field, ${\langle}p{\rangle}$. From the ${\langle}p{\rangle}$ distribution, we defined a ${\langle}p{\rangle}$ threshold at 5\% and considered two samples of GCC fields with ${\langle}p{\rangle} \leq 5\%$ and ${\langle}p{\rangle} > 5\%$.

We extended our analysis by testing various combinations of the parameters and thresholds described above. This allowed us to refine the selection criteria for each sample of GCC fields in order to better understand their similarities and differences and how these relate to the observed HROs.

\subsection{Filaments}
\label{sec:properties_filament}
In the first part of our study, we consider all the filaments detected in the GCC fields and we examine their properties. We focus on two filament properties: the column density, $N_{\rm{H_2}}$, and the filament bar width, $W_{\rm{b}}$. Here, filament pixels are treated individually, with $N_{\rm{H_2}}$ and $W_{\rm{b}}$ varying from pixel to pixel.

\begin{figure}
   \centering
   \includegraphics[width=\hsize]{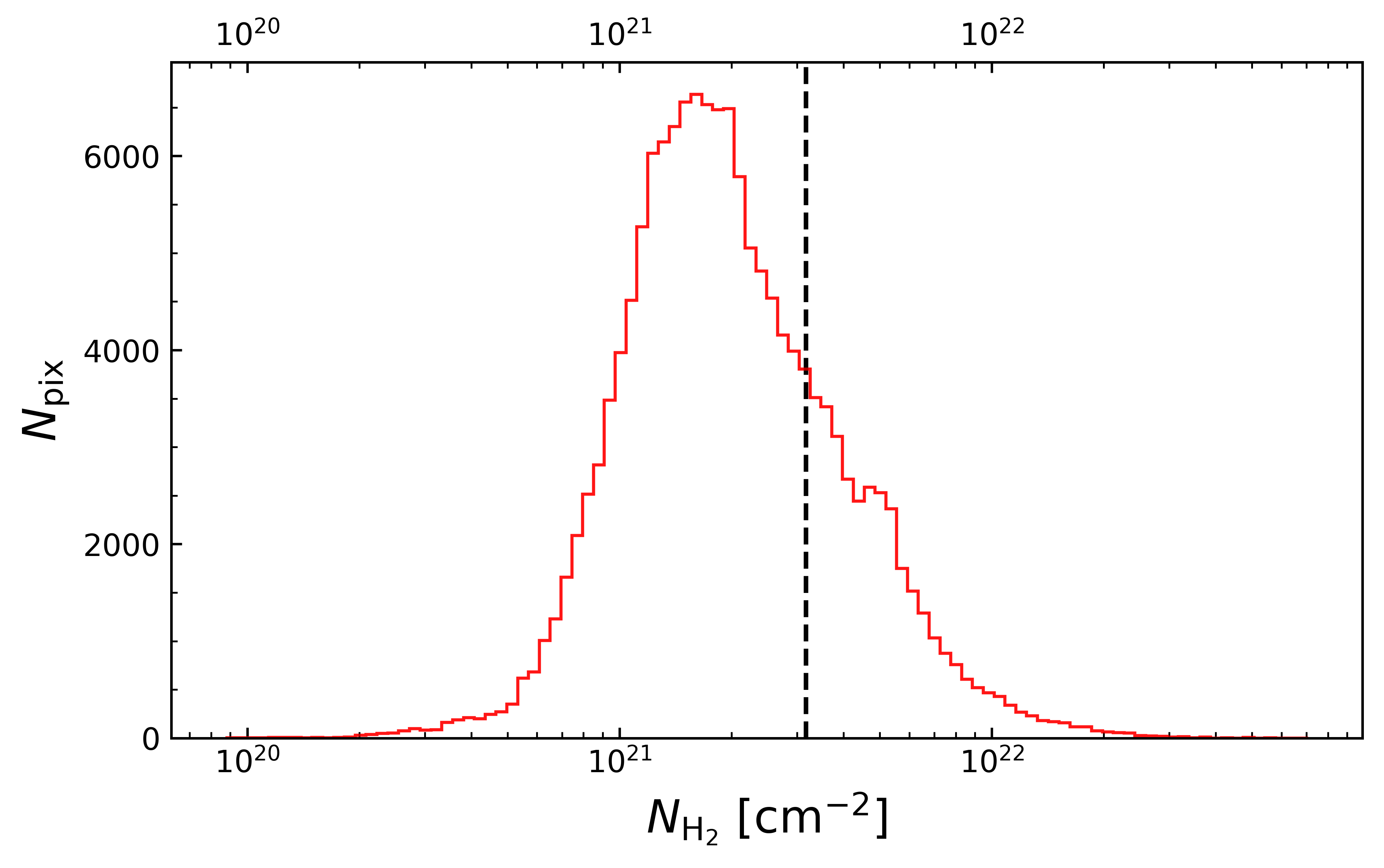}
   \includegraphics[width=\hsize]{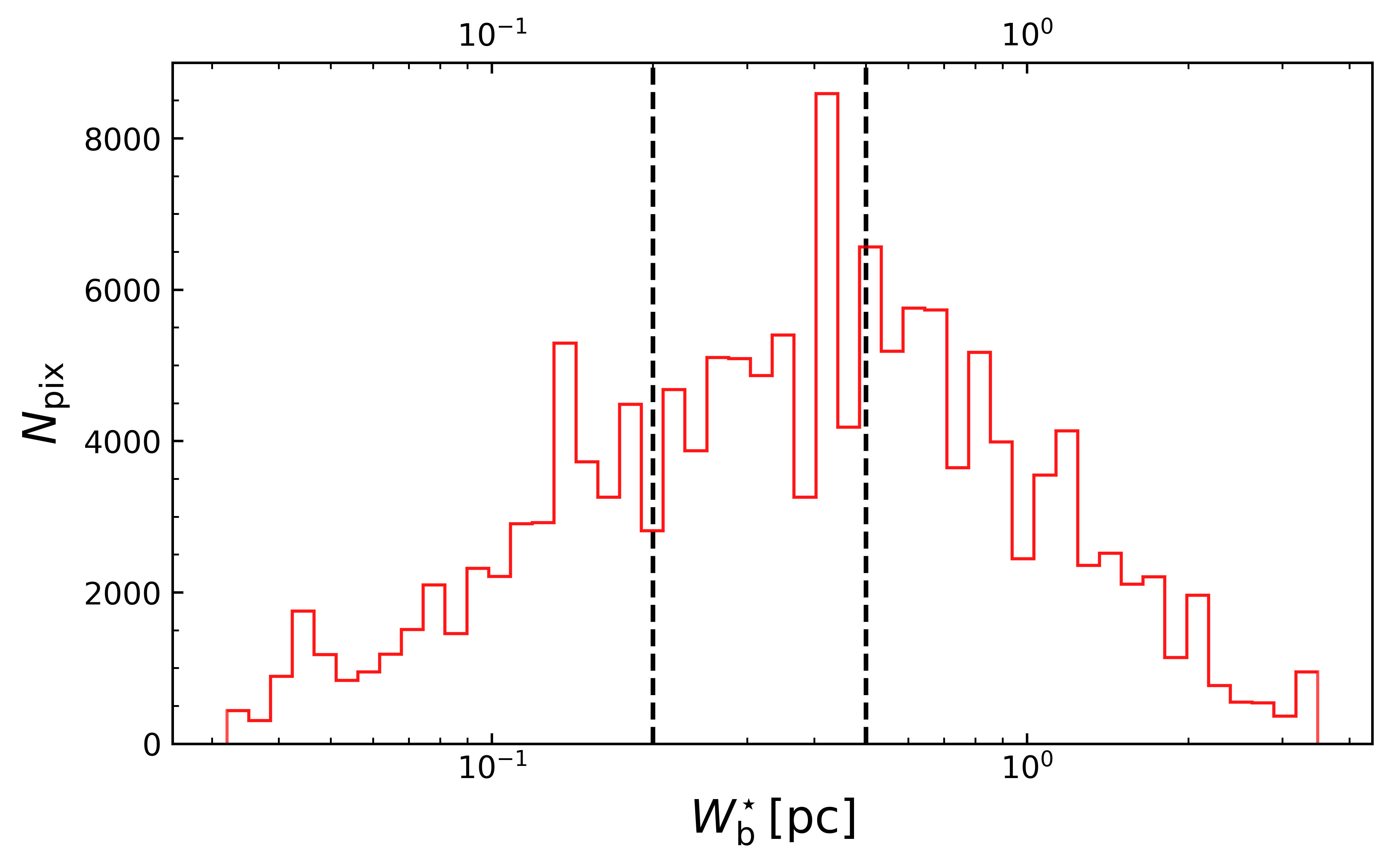}
      \caption{Histograms of filament properties derived from values in pixels belonging to filaments in all the GCC fields. \textbf{Top:} Histogram of the H$_2$ column density, $N_{\rm{H_2}}$. The dashed line represents the threshold at $10^{21.5}\,\rm{cm^{-2}}$ introduced in Sect.~\ref{sec:properties_filament}. \textbf{Bottom:} Histogram of the bar width of the most significant filament, $W_{\rm{b}}^{\star}$. The two dashed lines represent the thresholds at $0.2\,\rm{pc}$ and $0.5\,\rm{pc}$, respectively (Sect~\ref{sec:properties_filament}).}
         \label{fig:properties_filament_NH2_Wb_distribution}
   \end{figure}

To characterize the filament column density, we examine the distribution of $N_{\rm{H_2}}$ for filament pixels in all GCC fields shown in the top panel of Fig.~\ref{fig:properties_filament_NH2_Wb_distribution}. The distribution of filament $N_{\rm{H_2}}$ follows a log-normal shape, but with a pronounced tail toward high $N_{\rm{H_2}}$. The distribution ranges from 8.7$\,\times\,10^{19}\,{\rm{cm^{-2}}}$ to 7.1$\,\times\,10^{22}\,{\rm{cm^{-2}}}$, with a peak around 1.6$\,\times\,10^{21}\,{\rm{cm^{-2}}}$. We use $N_{\rm{H_2}}$ to split pixels in two different ways. First, we define a threshold at $10^{21.5}\,{\rm{cm^{-2}}}$ and consider two $N_{\rm{H_2}}$ bins, with $N_{\rm{H_2}} \leq 10^{21.5}\,{\rm{cm^{-2}}}$ and $N_{\rm{H_2}} > 10^{21.5}\,{\rm{cm^{-2}}}$. Second, as we described in Sect.~\ref{sec:method_hro}, we split pixels into equal-sized bins of $N_{\rm{H_2}}$ to closely examine how relative orientations evolve with $N_{\rm{H_2}}$.

Additionally, we used the bar width $W_{\rm{b}}^{\star}$ of the most significant filament passing through each pixel together with the distance to the corresponding GCC field to compute the effective physical width of the filament in \jonLEt{parsecs}. For this parameter, we examined the overall $W_{\rm{b}}^{\star}$ distribution as well as the $W_{\rm{b}}^{\star}$ distributions for nearby ($d \leq 500\,{\rm pc}$) and distant ($d > 500\,{\rm pc}$) GCC fields. We defined two thresholds at 0.2\,pc and 0.5\,pc. As such, we grouped the pixels into three samples of $W_{\rm{b}}^{\star}$, with $W_{\rm{b}}^{\star} \leq 0.2\,{\rm{pc}}$, $0.2\,{\rm{pc}} < W_{\rm{b}}^{\star} \leq 0.5\,{\rm{pc}}$, and $W_{\rm{b}}^{\star} > 0.5\,{\rm{pc}}$.

\subsection{Cores}
\label{sec:properties_source}
In the second part of our study, we focus on filaments with embedded cores from the GCC catalog \citep{Montillaud2015}. We aim to examine how the relative orientations vary in the vicinity of cores embedded within filaments, as well as possible correlations with the orientations of embedded cores. The catalog includes a total of 4466 cores. \cite{Montillaud2015} described the extraction process and provided a wide array of physical properties, which we use to categorize the cores and the corresponding pixels of their parent filaments. Out of the 4466 cores, 222 were identified as potential extra-galactic sources and were therefore excluded from our analysis. Hence, we retain the remaining 4244 cores in our analysis.

The catalog provides estimates of core masses, $M_{\rm c}$, which are useful to identify potentially massive cores associated with regions forming high-mass stars. This allowed us to investigate possible correlations between the relative orientations and high-mass or low-mass star formation.
We find a positive correlation between the masses and distances of cores (Fig.~\ref{fig_appendix:results_source_logmass_d_hist}), which is due to the physical sizes of the cores, beam dilution and integration along the LoS, all scaling with distance. Therefore we analyze the core mass distribution for the whole sample as well as for nearby ($d \leq 500\,{\rm{pc}}$) and distant ($d > 500\,{\rm{pc}}$) samples separately. We define a different mass threshold to split the pixels depending on the distance category. $M_{\rm c}$ range from 0.01 to 481$\,M_{\odot}$. For cores in nearby fields, we adopt a mass threshold at 0.5$\,M_{\odot}$. Otherwise, if the samples also include distant cores, we adopt a mass threshold at 5$\,M_{\odot}$.

We estimated the mean hydrogen volume density of cores, $n_{\rm{H,c}}$, by combining their masses and volumes. Cores in the catalog are described as ellipses, with measured angular sizes of the major and minor axes in the PoS. We estimated the core depths along the LoS by taking the root mean square of the major and minor axes, and use the LoS depths to estimate $n_{\rm{H,c}}$. We show the distribution of resulting $n_{\rm{H,c}}$ in Fig.~\ref{fig_appendix:sourcedensity_hist}. We defined a threshold at $10^{3.5}\,\rm{cm^{-3}}$ and considered two samples of cores with $n_{\rm{H,c}} \leq 10^{3.5}\,\rm{cm^{-3}}$ and $n_{\rm{H,c}} > 10^{3.5}\,\rm{cm^{-3}}$.

The catalog further provides a classification of cores as gravitationally bound or gravitationally unbound, based on a virial analysis (see \citealt{Montillaud2015} for more details). We use this classification to separate bound cores from unbound cores and perform our HRO analysis on each sample. Additionally, the catalog includes an estimate of the evolutionary stage of each core, distinguishing between starless and protostellar cores. This classification is derived by combining two methods, one using mid-IR data and the other using submillimeter data (see \citealt{Montillaud2015} for more details). We use this classification of the evolutionary stage and perform our HRO analysis on each sample.

We investigated possible correlations between the properties of cores, filaments, and GCC fields separately. We also extended our statistical analysis by testing each parameter described in this subsection as well as by combining them with those discussed in Sects.~\ref{sec:properties_field} and~\ref{sec:properties_filament}.


\section{Results}
\label{sec:results}

\begin{figure*}
\centering
\resizebox{\hsize}{!}{
\includegraphics{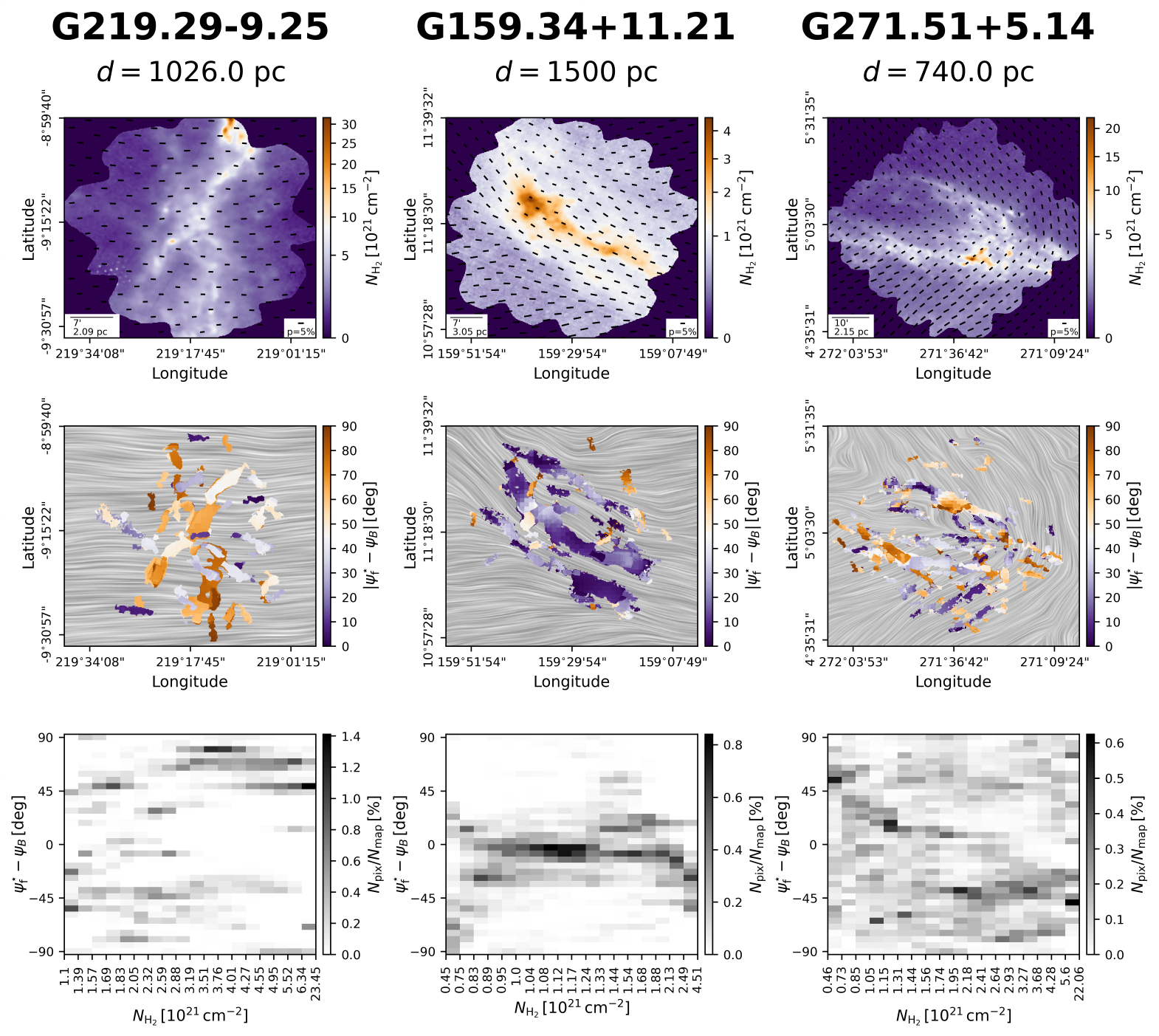}
}
\caption{Summary of the main results of our HRO analysis of three typical GCC fields: left column - G219.29-9.25; center column - G159.34+11.21; right column - G271.51+5.14. Top row: \textit{Herschel} $N_{\rm{H_2}}$ maps with \textbf{\textit{B}$_{\rm{PoS}}$} visualized using half-vectors with lengths proportional to the polarization fraction. Center row: Absolute relative orientation angle between the most significant filament and \textbf{\textit{B}$_{\rm{PoS}}$}, $\lvert\psi_{\rm{f}}^{\star} - \psi_{B}\rvert$. The \textbf{\textit{B}$_{\rm{PoS}}$} orientation is visualized using LIC in grayscale in the background. \textbf{Bottom row}: Two-dimensional HROs of ($\psi_{\rm{f}}^{\star} - \psi_{B}$) as functions of $N_{\rm{H_2}}$.}
\label{fig:results_indiv_stat}
\end{figure*}

In Sect.~\ref{sec:results_individual-fields}, we summarize our analysis of the individual GCC fields and present an overview of the results. In Sect.~\ref{sec:results2_FilamentHRO}, we present the statistical analysis, over all the GCC fields, of the filament-\textbf{\textit{B}$_{\rm{PoS}}$} relative orientations (see Sect.~\ref{sec:method_hro}) for each filament sample described in Sect.~\ref{sec:properties}. In Sect.~\ref{sec:results3_SourceHRO}, we present a similar statistical analysis, this time focusing on filaments with embedded cores from the GCC catalog \citep{Montillaud2015}.

\subsection{Individual GCC fields}
\label{sec:results_individual-fields}

We applied {\tt FilDReaMS} to the 116 GCC fields and performed our analysis of the relative orientations (Sect.~\ref{sec:method_hro}) for each GCC field individually. In Sect.~\ref{sec:results_3fields}, we present the results obtained for a sample of three GCC fields, chosen to illustrate different behaviors in filaments, \textbf{\textit{B}$_{\rm{PoS}}$}, and their relative orientations. In Sect.~\ref{sec:results_overview-indivfields}, we present an overview of the results obtained for the 116 GCC fields separately.

\subsubsection{Sample of three typical GCC fields}
\label{sec:results_3fields}

Figure~\ref{fig:results_indiv_stat} is a summary of the main results of our analysis on a sample of three GCC fields: G219.29-9.25 (hereby G219), in the left column; G159.34+11.21 (hereby G159), in the center column; and G271.51+5.14 (hereby G271), in the right column. In the top row, we find the \textit{Herschel} $N_{\rm{H_2}}$ map, with the orientation of \textbf{\textit{B}$_{\rm{PoS}}$} indicated using black half-vectors with lengths proportional to the polarization fraction, \textit{p}. The center row shows the most significant filaments reconstructed with {\tt FilDReaMS} and their absolute orientation angles relative to \textbf{\textit{B}$_{\rm{PoS}}$}, $\lvert\psi_{\rm{f}}^{\star} - \psi_{B}\rvert$, with the orientation of \textbf{\textit{B}$_{\rm{PoS}}$} visualized in the background using Line Integral Convolution (LIC) in grayscale. The bottom row shows the 2D HROs of ($\psi_{\rm{f}}^{\star} - \psi_{B}$) as functions of $N_{\rm{H_2}}$. Here, we build the 2D HROs in 36 angle bins of equal size from $-90^{\circ}$ to $90^{\circ}$ and in 18 $N_{\rm{H_2}}$ bins with equal number of pixels.

In G219 (Fig.~\ref{fig:results_indiv_stat}, left column), the top panel reveals a main filament spanning the whole region from north to south. \textbf{\textit{B}$_{\rm{PoS}}$} is well ordered, roughly orthogonal to the main filament, with ${\langle}p{\rangle}\,\approx\,6\%$. The center panel shows that most of the main filament was properly reconstructed and is oriented $\gtrsim 50^{\circ}$ relative to \textbf{\textit{B}$_{\rm{PoS}}$}. A few narrower filaments were detected around the main filament, most of which are roughly parallel or oblique to \textbf{\textit{B}$_{\rm{PoS}}$} ($\lvert\psi_{\rm{f}}^{\star} - \psi_{B}\rvert \lesssim 45^{\circ}$), and oriented close to perpendicular to the main filament. In the bottom panel, filaments have $N_{\rm{H_2}}$ ranging from $1.1\,\times\,10^{21}\,\rm{cm^{-2}}$ to $2.3\,\times\,10^{22}\,\rm{cm^{-2}}$. Lower-$N_{\rm{H_2}}$ filaments have a slight preference for relative orientations $\lesssim 45^{\circ}$, while higher-$N_{\rm{H_2}}$ filaments have relative orientations $\gtrsim 50^{\circ}$. The transition in relative orientations occurs at $N_{\rm{H_2}} \simeq\,2.9\,\times\,10^{21}\,\rm{cm^{-2}}$.

In G159 (center column), the top panel reveals a filament spanning the region from north-east to south-west. Here, \textbf{\textit{B}$_{\rm{PoS}}$} is ordered and well aligned with the main filament, with ${\langle}p{\rangle}\,\approx\,6.1\%$. The center panel shows that the main filament was properly reconstructed as well as two slightly narrower secondary filaments and a few narrower filaments in between. Most of the network of filaments is nearly parallel to \textbf{\textit{B}$_{\rm{PoS}}$} ($\lvert\psi_{\rm{f}}^{\star} - \psi_{B}\rvert \lesssim 30^{\circ}$). The 2D HRO (bottom panel) also shows that most filaments, with $N_{\rm{H_2}}$ ranging from $0.45\,\times\,10^{21}\,\rm{cm^{-2}}$ to $4.5\,\times\,10^{21}\,\rm{cm^{-2}}$, are oriented close to parallel to \textbf{\textit{B}$_{\rm{PoS}}$}.

G271 (right column) has a more complex morphology. $N_{\rm{H_2}}$ structures in this region are shaped as a horseshoe and extend westward, past the toe of the horseshoe. $N_{\rm{H_2}}$ peaks around the southern toe of the horseshoe, where the structures meet. Here, \textbf{\textit{B}$_{\rm{PoS}}$} is ordered in regions outside of the horseshoe structure and in the southern part of the horseshoe. \textbf{\textit{B}$_{\rm{PoS}}$} appears more tangled in the inner and northern parts of the horseshoe, and the polarization fraction drops compared to the two previous examples, with ${\langle}p{\rangle}\,\approx\,3.3\%$. The center panel shows that most of the detected filaments are aligned with each other. However, their orientations relative to \textbf{\textit{B}$_{\rm{PoS}}$} appear to be random, regardless of the filament size or the considered region. The bottom panel shows that filament $N_{\rm{H_2}}$ range from $0.46\,\times\,10^{21}\,\rm{cm^{-2}}$ to $2.2\,\times\,10^{22}\,\rm{cm^{-2}}$. Filaments at most $N_{\rm{H_2}}$ have random relative orientations, with a slight preference for $(\psi_{\rm{f}}^{\star} - \psi_{B}) \approx\,-45^{\circ}$ for $N_{\rm{H_2}} \gtrsim 1.74 \times 10^{21}\,{\rm cm^{-2}}$.

\subsubsection{Overview of the 116 GCC fields}
\label{sec:results_overview-indivfields}
We extend this analysis to each of the 116 GCC fields individually. The results are displayed in the form of an "identity card" (ID card) for each GCC field. In Appendix~\ref{sec:appendix_A}, we present the ID cards of the GCC fields analyzed in Sect.~\ref{sec:results_3fields}, while the rest of the ID cards can be found \jonLEt{on Zenodo}.

In this section, we present an overview of the results that emerge from the extended analysis. In each of the GCC fields, {\tt FilDReaMS} detected a network of filaments and substructures of different physical scales and $N_{\rm{H_2}}$. While the physical scales of filaments differ from one GCC field to another due to differences in distance, size of the map, or shape of the $N_{\rm H_2}$ structures, a few general trends emerged, with other less recurring trends depending on the considered GCC fields.

When numerous filaments are detected in a wide range of physical scales, they are intertwined and form a clear network. The widest detected structures in the GCC fields form either filaments or elongated portions of a cloud depending on the observed region. They are among the highest-$N_{\rm{H_2}}$ (i.e., highest-$N_{\rm{H_2}}$ bin) structures in each GCC field. If two or more wide structures are detected in the same GCC field, they are usually parallel. Medium-sized structures are detected at the edges of the wide structures, and they have lower $N_{\rm{H_2}}$. These medium-sized structures seem to connect with wide filaments at one end, and narrow, lower-$N_{\rm{H_2}}$ filaments at the other end. They are also oriented oblique or perpendicular to the wider structures.

We detect two main types of narrow filaments. The first type of narrow filaments appear around the wider structures, all over each of the GCC fields. They are either completely isolated or seem to connect with medium-sized filaments. They have lower $N_{\rm{H_2}}$ than the medium-sized filaments and are more blended into the background. While most of these narrow filaments seem to have random orientations, there is a tendency toward oblique or perpendicular orientations relative to the wider structures that they surround. In a few regions, these narrow filaments seem to line up and form striation patterns. The second type of narrow filaments is comprised of crests, which we define as the high-$N_{\rm{H_2}}$ spines of wide filaments, or clumps, which we define as high-$N_{\rm{H_2}}$, compact structures that can either be isolated or inside a wider filament. Crests are roughly aligned with their parent filaments. Clumps inside filaments are mostly aligned with their parent filaments, though with a higher dispersion in orientations, and isolated clumps have more random orientations.

Furthermore, in agreement with previous studies, we confirm that in most of the GCC fields, small-$W_{\rm{b}}^{\star}$, low-$N_{\rm{H_2}}$ filaments (lower $N_{\rm{H_2}}$ bins of a GCC field) are mostly parallel to \textbf{\textit{B}$_{\rm{PoS}}$}, while high-$N_{\rm{H_2}}$ filaments (higher $N_{\rm{H_2}}$ bins of a GCC field) at all scales range from mostly oblique to perpendicular to \textbf{\textit{B}$_{\rm{PoS}}$} (see, e.g., G219, left column of Figs.~\ref{fig:results_indiv_stat} and~\ref{fig_appendix:G219}). In most cases with this dual distribution of relative orientations, we observe a transition in relative orientations from roughly parallel at low-$N_{\rm{H_2}}$ to roughly perpendicular at high-$N_{\rm{H_2}}$ by checking whether $\mathfrak{r}_{\parallel}\,>\,\mathfrak{r}_{\perp},\mathfrak{r}_{\rm int}$ at low $N_{\rm{H_2}}$ and $\mathfrak{r}_{\perp}\,>\,\mathfrak{r}_{\parallel},\mathfrak{r}_{\rm int}$ at high $N_{\rm{H_2}}$. This $N_{\rm{H_2}}$ transition, $(N_{\rm{H_2}})_{\rm{t}}$, is clearly identified in 40$\%$ of the GCC fields, and the corresponding $(N_{\rm{H_2}})_{\rm{t}}$ typically lie in the range $[0.8$, $8]\,\times\,10^{21}\,\rm{cm^{-2}}$. In about 20$\%$ of cases, the $N_{\rm{H_2}}$ range is possibly not suitable to clearly observe a transition in relative orientations.

However, nearly 40$\%$ of the GCC fields do not show this usual trend between filaments and \textbf{\textit{B}$_{\rm{PoS}}$}. About $15\%$ of the GCC fields exhibit the reverse trend, where small-$W_{\rm{b}}^{\star}$, low-$N_{\rm{H_2}}$ filaments are mostly perpendicular to \textbf{\textit{B}$_{\rm{PoS}}$} ($\mathfrak{r}_{\perp}\,>\,\mathfrak{r}_{\parallel},\mathfrak{r}_{\rm int}$), while high-$N_{\rm{H_2}}$ filaments at all scales are mostly parallel to \textbf{\textit{B}$_{\rm{PoS}}$} ($\mathfrak{r}_{\parallel}\,>\,\mathfrak{r}_{\perp},\mathfrak{r}_{\rm int}$; e.g., G92.04+3.93, see Fig.~\ref{fig_appendix:G92_04}). In most of these cases, this reverse transition occurs for $(N_{\rm{H_2}})_{\rm{t}}$ in the range $[1, 6]\,\times\,10^{21}\,\rm{cm^{-2}}$, which matches the $(N_{\rm{H_2}})_{\rm{t}}$ range for the more common parallel to perpendicular transition. About $15\%$ of GCC fields show another trend where all filaments at all scales are mostly parallel to \textbf{\textit{B}$_{\rm{PoS}}$} (e.g., G159, center column of Fig.~\ref{fig:results_indiv_stat} and Fig.~\ref{fig_appendix:G159}). Finally, another $10\%$ of GCC fields have more random relative orientations (see, e.g., G271, right column of Fig.~\ref{fig:results_indiv_stat} and Fig.~\ref{fig_appendix:G271}).

The various behaviors we see in our analysis suggest that magnetic fields play a role in filament formation, but the interplay between magnetic fields and filament evolution is not as simple. Understanding each behavior and its origin is key to gaining insight into star formation. This motivates the statistical analysis we present in Sects.~\ref{sec:results2_FilamentHRO} and~\ref{sec:results3_SourceHRO}, where we strive to identify possible correlations between the relative orientations and the physical properties introduced in Sect.~\ref{sec:properties}.

\subsection{Statistical analysis of all filaments over all the GCC fields}
\label{sec:results2_FilamentHRO}

\begin{figure}
   \centering
   \includegraphics[width=\hsize]{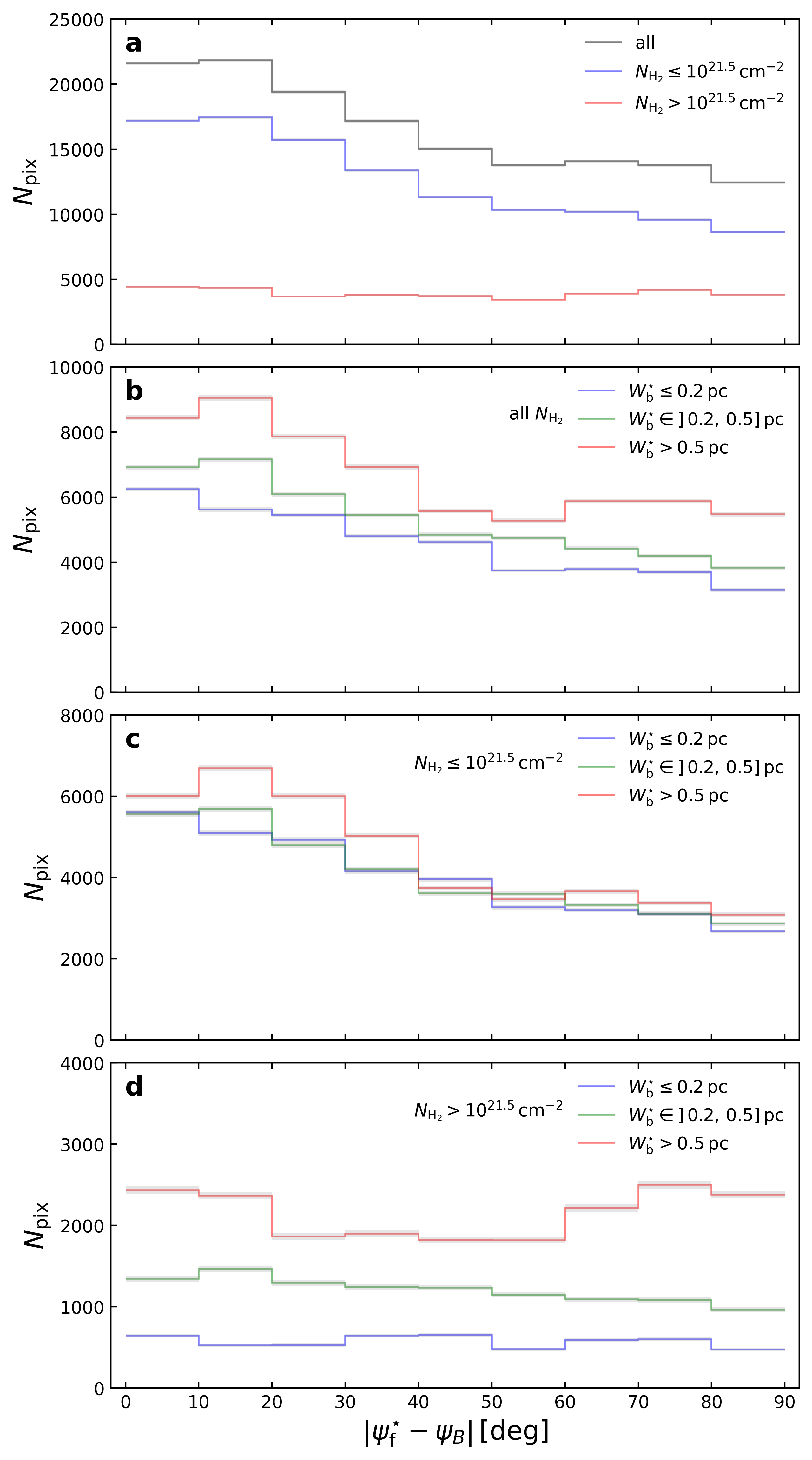}
      \caption{Histograms of relative orientations for filaments in all GCC fields. \textbf{(a)} HROs of all pixels (black), pixels with $N_{\rm{H_2}} \leq 10^{21.5}\,\rm{cm^{-2}}$ (blue), and pixels with $N_{\rm{H_2}} > 10^{21.5}\,\rm{cm^{-2}}$ (red) (see Sect.~\ref{sec:properties_filament}). \textbf{(b)}, \textbf{(c)}, and \textbf{(d)} show a decomposition of the three HROs of \textbf{(a)} for all $N_{\rm{H_2}}$, $N_{\rm{H_2}} \leq 10^{21.5}\,\rm{cm^{-2}}$, and $N_{\rm{H_2}} > 10^{21.5}\,\rm{cm^{-2}}$, respectively. The HROs are decomposed into the three $W_{\rm{b}}^{\star}$ bins introduced in Sect.~\ref{sec:properties_filament}: $W_{\rm{b}}^{\star} \leq 0.2\,\rm{pc}$ (blue), $0.2\,\rm{pc} < W_{\rm{b}}^{\star} \leq 0.5\,\rm{pc}$ (green), and $W_{\rm{b}}^{\star} > 0.5\,\rm{pc}$ (red). The uncertainties in all HROs are shown in gray shaded areas.}
         \label{fig:results_filamentHRO_all}
   \end{figure}

We aim to determine what physical parameters have an impact on filament orientations relative to \textbf{\textit{B}$_{\rm{PoS}}$} in order to better explain the different trends identified in Sect.~\ref{sec:results_overview-indivfields}. Hence, we perform a statistical analysis of the relative orientations of all filaments extracted from the 116 GCC fields and sample them according to the different parameters introduced in Sect.~\ref{sec:properties}. In this part of the study, we degraded the resolution of our sample to 1 pixel every 36$^{\prime\prime}$ (down from 1 pixel every 12$^{\prime\prime}$) to properly estimate the uncertainties in our HROs (see Sect.~\ref{sec:method_hro}). The values of both sets of ratios defined in Eqs.~\eqref{eq:perpara_ratios} and~\eqref{eq:perpara_ratios_bis}, as well as the associated uncertainties (see Sect.~\ref{sec:method_ratios}) for the samples presented in this section are summarized in Table~\ref{tab:perpara_ratios_values}, which is used to directly identify and compare trends between samples.

\begin{figure}
   \centering
   \includegraphics[width=\hsize]{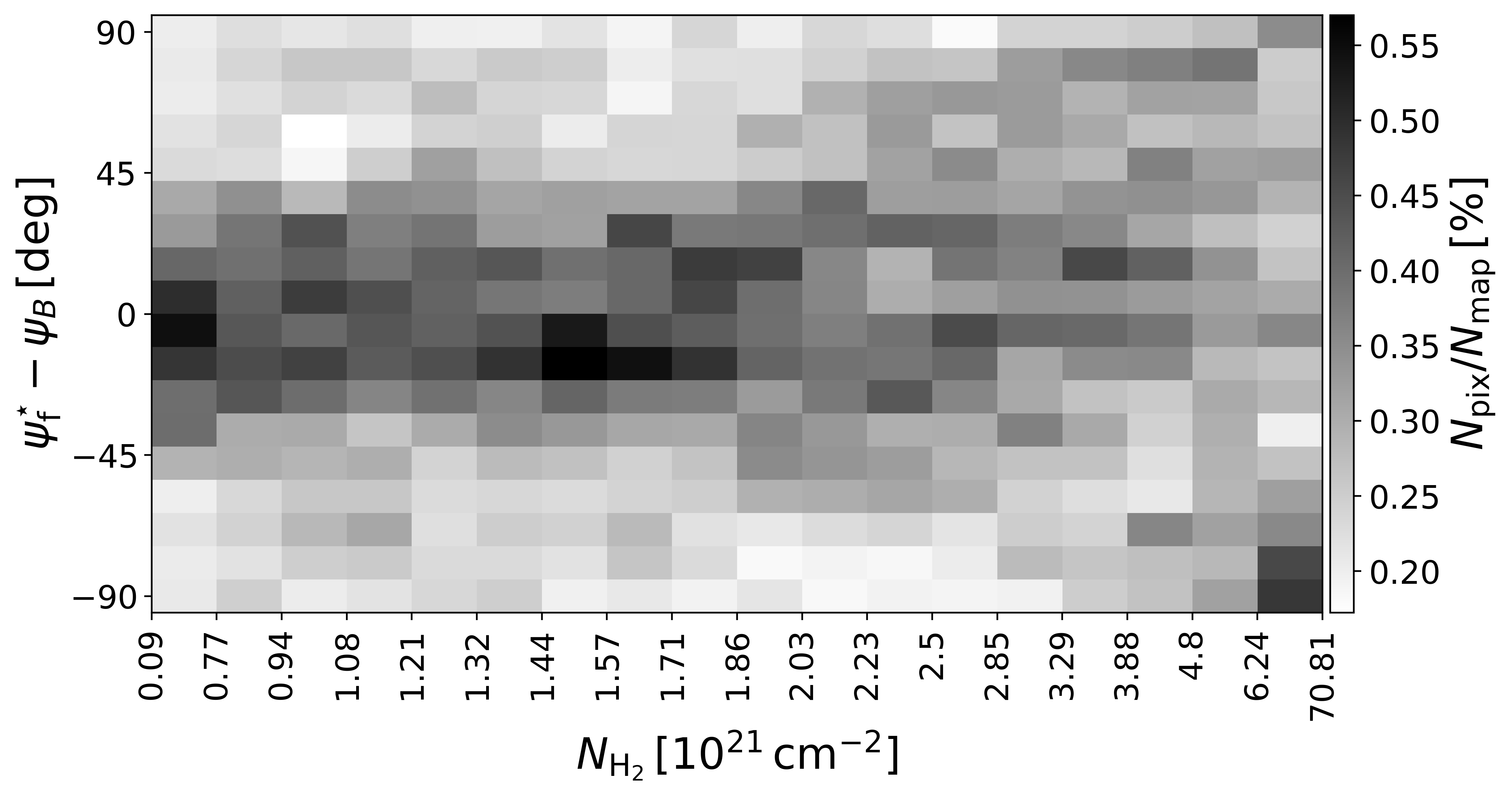}
   \includegraphics[width=\hsize]{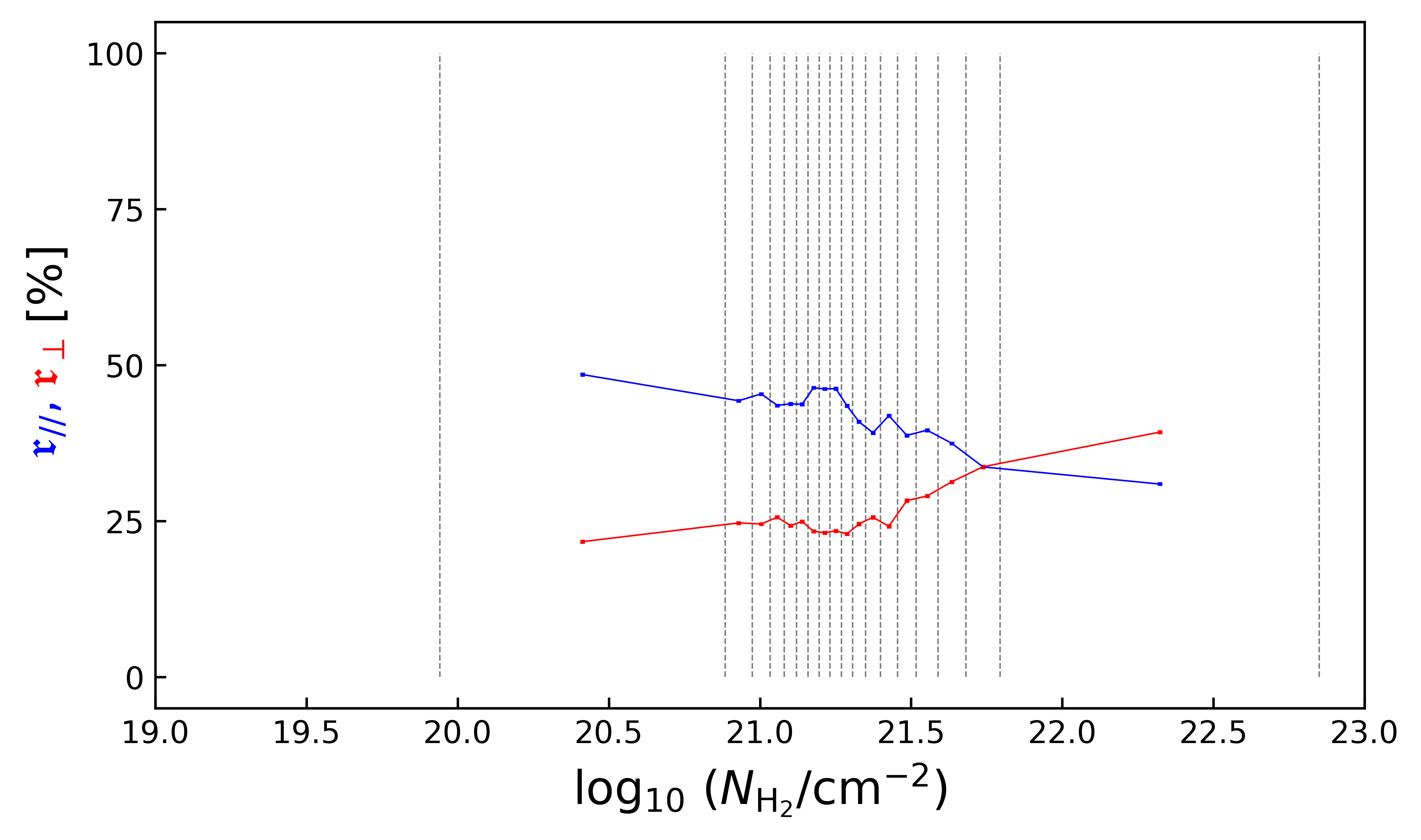}
      \caption{Measures of the relative orientations in 18 $N_{\rm{H_2}}$ bins with an equal number of pixels for the full sample of GCC fields. \textbf{Top}: Two-dimensional HRO of ($\psi_{\rm{f}}^{\star} - \psi_{B}$) in 18 angle bins. \textbf{Bottom:} Values of $\mathfrak{r}_{\parallel}$ and $\mathfrak{r}_{\perp}$ (defined in Eq.~\eqref{eq:perpara_ratios_bis}) in the 18 $N_{\rm{H_2}}$ bins. The $N_{\rm{H_2}}$ bins are delimited by the dashed vertical lines.}
         \label{fig:results_filament2DHRO_all}
   \end{figure}

Figure~\ref{fig:results_filamentHRO_all} shows histograms of the relative orientations between filaments and \textbf{\textit{B}$_{\rm{PoS}}$} for the entire sample of GCC fields, and for samples depending on filament $N_{\rm H_2}$ and $W_{\rm{b}}^{\star}$. Panel (a) shows the HRO for all filaments, as well as the HROs for filaments subdivided into two $N_{\rm{H_2}}$ bins with a threshold at $10^{21.5}\,\rm{cm^{-2}}$. Overall, the relative orientations are not uniform, with a preference for nearly parallel relative orientations. The lower-$N_{\rm{H_2}}$ filaments ($N_{\rm{H_2}}\,\leq\,10^{21.5}\,{\rm cm^{-2}}$) have a similar distribution, while the higher-$N_{\rm{H_2}}$ filaments ($N_{\rm{H_2}}\,>\,10^{21.5}\,{\rm cm^{-2}}$) have a nearly flat HRO. Panel (b) shows the HROs for filaments subdivided into three $W_{\rm{b}}^{\star}$ bins with thresholds at 0.2\,pc and 0.5\,pc (see Fig.~\ref{fig:properties_filament_NH2_Wb_distribution}). We see that $\mathfrak{r}_{\perp}$ increases as $W_{\rm{b}}^{\star}$ increases, implying that perpendicular orientations become relatively more prevalent.
Panels (c) and (d) show a combination of both criteria: the three $W_{\rm{b}}^{\star}$ bins for $N_{\rm{H_2}} \leq 10^{21.5}\,\rm{cm^{-2}}$ (panel (c)) and for $N_{\rm{H_2}} > 10^{21.5}\,\rm{cm^{-2}}$ (panel (d)), respectively. For the lower-$N_{\rm{H_2}}$ filaments, there is a preference for nearly parallel orientations regardless of the $W_{\rm{b}}^{\star}$ range. However, for the higher-$N_{\rm{H_2}}$ filaments, results depend on $W_{\rm{b}}^{\star}$: high-$N_{\rm{H_2}}$, small-$W_{\rm{b}}^{\star}$ filaments ($W_{\rm{b}}^{\star}\,\leq\,0.2\,{\rm pc}$) have a flat HRO; high-$N_{\rm{H_2}}$, intermediate-$W_{\rm{b}}^{\star}$ filaments are slightly more parallel to \textbf{\textit{B}$_{\rm{PoS}}$}; and high-$N_{\rm{H_2}}$, large-$W_{\rm{b}}^{\star}$ filaments ($W_{\rm{b}}^{\star}\,>\,0.5\,{\rm pc}$) have a slightly bimodal parallel-perpendicular distribution.

Figure~\ref{fig:results_filament2DHRO_all} displays the relative orientations in 18 $N_{\rm{H_2}}$ bins of equal number of pixels (see Sect.~\ref{sec:method_hro}). The top panel exhibits the 2D HRO of ($\psi_{\rm{f}}^{\star} - \psi_{B}$) as a function of $N_{\rm{H_2}}$ (see Sect.~\ref{sec:method_hro}). We see that low-$N_{\rm{H_2}}$ filaments are mostly parallel to \textbf{\textit{B}$_{\rm{PoS}}$}, and the distribution of ($\psi_{\rm{f}}^{\star} - \psi_{B}$) fans out to all orientations for $N_{\rm{H_2}}\,\gtrsim\,2\times10^{21}\,{\rm{cm^{-2}}}$, with a preference toward $\approx\,-90^{\circ}$ for $N_{\rm{H_2}}\,\geq\,6.2\times10^{21}\,{\rm{cm^{-2}}}$. The bottom panel shows the values of $\mathfrak{r}_{\parallel}$ and $\mathfrak{r}_{\perp}$ (Eq.~\eqref{eq:perpara_ratios_bis}) in each $N_{\rm{H_2}}$ bin. Again, we see that in lower-$N_{\rm{H_2}}$ bins, filaments are mostly parallel to \textbf{\textit{B}$_{\rm{PoS}}$}, with $\mathfrak{r}_{\parallel}$ reaching $\gtrsim\,47\%$ and $\mathfrak{r}_{\perp}\,\lesssim\,24\%$ for $N_{\rm{H_2}}\,\leq\,10^{21}\,{\rm{cm^{-2}}}$. $\mathfrak{r}_{\parallel}$ decreases and $\mathfrak{r}_{\perp}$ increases with increasing $N_{\rm{H_2}}$, and filaments transition from being mostly parallel to mostly perpendicular to \textbf{\textit{B}$_{\rm{PoS}}$} where $\mathfrak{r}_{\parallel}$ and $\mathfrak{r}_{\perp}$ cross, at $(N_{\rm{H_2}})_{\rm{t}}\,\simeq\,10^{21.7}\,{\rm{cm^{-2}}}$.

We repeat the above analysis subsequently for the samples of filaments that meet the various criteria and combinations of criteria listed in Sect.~\ref{sec:properties}. When combining criteria, we also check for possible correlations between the criteria to better compare their impact on the relative orientations. Here, we only present the main results and the main trends that emerge from this statistical analysis.

We separated our GCC fields into two samples according to distance. Overall, distant fields ($d > 500\,$pc) have higher ${\langle}N_{\rm{H_2}}{\rangle}$ than nearby fields ($d \leq 500\,$pc). In both samples, we find that low-$N_{\rm{H_2}}$ filaments at all scales are mostly parallel to \textbf{\textit{B}$_{\rm{PoS}}$}, but the trend is more pronounced for nearby fields. However, we do not recover the $N_{\rm{H_2}}$ transition from parallel to perpendicular orientations for nearby fields. In distant fields, high-$N_{\rm{H_2}}$, large-$W_{\rm{b}}^{\star}$ filaments have a bimodal distribution, with a clear transition from parallel to perpendicular orientations with increasing $N_{\rm{H_2}}$. High-latitude fields exhibit similar trends to nearby fields.

Next, we built five samples of GCC fields with different ranges of ${\langle}N_{\rm{H_2}}{\rangle}$, as explained in Sect.~\ref{sec:properties_field}. We find a positive correlation between ${\langle}N_{\rm{H_2}}{\rangle}$ and the number of high-$N_{\rm{H_2}}$ ($N_{\rm{H_2}}\,>\,10^{21.5}\,{\rm cm^{-2}}$) pixels within filaments.
Additionally, low-$N_{\rm{H_2}}$ filaments are mostly parallel to \textbf{\textit{B}$_{\rm{PoS}}$} while high-$N_{\rm{H_2}}$ filaments have a flat or slightly more perpendicular distribution. However with increasing ${\langle}N_{\rm{H_2}}{\rangle}$, the HROs become increasingly flat with a slight parallel-perpendicular bi-modality. The trends are more pronounced in the lowest ${\langle}N_{\rm{H_2}}{\rangle}$ range, where high-$N_{\rm{H_2}}$ filaments are more contrasted with respect to their environment. As ${\langle}N_{\rm{H_2}}{\rangle}$ increases, there are more filaments perpendicular to \textbf{\textit{B}$_{\rm{PoS}}$}, but the preferred orientations are less clear. In each ${\langle}N_{\rm{H_2}}{\rangle}$ range, we find a transition from parallel to perpendicular orientations for $(N_{\rm{H_2}})_{\rm{t}}$ between $10^{21.5}$ and $10^{21.8}\,{\rm{cm^{-2}}}$ (see Figs.~\ref{fig_appendix:results_filamentperparaHRO_NH2avg-},~\ref{fig_appendix:results_filamentperparaHRO_NH2avg+} and~\ref{fig_appendix:results_filamentperparaHRO_NH2avg++} in Appendix~\ref{sec:appendix_C}).

\begin{figure}
   \centering
   \includegraphics[width=\hsize]{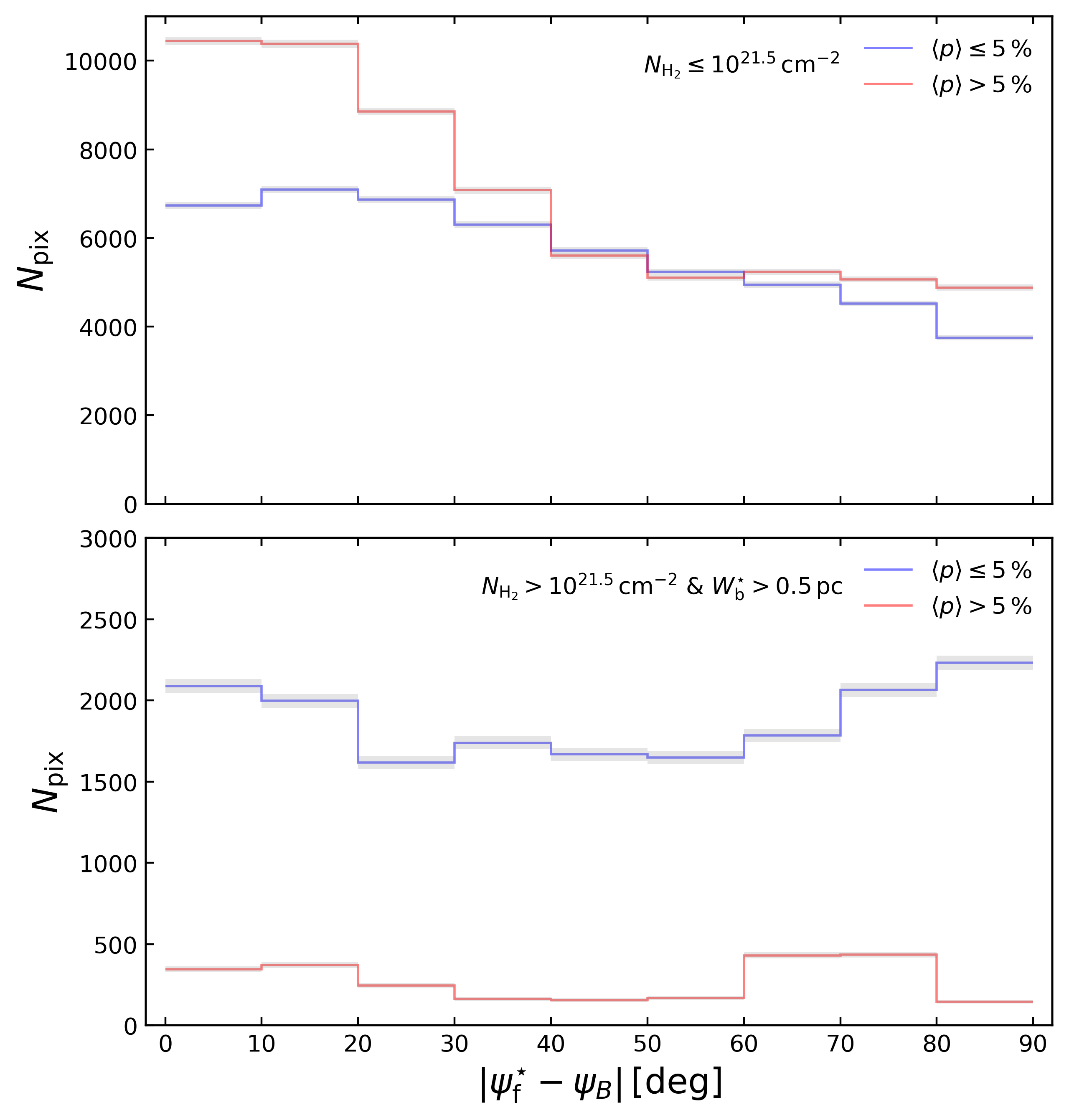}
      \caption{Histograms of relative orientations for filaments in two samples of GCC fields with ${\langle}p{\rangle} \leq 5\%$ (blue) and ${\langle}p{\rangle} > 5\%$ (red). The uncertainties are shown in gray shaded areas. \textbf{Top:} HROs of pixels with $N_{\rm{H_2}} \leq 10^{21.5}\,\rm{cm^{-2}}$. \textbf{Bottom:} HROs of pixels with $N_{\rm{H_2}} > 10^{21.5}\,\rm{cm^{-2}}$ and $W_{\rm{b}}^{\star} > 0.5\,\rm{pc}$.}
         \label{fig:results_filamentHRO_bothp}
   \end{figure}

Similarly, we built two samples of GCC fields according to ${\langle}p{\rangle}$ with a threshold at $5\%$. Figure~\ref{fig:results_filamentHRO_bothp} shows HROs for filaments in GCC fields with either low ${\langle}p{\rangle}$ (${\langle}p{\rangle} \leq 5\%$) or high ${\langle}p{\rangle}$ (${\langle}p{\rangle} > 5\%$. In the top panel, we see HROs of pixels with $N_{\rm{H_2}} \leq 10^{21.5}\,\rm{cm^{-2}}$. We recover the trend that low-$N_{\rm{H_2}}$ filaments are mostly parallel to \textbf{\textit{B}$_{\rm{PoS}}$}, yet the trend is weaker in low-${\langle}p{\rangle}$ GCC fields, with $\mathfrak{r}_{\parallel} = 40.5\%$, and more pronounced in high-${\langle}p{\rangle}$ GCC fields, with $\mathfrak{r}_{\parallel} = 46.5\%$ (see Table~\ref{tab:perpara_ratios_values}). In the bottom panel, we see HROs of pixels with $N_{\rm{H_2}} > 10^{21.5}\,\rm{cm^{-2}}$ and $W_{\rm{b}}^{\star} > 0.5\,\rm{pc}$. We recover the trend that high-$N_{\rm{H_2}}$, large-$W_{\rm{b}}^{\star}$ filaments in low-${\langle}p{\rangle}$ GCC fields have a nearly flat HRO. However, we derive slightly different trends for high-$N_{\rm{H_2}}$, large-$W_{\rm{b}}^{\star}$ filaments in high-${\langle}p{\rangle}$ GCC fields depending on the set of ratios used. With the first set of ratios (Eq.~\eqref{eq:perpara_ratios}), we find $r_{\parallel} = 39.1\%$ and $r_{\perp} = 41.1\%$ (see Table~\ref{tab:perpara_ratios_values}), which implies a bimodal parallel-perpendicular distribution. With the second set of ratios (Eq.~\eqref{eq:perpara_ratios_bis}), we find $\mathfrak{r}_{\parallel} = 37.8\%$ and $\mathfrak{r}_{\perp} = 34.9\%$ (see Table~\ref{tab:perpara_ratios_values}), which implies no preference for perpendicular relative orientations. The difference between the two sets of ratios is due to the lack of filaments with $|\psi_{\rm{f}}^{\star} - \psi_{B}| \in [80^{\circ}, 90^{\circ}]$. Overall, each sample of filaments inside high-${\langle}p{\rangle}$ GCC fields exhibits similar results to the full sample, but trends are usually more pronounced, with more filaments being either parallel or perpendicular to \textbf{\textit{B}$_{\rm{PoS}}$}.

\begin{figure}
   \centering
   \includegraphics[width=\hsize]{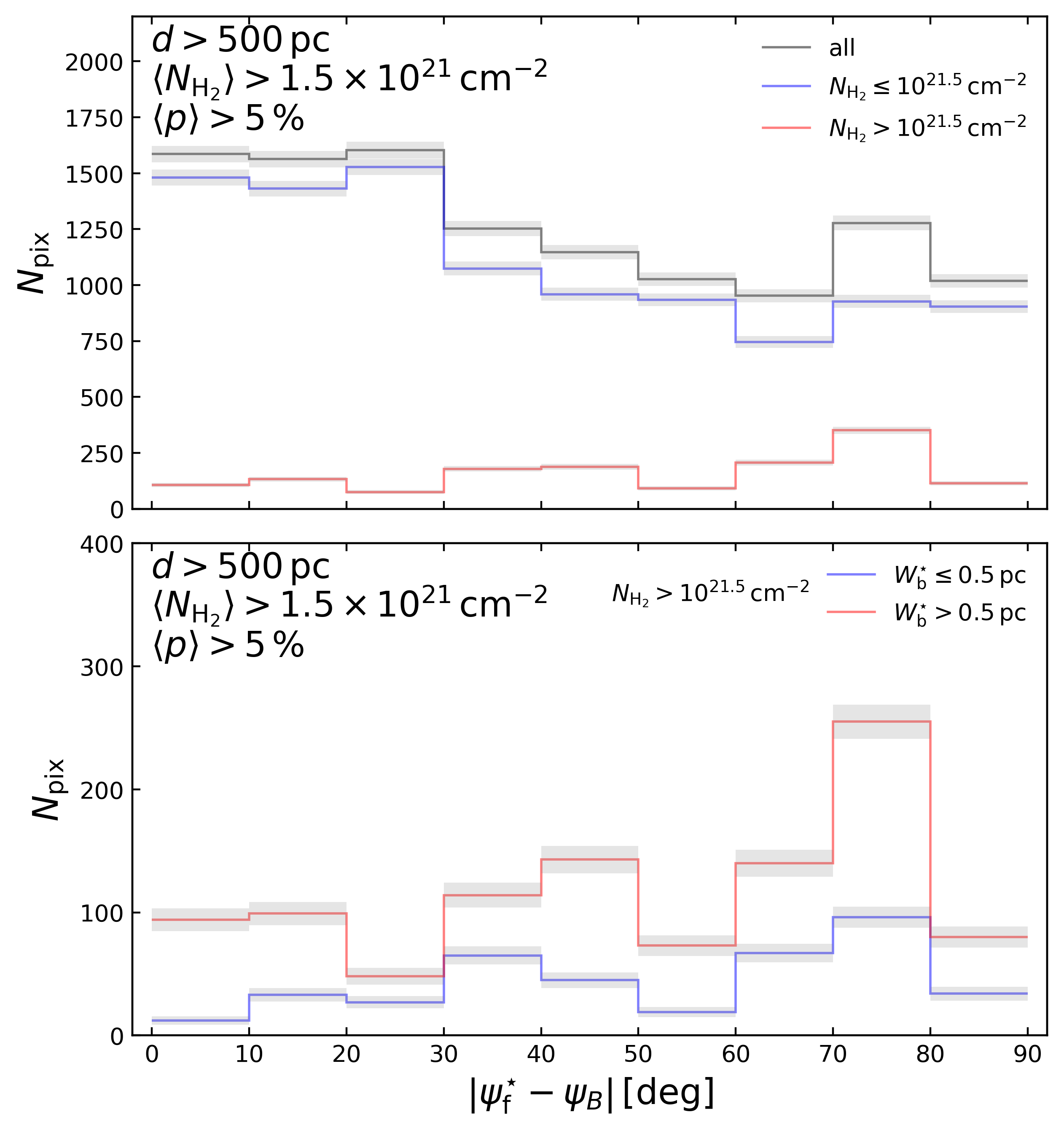}
      \caption{Histograms of relative orientations for filaments in GCC fields with $d > 500\,{\rm{pc}}$, ${\langle}N_{\rm{H_2}}{\rangle} > 1.5 \times 10^{21}\,{\rm{cm^{-2}}}$, and ${\langle}p{\rangle} > 5\%$. The uncertainties are shown in gray shaded areas. \textbf{Top:} Similar HROs to Fig.~\ref{fig:results_filamentHRO_all}a. \textbf{Bottom:} Similar HROs to Fig.~\ref{fig:results_filamentHRO_all}d, except that the HRO is decomposed into two $W_{\rm{b}}^{\star}$ bins: $W_{\rm{b}}^{\star} \leq 0.5\,\rm{pc}$ (blue) and $W_{\rm{b}}^{\star} > 0.5\,\rm{pc}$ (red).}
         \label{fig:results_filamentHRO_d+_NH2+_p+}
   \end{figure}

We then combined the different criteria to examine their possible joint impact on the HROs, while keeping sufficiently large samples for statistical significance. For example, Fig.~\ref{fig:results_filamentHRO_d+_NH2+_p+} displays HROs for filaments in GCC fields combining three criteria on GCC field properties: $d > 500\,{\rm{pc}}$, ${\langle}N_{\rm{H_2}}{\rangle} > 1.5 \times 10^{21}\,{\rm{cm^{-2}}}$ and ${\langle}p{\rangle} > 5\%$. The top panel shows the HRO for all filaments for this sample, as well as the HROs for filaments subdivided into two $N_{\rm H_2}$ bins with a threshold at $10^{21.5}\,{\rm{cm^{-2}}}$. We find that low-$N_{\rm H_2}$ filaments are mostly parallel to \textbf{\textit{B}$_{\rm{PoS}}$}, while high-$N_{\rm H_2}$ filaments are mostly perpendicular to \textbf{\textit{B}$_{\rm{PoS}}$}. The bottom panel exhibits the HRO for high-$N_{\rm H_2}$ filaments subdivided into two $W_{\rm{b}}^{\star}$ bins with a threshold at 0.5\,pc, set for distant fields. In both cases, we find that filaments are mostly perpendicular to \textbf{\textit{B}$_{\rm{PoS}}$}, with $\mathfrak{r}_{\perp} = 46.9\%$ and $\mathfrak{r}_{\perp} = 42.6\%$ for small-$W_{\rm{b}}^{\star}$ and large-$W_{\rm{b}}^{\star}$ filaments, respectively (see Table~\ref{tab:perpara_ratios_values}). The perpendicular trend for high-$N_{\rm H_2}$ filaments in $d > 500\,{\rm{pc}}$, ${\langle}N_{\rm{H_2}}{\rangle} > 1.5 \times 10^{21}\,{\rm{cm^{-2}}}$ and ${\langle}p{\rangle} > 5\%$ GCC fields is more pronounced compared to other combinations of criteria.

The parameters with the strongest effects on the HROs are the filament $N_{\rm{H_2}}$, $W_{\rm{b}}^{\star}$ and the GCC field ${\langle}p{\rangle}$, while other parameters play a smaller role. \jonLEt{Here, we summarize our key findings.}
\begin{itemize}
    \item Low-$N_{\rm{H_2}}$ filaments are mostly parallel to \textbf{\textit{B}$_{\rm{PoS}}$} at all scales, while high-$N_{\rm{H_2}}$ filaments usually have flat or bimodal parallel-perpendicular HROs.
    \item High-$N_{\rm{H_2}}$, small-$W_{\rm{b}}^{\star}$ filaments have flat, slightly parallel, or slightly perpendicular HROs, whereas high-$N_{\rm{H_2}}$, large-$W_{\rm{b}}^{\star}$ filaments have flat, bimodal parallel-perpendicular, or mostly perpendicular HROs.
    \item In high-${\langle}p{\rangle}$ GCC fields, both HROs and 2D HROs show more pronounced trends than in low-${\langle}p{\rangle}$ GCC fields.
    \item Criteria based on distance or latitude play a more indirect role: GCC fields at closer distances or higher latitudes usually have lower ${\langle}N_{\rm{H_2}}{\rangle}$ than GCC fields at longer distances or lower latitudes, which is reflected in the results of this combined analysis.
\end{itemize}
Combining criteria enables us to identify a few trends in the HROs more clearly (see, e.g., Fig.~\ref{fig:results_filamentHRO_d+_NH2+_p+}). Improving the estimates of the different criteria used in this study as well as adding criteria on other physical parameters could help improve the constraints and the analysis of the results.

\subsection{Statistical analysis of filaments hosting cores over all the GCC fields}
\label{sec:results3_SourceHRO}

Out of the final 4244 cores from the GCC catalog \citep{Montillaud2015} considered in our study, 93$\%$ are located inside filaments. We find a positive correlation between the number of high-$N_{\rm{H_2}}$ pixels within filaments and the number of cores in a GCC field, with most cores located within the highest-$N_{\rm{H_2}}$ filaments, confirming the strong link between filaments and star formation \citep{Andre2014}. We aim to investigate the link between the presence of a core and the relative orientations, as well as its dependence on physical properties (see Sect.~\ref{sec:properties}). Here, we focus on filaments with embedded cores from the GCC catalog, and we proceed with the same analysis as in Sect.~\ref{sec:results2_FilamentHRO}. Moreover, we have access to the core properties introduced in Sect.~\ref{sec:properties_source}, providing additional characterization of the sampled filaments and cores.

However, one must be cautious when analyzing the HROs for filaments with embedded cores. {\tt FilDReaMS} detects elongated structures that can be well modeled by a rectangular bar. Therefore, if a model bar of aspect ratio $r_{\rm{b}}\,=\,3$ can fit within the FWHM (full width at half maximum) ellipse of a core, that core can potentially be detected as a filament by {\tt FilDReaMS}, which artificially increases the number of cores parallel to their host filaments. To limit this effect, we retain the most significant filament that each core belongs to, as these filaments tend to have $W_{\rm{b}}^{\star}$ larger than the FWHM ellipses of the embedded cores.

In Sect.~\ref{sec:results3_SourceHRO_f-B}, we analyze the relative orientations between filaments hosting cores and \textbf{\textit{B}$_{\rm{PoS}}$}. In Sect.~\ref{sec:results3_SourceHRO_s}, we consider the position angles of cores, first analyzing their orientations relative to their host filaments, then their orientations relative to \textbf{\textit{B}$_{\rm{PoS}}$}. The values of both sets of ratios defined in Eqs.~\eqref{eq:perpara_ratios} and~\eqref{eq:perpara_ratios_bis}, as well as the associated uncertainties (see Sect.~\ref{sec:method_ratios}) for the samples presented in this section are summarized in Table~\ref{tab:perpara_ratios_values}.

\subsubsection{HROs of \texorpdfstring{$\lvert\psi_{\rm{f}}^{\star} - \psi_{B}\rvert$}{sourceFilvBHRO}}
\label{sec:results3_SourceHRO_f-B}

Here, we focus the HRO analysis on the relative orientations between filaments with embedded cores and \textbf{\textit{B}$_{\rm{PoS}}$}. Our results show that the HROs are similar to those of filaments with $N_{\rm{H_2}}\,>\,10^{21.5}\,{\rm{cm^{-2}}}$, but trends are comparatively less pronounced. This similarity is mostly due to cores being usually detected in high-$N_{\rm{H_2}}$ filaments (see Fig.~\ref{fig_appendix:filNH2_sourceNH2_hist}). Within samples where high-$N_{\rm{H_2}}$ filaments are mostly parallel to \textbf{\textit{B}$_{\rm{PoS}}$}, we also find that filaments with embedded cores are mostly parallel to \textbf{\textit{B}$_{\rm{PoS}}$}. However, in samples where high-$N_{\rm{H_2}}$ filaments have bimodal HROs, filaments hosting cores have different behaviors in their HROs depending on ${\langle}p{\rangle}$ and core properties.

\begin{figure}
   \centering
   \includegraphics[width=\hsize]{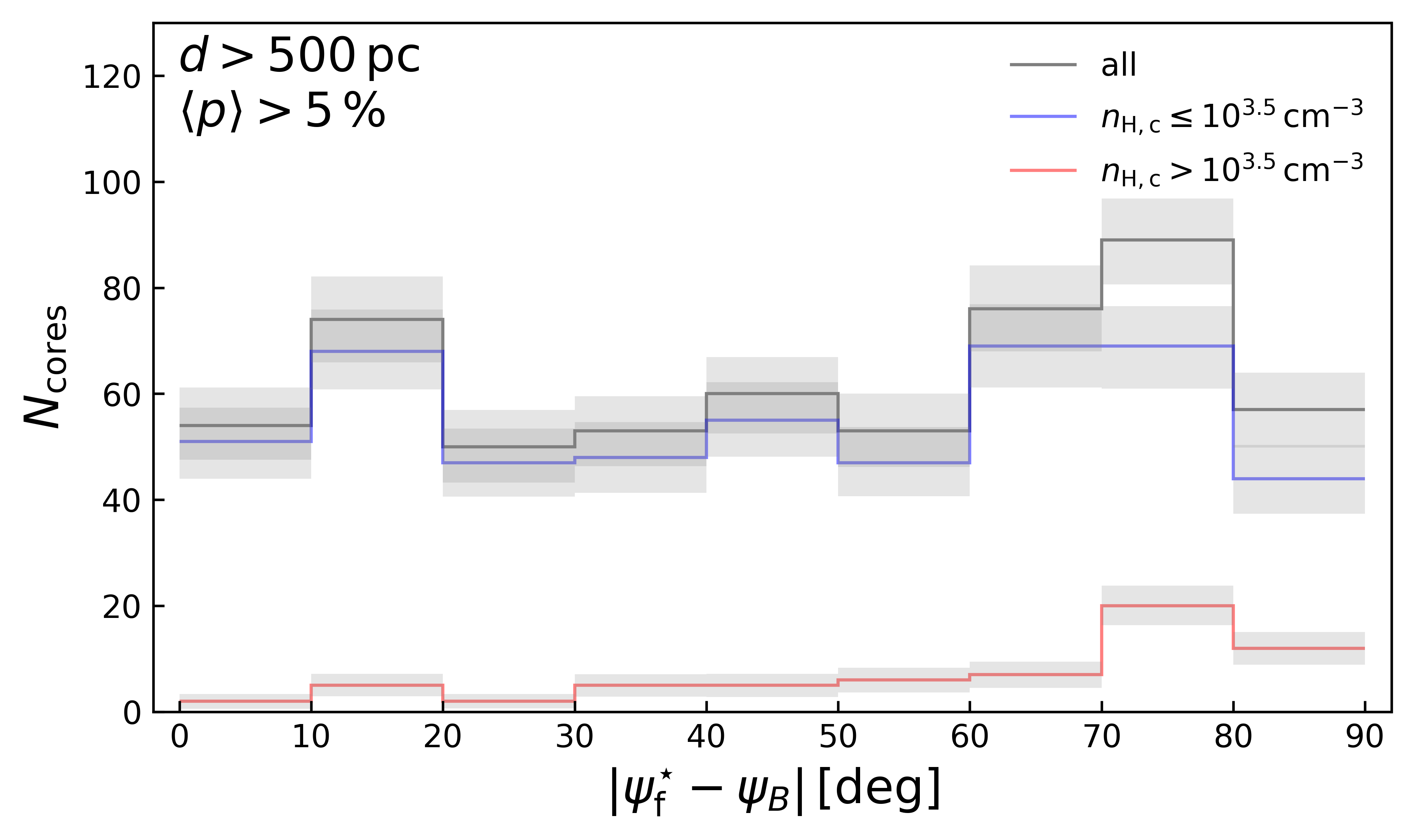}
      \caption{Histograms of relative orientations for filaments hosting cores in GCC fields with $d > 500\,{\rm{pc}}$ and ${\langle}p{\rangle} > 5\%$ for all cores (black), $n_{\rm{H,c}} \leq 10^{3.5}\,\rm{cm^{-3}}$ cores (blue), and $n_{\rm{H,c}} > 10^{3.5}\,\rm{cm^{-3}}$ cores (red). The uncertainties are shown in gray shaded areas.}
         \label{fig:results_sourceHRO_d+_p+_density}
   \end{figure}

Figure~\ref{fig:results_sourceHRO_d+_p+_density} displays HROs for filaments hosting cores in GCC fields with $d\,>\,500\,\rm{pc}$ and ${\langle}p{\rangle} > 5\%$, as well as the HROs for filaments subdivided into two core $n_{\rm{H,c}}$ bins with a threshold at $10^{3.5}\,\rm{cm^{-3}}$. We see that the HRO for filaments in the sample including all $n_{\rm{H,c}}$ shows a slight preference for perpendicular relative orientations. Additionally, the HRO for filaments with low-$n_{\rm{H,c}}$ cores is mostly flat, while the HRO for filaments with high-$n_{\rm{H,c}}$ cores displays a strong preference for perpendicular relative orientations, with $\mathfrak{r}_{\perp} = 58.6\%$ (see Table~\ref{tab:perpara_ratios_values}).

\begin{figure}
   \centering
   \includegraphics[width=\hsize]{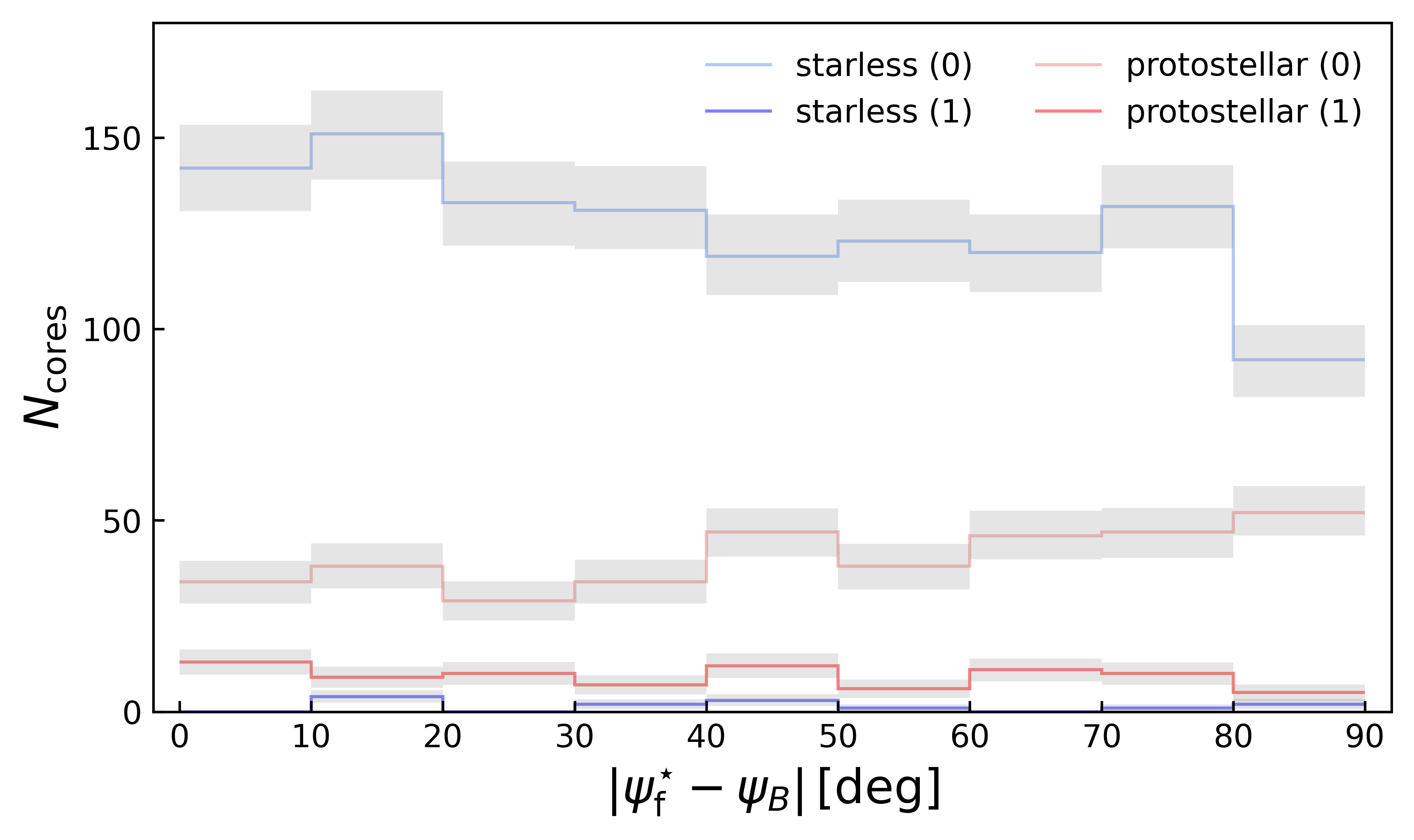}
      \caption{Histograms of relative orientations for filaments hosting cores in all GCC fields according to an estimation of their evolutionary stage and its reliability. The HRO for unreliable (0) starless cores is shown in light blue, the HRO for reliable (1) starless cores is in dark blue, the HRO for unreliable (0) protostellar cores is in light red, and the HRO for reliable protostellar (1) cores is in dark red. The uncertainties are shown in gray shaded areas.}
         \label{fig:results_sourceHRO_evolstage}
   \end{figure}

Overall, we find that filaments with high-$M_{\rm c}$ cores ($M_{\rm c}\,>\,5\,M_{\odot}$) tend to be slightly more perpendicular to \textbf{\textit{B}$_{\rm{PoS}}$}, whereas filaments with low-$M_{\rm c}$ cores have random orientations, with a slight preference toward orientations parallel to \textbf{\textit{B}$_{\rm{PoS}}$}. The parallel trends are reinforced if we only consider nearby fields ($d \leq 500\,{\rm pc}$), where the detected cores are less massive by at least an order of magnitude than distant ($d > 500\,{\rm pc}$) cores (see Fig.~\ref{fig_appendix:results_source_logmass_d_hist}). This supports the natural notion that filament $N_{\rm{H_2}}$ and core masses are correlated, with the most massive cores located in filaments with the highest $N_{\rm{H_2}}$. Additionally, we find that filaments with high-$n_{\rm{H,c}}$ cores or bound cores tend to be slightly more perpendicular to \textbf{\textit{B}$_{\rm{PoS}}$}, whereas filaments with low-$n_{\rm{H,c}}$ cores or unbound cores are slightly more parallel to \textbf{\textit{B}$_{\rm{PoS}}$}. These trends are stronger for cores in high-${\langle}p{\rangle}$ GCC fields.

The link between the evolutionary stage of cores and HROs is harder to determine. Out of all the cores in the sample, 1604 cores have a characterization of their evolutionary stage, with only 96 considered "reliable" by \cite{Montillaud2015}, of which 13 are starless and 83 are protostellar. Figure~\ref{fig:results_sourceHRO_evolstage} shows the HROs for filaments hosting cores in all GCC fields subdivided according to their evolutionary stage (starless or protostellar) and the reliability of the determination (unreliable or reliable). We find that the HRO for filaments with starless cores flagged as unreliable (0) displays a slight preference for more parallel relative orientations, with $\mathfrak{r}_{\parallel} = 37.7\%$ (see Table~\ref{tab:perpara_ratios_values}). Conversely, the HRO for filaments with protostellar cores flagged as unreliable (0) displays a slight preference for more perpendicular relative orientations, with $\mathfrak{r}_{\perp} = 39.6\%$ (see Table~\ref{tab:perpara_ratios_values}). However, the HRO for filaments with protostellar cores flagged as reliable (1) shows preferentially parallel relative orientations, with $\mathfrak{r}_{\parallel} = 39.5\%$, slightly higher than in the unreliable (0) starless case. We find this trend in almost every combination of parameters including the evolutionary stage of cores. Finally, the number of reliable starless cores is too low to carry out this analysis.

\subsubsection{HROs of \texorpdfstring{$\lvert\psi_{\rm{f}}^{\star} - \psi_{\rm{c}}\rvert$}{sourceFilvPAHRO} and \texorpdfstring{$\lvert\psi_{\rm{c}} - \psi_{B}\rvert$}{sourceBvPAHRO}}
\label{sec:results3_SourceHRO_s}

\begin{figure}
    \centering
    \includegraphics[width=\hsize]{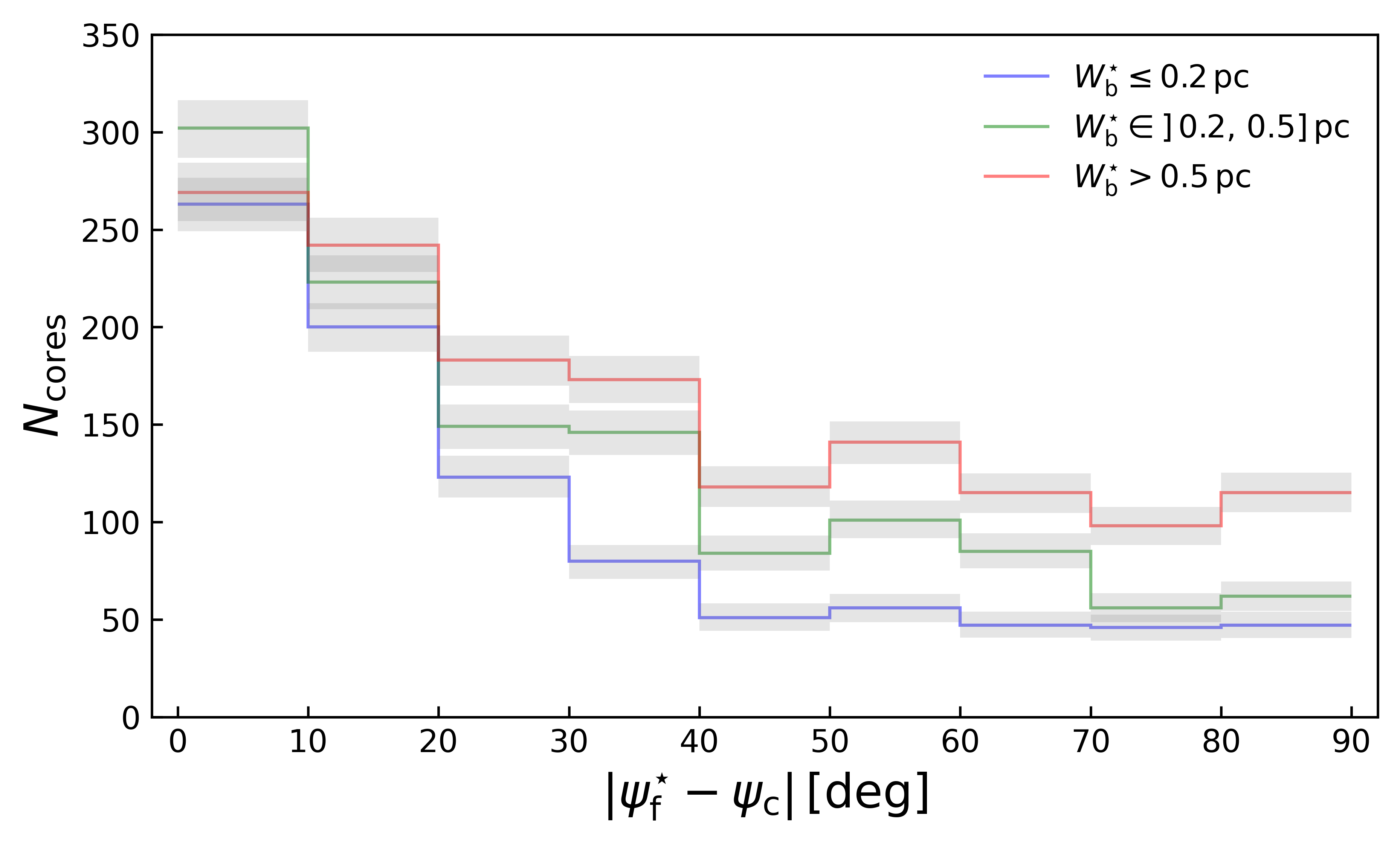}
    \caption{Histograms of $\lvert\psi_{\rm{f}}^{\star} - \psi_{\rm{c}}\rvert$ for filaments hosting cores in all GCC fields subdivided into three filament $W_{\rm{b}}^{\star}$ bins: $W_{\rm{b}}^{\star} \leq 0.2\,\rm{pc}$ (blue), $0.2\,\rm{pc} < W_{\rm{b}}^{\star} \leq 0.5\,\rm{pc}$ (green), and $W_{\rm{b}}^{\star} > 0.5\,\rm{pc}$ (red). The uncertainties are shown in gray shaded areas.}
    \label{fig:results_sourceHRO_sourcePA_f}
\end{figure}

\begin{figure}
    \centering
    \includegraphics[width=\hsize]{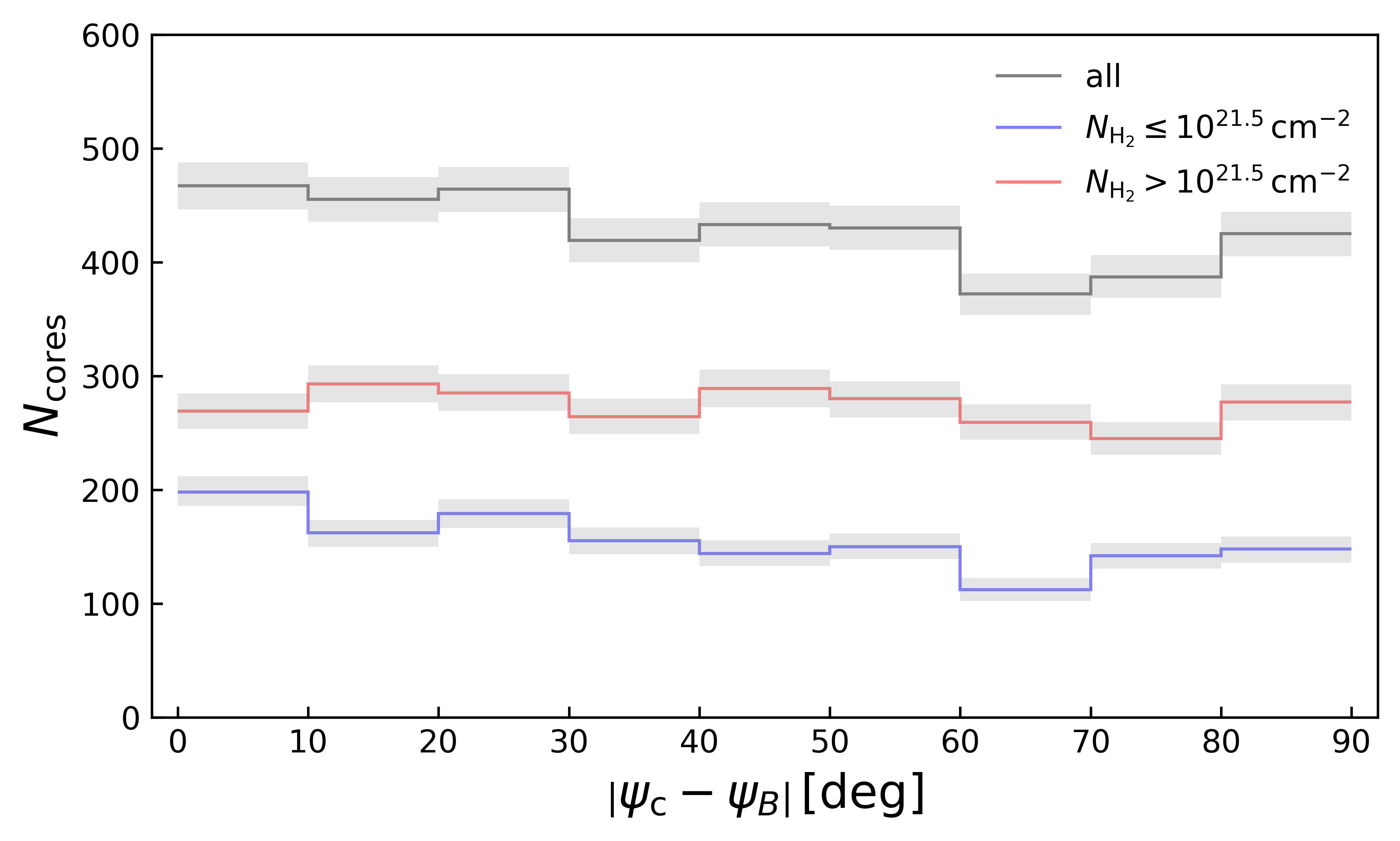}
    \caption{Histogram of $\lvert\psi_{\rm{c}} - \psi_{B}\rvert$ for filaments hosting cores in all GCC fields: for all cores (black), $N_{\rm{H_2}} \leq 10^{21.5}\,\rm{cm^{-2}}$ cores (blue), and $N_{\rm{H_2}} > 10^{21.5}\,\rm{cm^{-2}}$ cores (red). The uncertainties are shown in gray shaded areas.}
    \label{fig:results_sourceHRO_sourcePA_b}
\end{figure}

Here, we consider the position angles of cores, and analyze their orientations relative to their host filaments, then their orientations relative to \textbf{\textit{B}$_{\rm{PoS}}$}, and the dependence of these relative orientations on physical properties (see Sect.~\ref{sec:properties}).

Figure~\ref{fig:results_sourceHRO_sourcePA_f} shows histograms of $\lvert\psi_{\rm{f}}^{\star} - \psi_{\rm{c}}\rvert$ for filaments hosting cores in all GCC fields subdivided into three filament $W_{\rm{b}}^{\star}$ bins with thresholds at $0.2\,\rm{pc}$ and $0.5\,\rm{pc}$, respectively. We find that the core major axes are predominantly parallel to the parent filaments. However, this trend is slightly less pronounced for cores within large-$W_{\rm{b}}^{\star}$ filaments (see Table~\ref{tab:perpara_ratios_values}), or for protostellar cores flagged as reliable. These result are found for most of the combinations of criteria used to sample filaments and cores.

Figure~\ref{fig:results_sourceHRO_sourcePA_b} shows histograms of $\lvert\psi_{\rm{c}} - \psi_{B}\rvert$ for filaments hosting cores in all GCC fields, as well as the HROs for filaments subdivided into two $N_{\rm{H_2}}$ bins with a threshold at $10^{21.5}\,{\rm cm^{-2}}$. While preferred orientations seem more random and are harder to identify, there is a very small preference for parallel orientations for lower $N_{\rm{H_2}}$, while HROs are flatter for higher $N_{\rm{H_2}}$. We find similar trends for most of the combinations of criteria used to sample filaments and cores, with trends for parallel relative orientations slightly more pronounced in high-${\langle}p{\rangle}$ GCC fields.

In summary, the cores in our sample are mostly located inside high-$N_{\rm{H_2}}$ filaments. Filaments with embedded cores have histograms of $\lvert\psi_{\rm{f}}^{\star} - \psi_{B}\rvert$ similar to the HROs for high-$N_{\rm{H_2}}$ filaments from the entire sample. However, the trends are less pronounced for filaments with embedded cores, with a few samples having very low statistics. Cores tend to have their major axes nearly parallel to their host filaments, and more randomly oriented relative to \textbf{\textit{B}$_{\rm{PoS}}$}, with a slight preference toward parallel orientations. Protostellar cores seem to deviate from this trend, but due to very low statistics and unreliable estimates, the results for this criterion are inconclusive.


\section{Projection effects}
\label{sec:projection_effects}

Projection effects must be taken into account in the study of relative orientations. Through observations, we do not have direct access to LoS components, we only see what is projected onto the PoS. The apparent shape and orientation of an object may change drastically upon PoS projection depending on its inclination angle to the LoS, $\theta$. Numerical simulations by \cite{Seifried2020} showed that a transition in relative orientations from preferentially parallel to perpendicular with increasing $N_{\rm{H_2}}$ in 3D does not always show up in 2D projected maps depending on the chosen LoS. Hence, relative orientations in the 2D PoS are not always representative of true relative orientations in 3D.

To quantify projection effects on relative orientations, we took two complementary approaches. The first is purely analytical (Sect.~\ref{sec:projection_analytical}), and the second is based on a simple toy model (Sect.~\ref{sec:projection_model}).

\subsection{Analytical considerations}
\label{sec:projection_analytical}

We considered a cartesian coordinate system $(x,y,z)$ and a spherical coordinate system $(r,\theta,\psi)$ with the $z$-axis along the LoS toward the observer (see Fig.~\ref{fig:data_angles}a). We let $\hat{e}_{\rm f}$ and $\hat{e}_{B}$ be unit vectors along the local filament and the local $\boldvec{B}$, respectively; $(\theta_{\rm f},\psi_{\rm f})$ and $(\theta_{B},\psi_{B})$ be the angular coordinates of $\hat{e}_{\rm f}$ and $\hat{e}_{B}$, respectively; and $f_{\rm f} (\theta_{\rm f},\psi_{\rm f})$ and $f_{B} (\theta_{B},\psi_{B})$ be their respective distribution functions.
Here, for mathematical convenience, we do not restrict the ranges of $\psi_{\rm f}$ and $\psi_{B}$ to $[-90^\circ,90^\circ]$, as done in the rest of the paper, but we allowed $\hat{e}_{\rm f}$ and $\hat{e}_{B}$ to cover all possible directions.

The relative orientation angle between $\hat{e}_{\rm f}$ and $\hat{e}_{B}$ projected onto the PoS (see Fig.~\ref{fig:data_angles}b) is simply
\begin{equation}
\alpha = \psi_{\rm f} - \psi_{B} \ ,
\label{eq:alpha}
\end{equation}
whereas the angle between $\hat{e}_{\rm f}$ and $\hat{e}_{B}$ in 3D, called $\vartheta$ (see Fig.~\ref{fig:data_angles}a), is given by
\begin{equation}
\cos \vartheta = \sin \theta_{\rm f} \ \sin \theta_{B} \ \cos \alpha + \cos \theta_{\rm f} \ \cos \theta_{B} \ \cdot
\label{eq:vartheta}
\end{equation}
Equation~\eqref{eq:vartheta} shows that projection onto the PoS can make $\hat{e}_{\rm f}$ and $\hat{e}_B$ either "more parallel" or "more perpendicular" than in 3D.\footnote{By $\hat{e}_{\rm f}$ and $\hat{e}_B$ being "more parallel" or "more perpendicular" in projection than in 3D, we just mean that the acute angle between their projected orientations is smaller or larger than the acute angle between their orientations in 3D.}
To find out which situation prevails, we can temporarily assume, without loss of generality, that $\hat{e}_{\rm f}$ and $\hat{e}_B$ point away from the PoS ($\theta_{\rm f}, \theta_B \in [0^\circ,90^\circ]$). 
Then the condition for $\hat{e}_{\rm f}$ and $\hat{e}_B$ to look "more parallel" in projection than in 3D is simply $|\cos \alpha| > |\cos \vartheta|$, which is equivalent to
\begin{equation}
\cos \alpha > \frac
{\cos \theta_{\rm f} \ \cos \theta_B}
{1 - \sin \theta_{\rm f} \ \sin \theta_B} 
\quad {\rm or} \quad
\cos \alpha < - \frac
{\cos \theta_{\rm f} \ \cos \theta_B}
{1 + \sin \theta_{\rm f} \ \sin \theta_B} \ \cdot
\label{eq:more_parallel}
\end{equation}
From this, we can already note two important points: \\
\noindent
$\bullet$ If $\hat{e}_{\rm f}$ or $\hat{e}_B$ lies in the PoS ($\theta_{\rm f}$ or $\theta_B = 90^\circ$), $\hat{e}_{\rm f}$ and $\hat{e}_B$ always look "more parallel" in projection than in 3D (except in the particular case $\alpha = \pm 90^\circ$, where $\hat{e}_{\rm f}$ and $\hat{e}_B$ remain perpendicular) . \\
\noindent
$\bullet$
If $\hat{e}_{\rm f}$ and $\hat{e}_B$ differ only by their polar angles ($\theta_{\rm f} \not= \theta_B$, $\psi_{\rm f} = \psi_{B}$), then $|\alpha| = 0^\circ < |\vartheta|$, whereas if $\hat{e}_{\rm f}$ and $\hat{e}_B$ differ only by their azimuthal angles ($\theta_{\rm f} = \theta_B$, $\psi_{\rm f} \not= \psi_{B}$), then $|\alpha| > |\vartheta|$.
As a general rule, large $|\theta_{\rm f} - \theta_B|$ and small $|\psi_{\rm f} - \psi_B|$ lead to $|\alpha| < |\vartheta|$, which implies that $\hat{e}_{\rm f}$ and $\hat{e}_B$ look "more parallel" in projection than in 3D, whereas small $|\theta_{\rm f} - \theta_B|$ and large $|\psi_{\rm f} - \psi_B|$ lead to $|\alpha| > |\vartheta|$, which implies that $\hat{e}_{\rm f}$ and $\hat{e}_B$ look "more perpendicular" ["more parallel"] in projection than in 3D if $|\alpha|, |\vartheta| < 90^\circ$ [$|\alpha|, |\vartheta| > 90^\circ$].

\noindent
The fraction of pairs $(\hat{e}_{\rm f},\hat{e}_B)$ that satisfy Eq.~\eqref{eq:more_parallel} is
\begin{align}
{\rm f}_\parallel = &
\int_{0}^{\frac{\pi}{2}} d\theta_{\rm f} \ \sin \theta_{\rm f} \ 
\int_{0}^{\frac{\pi}{2}} d\theta_B \ \sin \theta_B \ \nonumber \\
&\times \frac{1}{\pi}
\left( 
\arccos \frac
{\cos \theta_{\rm f} \ \cos \theta_B}
{1 - \sin \theta_{\rm f} \ \sin \theta_B} 
+ \arccos \frac
{\cos \theta_{\rm f} \ \cos \theta_B}
{1 + \sin \theta_{\rm f} \ \sin \theta_B} 
\right) \ ,
\label{eq:fraction_parallel}
\end{align}
or, upon numerical integration, ${\rm f}_\parallel \simeq 68.7\,\%$. Hence, one can see that projection onto the PoS is more likely to make $\hat{e}_{\rm f}$ and $\hat{e}_B$ "more parallel" than "more perpendicular." This is consistent with the well-known fact that vectors that are (nearly) parallel in 3D remain (nearly) parallel in projection, whereas vectors that are (nearly) perpendicular in 3D can take on any relative orientation in projection.
The distribution function of $\alpha$ reads
\begin{equation}
f_\alpha(\alpha) = \iiint d\theta_{\rm f} \ d\psi_{\rm f} \ d\theta_{B} \ \sin \theta_{\rm f} \ \sin \theta_{B} \ f_{\rm f} (\theta_{\rm f},\psi_{\rm f}) \ f_{B} (\theta_{B},\psi_{B} \! = \! \psi_{\rm f} \! - \! \alpha) \ \cdot
\label{eq:f_alpha}
\end{equation}
The distribution function of $\vartheta$ can be written in the convenient form
\begin{equation}
f_\vartheta(\vartheta) = \sin \vartheta \ g(\vartheta) \ ,
\label{eq:f_vartheta}
\end{equation}
which factors out the geometric factor $\sin \vartheta$ arising from the differential solid angle, $d\Omega = \sin \vartheta \ d\vartheta \ d\varphi$.
Thus, the function $g(\vartheta)$ describes departures (along $\vartheta$) from a purely isotropic 3D distribution (for which $g(\vartheta) = \frac{1}{2}$).

A first obvious conclusion, which is not always fully appreciated, is the following: if $\hat{e}_{\rm f}$ and $\hat{e}_{B}$ have totally isotropic distributions in 3D ($f_{\rm f} (\theta_{\rm f},\psi_{\rm f}) = f_{B} (\theta_{B},\psi_{B}) = \frac{1}{4\pi}$), then $\alpha$ also has a totally isotropic distribution, i.e., a flat histogram ($f_\alpha(\alpha) = \frac{1}{2\pi}$). This can be understood as the net result of two opposing trends. On the one hand, the probability that two random vectors make an angle $\vartheta$ in 3D increases as $\sin \vartheta$ (see Eq.~\eqref{eq:f_vartheta}); in other words, in 3D, more perpendicular is more likely. On the other hand, projection onto the PoS introduces a statistical trend toward "more parallel" (see comment below Eq.~\eqref{eq:fraction_parallel}).

A second important conclusion can be obtained by considering the situation where the direction of $\hat{e}_{B}$ is uniform with, say, $\theta_{B} = \theta_{B0}$ and $\psi_{B} = 0$ (the exact value of $\psi_{B}$ is irrelevant),
\begin{equation}
f_{B} (\theta_{B},\psi_{B}) = \frac{1}{\sin \theta_{B0}} \ \delta(\theta_{B} \! - \! \theta_{B0}) \ \delta(\psi_{B}) \ ,
\label{eq:fB}
\end{equation}
and $\hat{e}_{\rm f}$ has an isotropic distribution around $\hat{e}_{B}$,
\begin{equation}
f_{\rm f} (\theta_{\rm f},\psi_{\rm f}) = \frac{1}{2\pi} \ g(\vartheta) \ ,
\label{eq:ff}
\end{equation}
with $\int d\vartheta \ \sin \vartheta \ g(\vartheta) = 1$.
Inserting Eqs.~\eqref{eq:fB} and~\eqref{eq:ff} into Eq.~\eqref{eq:f_alpha} leads to
\begin{equation}
f_\alpha(\alpha) = \frac{1}{2\pi} \ \int d\theta_{\rm f} \ \sin \theta_{\rm f} \ g(\vartheta) \ ,
\label{eq:f_alpha_Bunif}
\end{equation}
where $\vartheta$ is given by Eq.~\eqref{eq:vartheta} with $\theta_{B} = \theta_{B0}$.

\noindent
Next, we look into three representative cases. The first concerns uniform distribution. If $\hat{e}_{\rm f}$ is uniformly distributed in 3D ($g(\vartheta) = \frac{1}{2}$), then $\alpha$ is also uniformly distributed (flat histogram, $f_\alpha(\alpha) = \frac{1}{2\pi}$).

The second case is regarding preferentially parallel distribution.
If $\hat{e}_{\rm f}$ tends to be more parallel than perpendicular to $\hat{e}_{B}$ in 3D, the parallel trend persists in the PoS, i.e., the histogram of $\alpha$ rises toward $0^\circ$ and $180^\circ$.
This can be verified with the analytically tractable function $g(\vartheta) = \frac{3}{2} \, \cos^2 \vartheta$, for which we obtained
\begin{equation}
f_\alpha(\alpha) = \frac{1}{\pi} \,
\left(
\sin^2 \theta_{B0} \ \cos^2 \alpha
+ \frac{1}{2} \, \cos^2 \theta_{B0}
\right) \ ,
\label{eq:f_alpha_Bunif_2}
\end{equation}
and, upon averaging over $\theta_{B0}$, it changes to
\begin{equation}
\langle f_\alpha(\alpha) \rangle = \frac{1}{3\pi} \,
\left(
2 \, \cos^2 \alpha + \frac{1}{2}
\right) \ \cdot
\label{eq:f_alpha_Bunif_2_average}
\end{equation}
Equation~\eqref{eq:f_alpha_Bunif_2} indicates that the trend toward parallel orientations in the PoS is most pronounced (and in this case exactly equal to the 3D trend) when the magnetic field lies in the PoS ($\theta_{B0} = 90^\circ$), while the trend disappears when the magnetic field is along the LoS ($\theta_{B0} = 0^\circ$ or $180^\circ$).

The third case is preferentially perpendicular distribution. Similarly, if $\hat{e}_{\rm f}$ tends to be more perpendicular than parallel to $\hat{e}_{B}$ in 3D, the perpendicular trend persists in the PoS, i.e., the histogram of $\alpha$ rises toward $\pm 90^\circ$.
This can be verified with the function $g(\vartheta) = \frac{3}{4} \, \sin^2 \vartheta$, for which we obtained
\begin{equation}
f_\alpha(\alpha) = \frac{1}{2\pi} \,
\left(
\sin^2 \theta_{B0} \ \sin^2 \alpha
+ \frac{1}{2} \, \sin^2 \theta_{B0} + \cos^2 \theta_{B0}
\right) \ ,
\label{eq:f_alpha_Bunif_3}
\end{equation}
and, upon averaging over $\theta_{B0}$, it becomes
\begin{equation}
\langle f_\alpha(\alpha) \rangle = \frac{1}{3\pi} \,
\left(
\sin^2 \alpha + 1
\right) \ \cdot
\label{eq:f_alpha_Bunif_3_average}
\end{equation}
Here, too, the perpendicular trend in the PoS is most pronounced when the magnetic field lies in the PoS ($\theta_{B0} = 90^\circ$), and it disappears when the magnetic field is along the LoS ($\theta_{B0} = 0^\circ$ or $180^\circ$). There is, however, an important difference with the preferentially parallel case: here, the perpendicular trend is always reduced by projection onto the PoS, even when the magnetic field lies in the PoS; this is because, as mentioned above, projection onto the PoS tends to make vectors "more parallel."

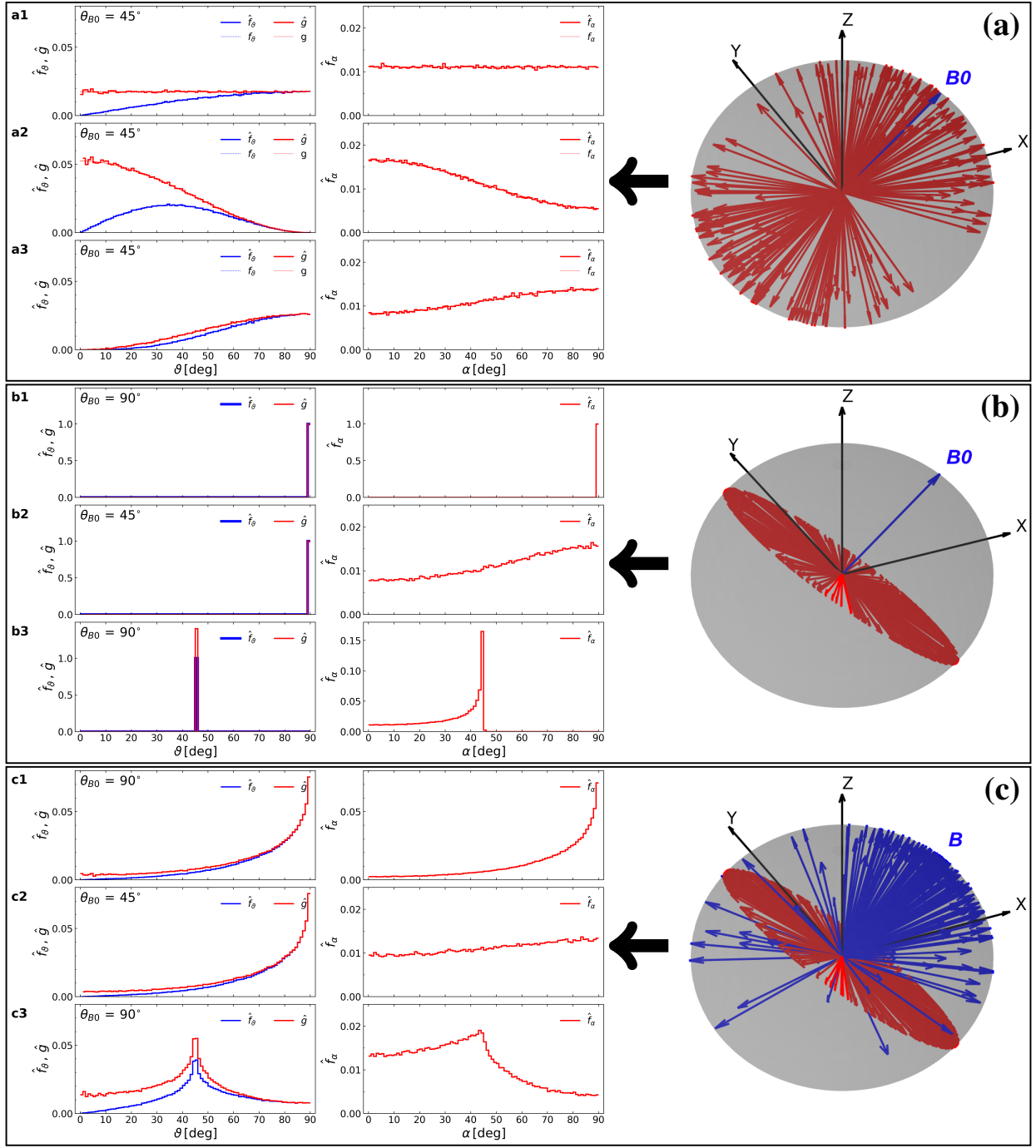
\begin{figure*}[h!]
   \centering
   \resizebox{0.9\hsize}{!}{
   \begin{tikzpicture}
   \node[anchor=south west,inner sep=0] (img1) at (0,0) {\includegraphics[height=4.1cm,keepaspectratio]{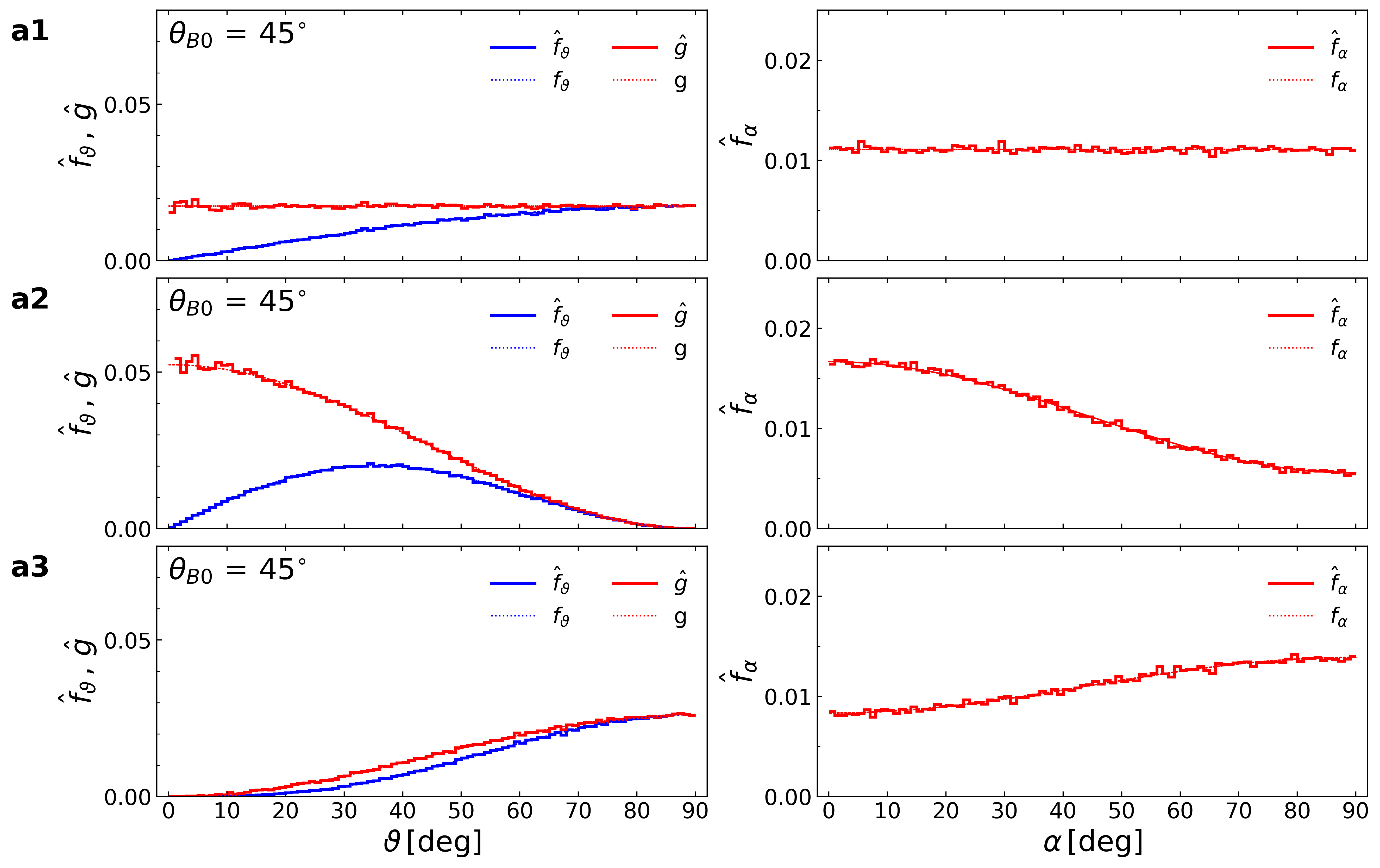}};
   \node[anchor=south west,inner sep=0] (img2) at (7,0) {\includegraphics[height=4.1cm,keepaspectratio]{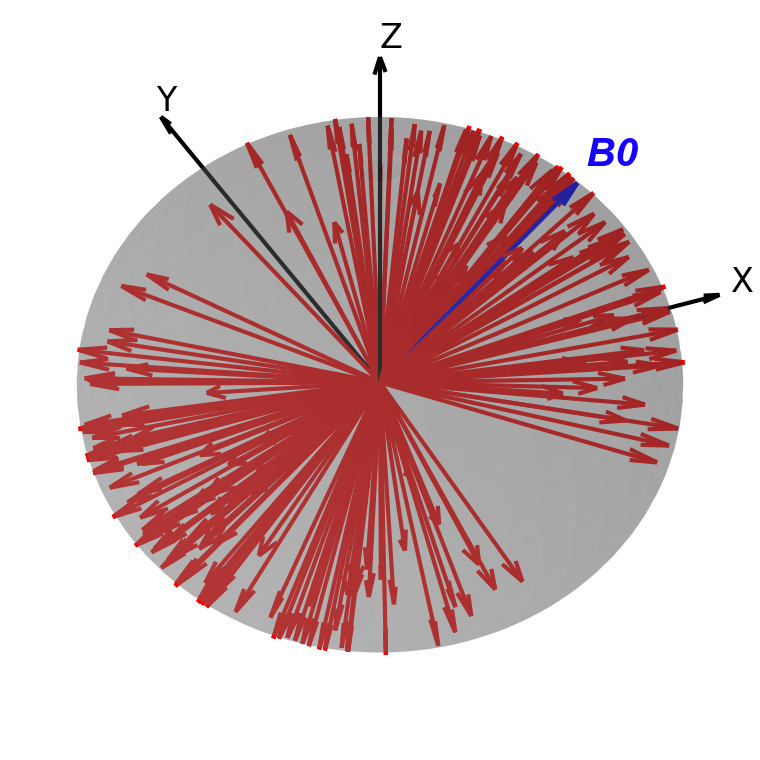}};
   \node [below left,text width=3cm,align=right,font=\bf] at (img2.north east){(a)};
   \draw[->, line width=4pt, black] ([xshift=5pt, yshift=3pt]img2.west) -- ([yshift=3pt]img1.east);
   \draw[draw=black, line width=0.5pt] (0,0) rectangle ++(11.1,4.1);
   \end{tikzpicture}
   }
   \resizebox{0.9\hsize}{!}{
   \begin{tikzpicture}
   \node[anchor=south west,inner sep=0] (img1) at (0,0) {\includegraphics[height=4.1cm,keepaspectratio]{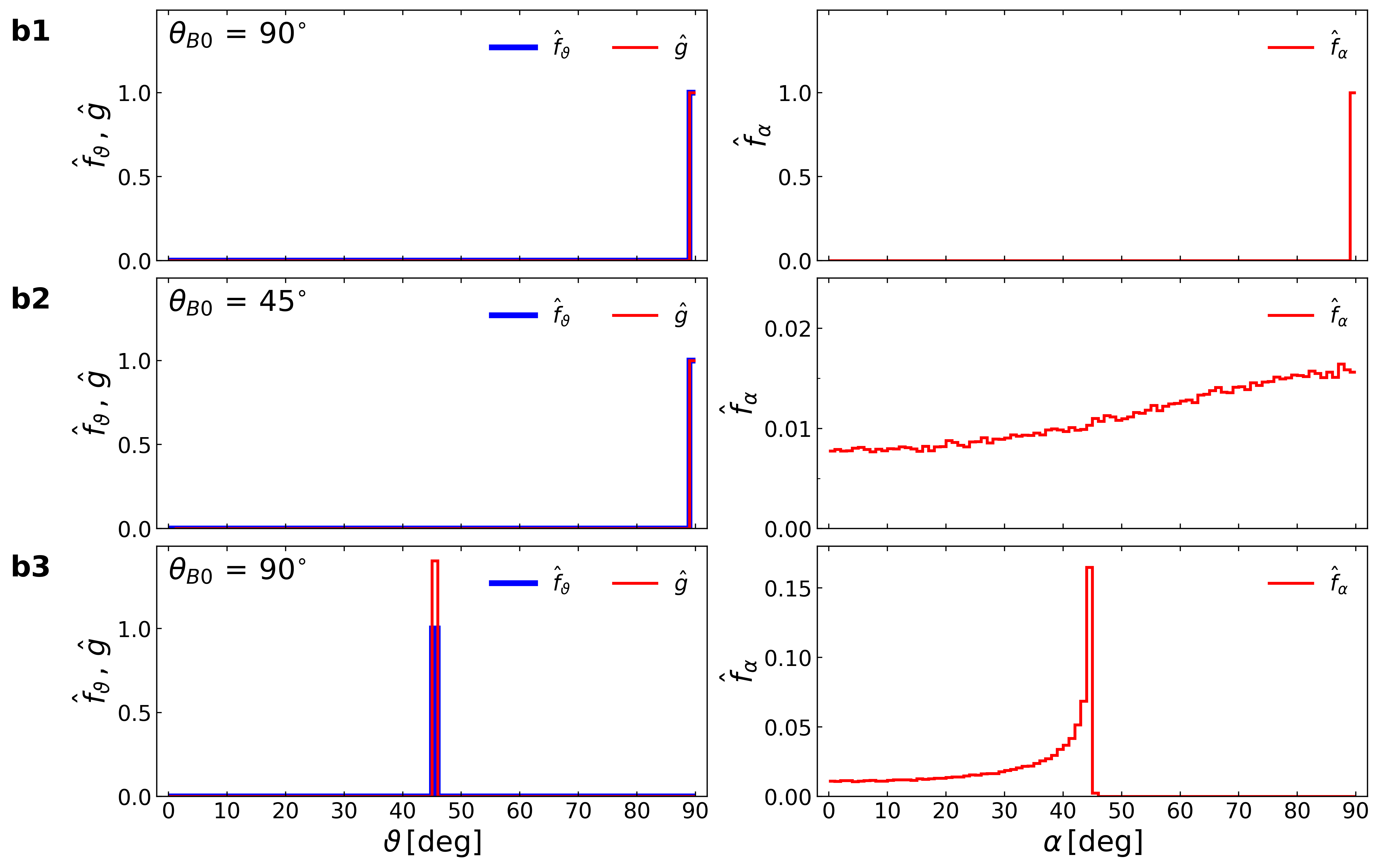}};
   \node[anchor=south west,inner sep=0] (img2) at (7,0) {\includegraphics[height=4.1cm,keepaspectratio]{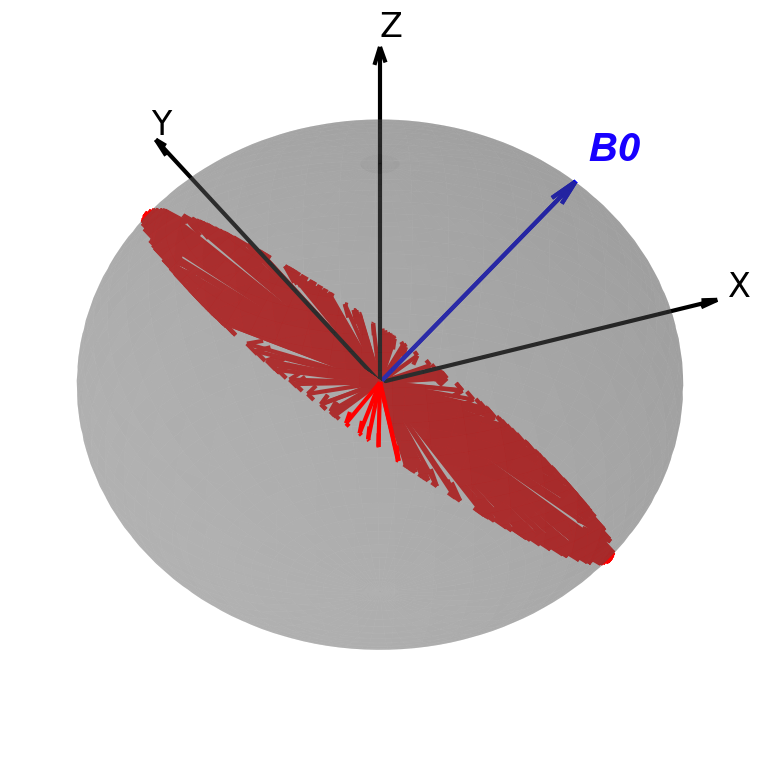}};
   \node [below left,text width=3cm,align=right,font=\bf] at (img2.north east){(b)};
   \draw[->, line width=4pt, black] ([xshift=5pt, yshift=3pt]img2.west) -- ([yshift=3pt]img1.east);
   \draw[draw=black, line width=0.5pt] (0,0) rectangle ++(11.1,4.1);
   \end{tikzpicture}
   }
   \resizebox{0.9\hsize}{!}{
   \begin{tikzpicture}
   \node[anchor=south west,inner sep=0] (img1) at (0,0) {\includegraphics[height=4.1cm,keepaspectratio]{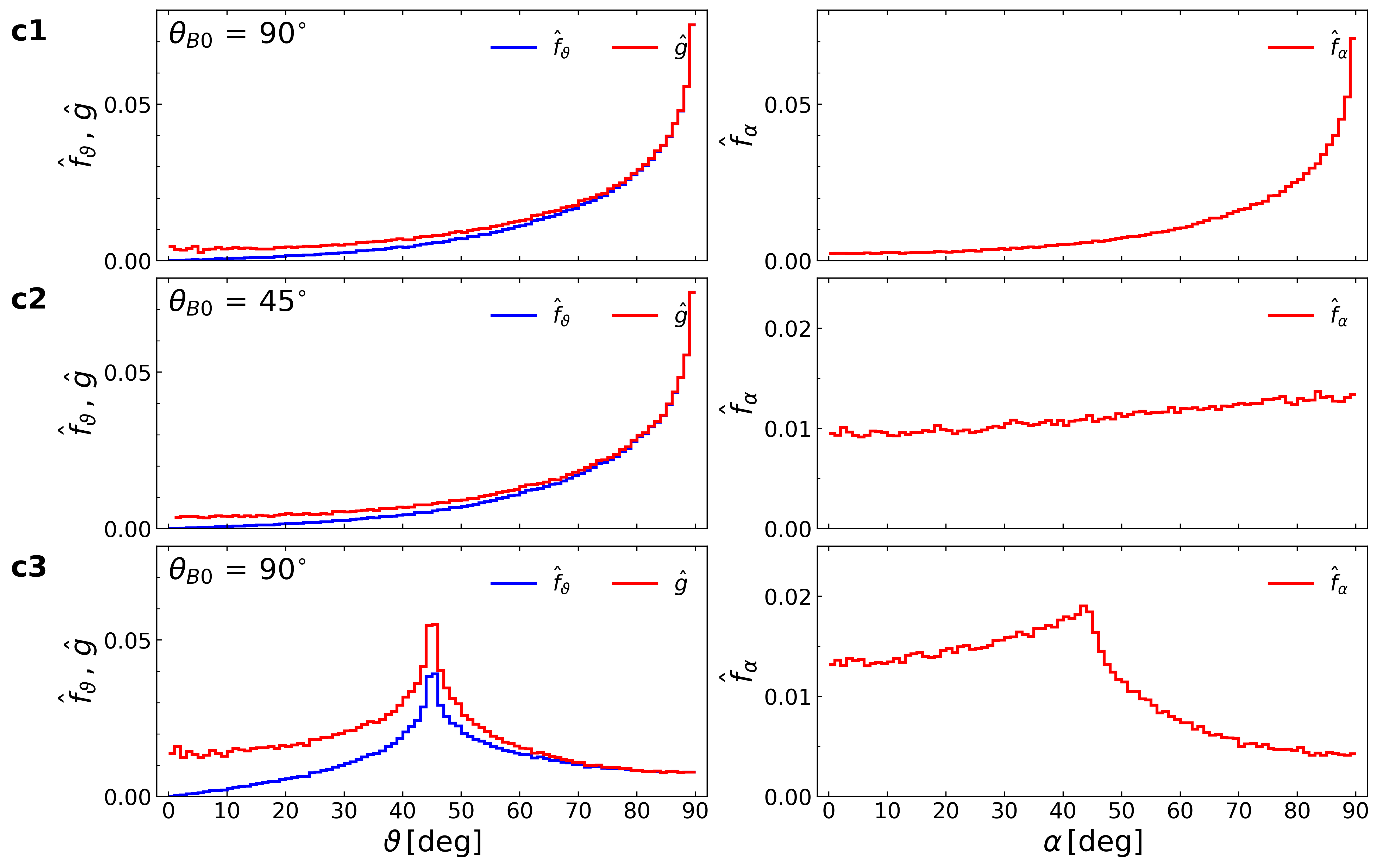}};
   \node[anchor=south west,inner sep=0] (img2) at (7,0) {\includegraphics[height=4.1cm,keepaspectratio]{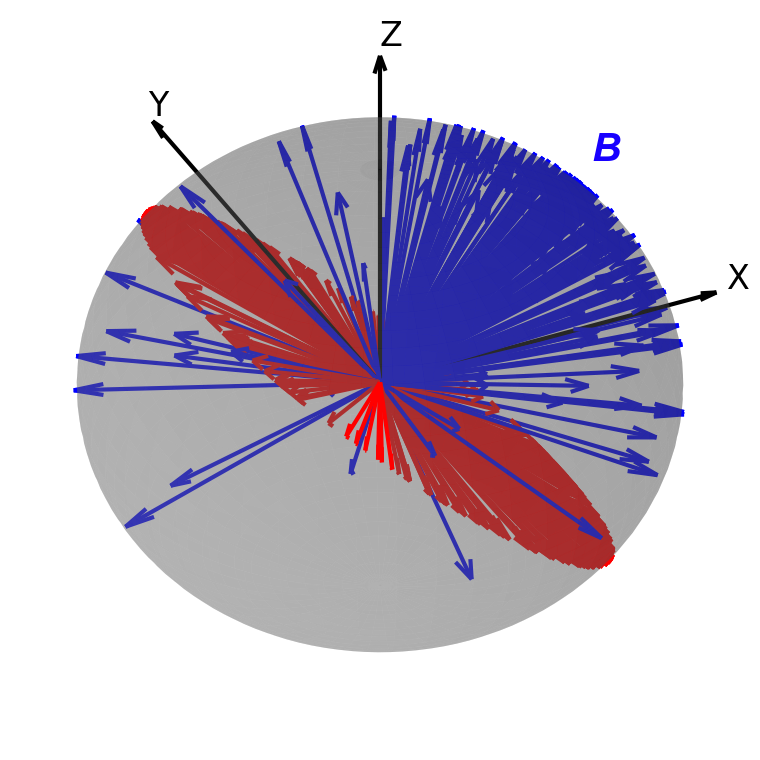}};
   \node [below left,text width=3cm,align=right,font=\bf] at (img2.north east){(c)};
   \draw[->, line width=4pt, black] ([xshift=5pt, yshift=3pt]img2.west) -- ([yshift=3pt]img1.east);
   \draw[draw=black, line width=0.5pt] (0,0) rectangle ++(11.1,4.1);
   \end{tikzpicture}
   }
   \caption{Toy model of relative orientation distributions in nine different configurations of $\theta_{B}$ and $\vartheta$ built from $\theta_{B0}$ and $\vartheta_0$ values, respectively, both in 3D (left column) and 2D PoS projection (center column). \textbf{Left column:} Histograms of $\vartheta$, $\hat{f}_{\vartheta}$ (thick, solid blue line), and corresponding HROs divided by $\sin \vartheta$, $\hat{g}$ (thick, solid red line). \textbf{Center column:} Histograms of $\alpha$, $\hat{f}_{\alpha}$ (thick solid red line). \textbf{Right column:} Three-dimensional views of the orientations of filaments (red arrows) and $\textbf{\textit{B}}$ (blue arrows) following the distributions used in panels {\bf a2}, {\bf b2}, and {\bf c2}, respectively. The gray shape corresponds to the sphere containing $\hat{e}_{{\rm f}}$ and $\hat{e}_{B}$, and the ($\hat{x}$, $\hat{y}$) plane represents the PoS.
   \textbf{Panels a:} Histograms of relative orientations for the three representative cases from Section~\ref{sec:projection_analytical} with $\theta_{B} = \theta_{B0} = 45\,\degr$ (uniform in {\bf a1}, preferentially parallel in {\bf a2}, and preferentially perpendicular in {\bf a3}) together with their respective analytical distribution functions $f_\alpha(\alpha)$, $f_\vartheta(\vartheta)$, and $g(\vartheta) = f_\vartheta(\vartheta) / \sin \vartheta$ (thin dotted red and blue lines). \textbf{Panels b:} Histograms of relative orientations for $\textbf{\textit{B}} = \textbf{\textit{B}}_0$, with varying $\theta_{B0}$ and $\vartheta_0$. \textbf{Panels c:} Histograms of relative orientations for $\textbf{\textit{B}} = \textbf{\textit{B}}_0 + \delta \textbf{\textit{B}}$, with varying $\theta_{B0}$ and $\vartheta_0$.
   }
   \label{fig:projectionfx_3by3fig}
   \end{figure*}

\subsection{Toy model}
\label{sec:projection_model}

To complement the analytical approach of Sect.~\ref{sec:projection_analytical}, which is necessarily limited, we developed a simple numerical toy model that can handle any given distributions of $\hat{e}_{\rm f}$ and $\hat{e}_B$. In this model, the local filaments and local $\boldvec{B}$ are represented by two sets of vectors $\hat{e}_{{\rm f}}$ and $\hat{e}_{B}$ with angular coordinates drawn from their respective distribution functions $f_{\rm f} (\theta_{\rm f},\psi_{\rm f})$ and $f_{B} (\theta_{B},\psi_{B})$.
The code first computes the 2D (PoS) and 3D relative orientation angles, $\alpha$ and $\vartheta$, respectively, using Eqs.~\eqref{eq:alpha} and \eqref{eq:vartheta}.
It then builds the histograms of $\alpha$ and $\vartheta$, folded over the range $[0^{\circ}, 90^{\circ}]$, with $\simeq\,0^{\circ}$ indicating nearly parallel and $\simeq\,90^{\circ}$ nearly perpendicular relative orientations.
The number of counts in each bin is divided by the bin size and by the total number of counts to make the HROs more directly comparable to the corresponding distribution functions (also folded over the range $[0^{\circ}, 90^{\circ}]$).

In the following, the renormalized HROs corresponding to the distribution functions $f_\alpha(\alpha)$, $f_\vartheta(\vartheta)$, and $g(\vartheta) = f_\vartheta(\vartheta) / \sin \vartheta$ (see Eq.~\eqref{eq:f_vartheta}) are denoted $\hat{f}_\alpha(\alpha)$, $\hat{f}_\vartheta(\vartheta)$, and $\hat{g}(\vartheta) = \hat{f}_\vartheta(\vartheta) / \sin \vartheta$, respectively. The reason why we consider $\hat{g}(\vartheta)$ here is because it is $g(\vartheta)$ rather than $f_\vartheta(\vartheta)$ which truly indicates the parallel versus perpendicular trends.
For instance, in the case of fully isotropic distributions, $f_\alpha(\alpha)$ and $g(\vartheta)$ are constant, while $f_\vartheta(\vartheta) \propto \sin \vartheta$ (see Sect.~\ref{sec:projection_analytical}).

Figure~\ref{fig:projectionfx_3by3fig} shows three sets of plots (labeled a, b and c), with three configurations of magnetic fields and filaments each (labeled 1, 2 and 3). The left column shows the histograms of the 3D relative orientation, $\hat{f}_\vartheta(\vartheta)$, and these HROs divided by $\sin \vartheta$, $\hat{g}(\vartheta)$, the center column shows the histograms of the 2D relative orientation, $\hat{f}_\alpha(\alpha)$, and the right column shows a 3D view of the orientations of filaments and $\boldvec{B}$ following the distribution used in the second configuration of each set of plots (labeled a2, b2 and c2).

We first tested the different distributions considered in Sect.~\ref{sec:projection_analytical}, for $\theta_{B} = \theta_{B0} = 45^\circ$.
The results are shown in Fig.~\ref{fig:projectionfx_3by3fig}a for the "Uniform" (panels a1), "Preferentially parallel" (panels a2), and "Preferentially perpendicular" (panels a3) distributions. In the three cases of Fig.~\ref{fig:projectionfx_3by3fig}a, the analytical distribution functions $f_\alpha(\alpha)$, $f_\vartheta(\vartheta)$, and $g(\vartheta) = f_\vartheta(\vartheta) / \sin \vartheta$, renormalized to the reduced range $[0^{\circ}, 90^{\circ}]$, are over-plotted. The very good agreement between the analytical functions and the simulated HROs lends support to our numerical setup and confirms the trends identified in Sect.~\ref{sec:projection_analytical}.

We then explored a wide range of configurations and more complex angular distributions that are not tractable analytically. Here, we discuss a few cases which are representative of our conclusions. Figure~\ref{fig:projectionfx_3by3fig}b shows the results obtained for three simple configurations where $\boldvec{B} = \boldvec{B}_0$ ($\theta_{B} = \theta_{B0}$) and filaments are uniformly distributed on a cone around $\boldvec{B}_0$ with opening angle $\vartheta_0$ ($\vartheta = \vartheta_0$).\\
\noindent
- In panels b1, $\boldvec{B}_0$ lies in the PoS ($\theta_{B0} = 90^\circ$) and filaments are distributed in a disk perpendicular to $\boldvec{B}_0$ ($\vartheta_0 = 90^\circ$); this disk is therefore observed edge on. We see that all three HROs have the exact same shape, characterized by a sharp peak at $90^\circ$, which means that relative orientations are unaffected by projection onto the PoS. \\
\noindent
- In panels b2 ($\theta_{B0} = 45^\circ$, $\vartheta_0 = 90^\circ$), $\hat{f}_\alpha(\alpha)$ (center panel) is very different from both $\hat{f}_\vartheta(\vartheta)$ and $\hat{g}(\vartheta)$ (left panel). Perpendicular relative orientations are still favored in projection, but only slightly so; all possible values for $\alpha$ are now observed. This can be explained by the projection of both $\boldvec{B}$ and filaments onto the PoS (see comment below Eq.~\eqref{eq:fraction_parallel}). Interestingly, $\hat{f}_\alpha(\alpha)$ in panel b2 has almost the same profile as $\hat{f}_\alpha(\alpha)$ in panel a3, although the corresponding $\hat{g}(\vartheta)$ are completely different. Here, what appears to govern the shape of $\hat{f}_\alpha(\alpha)$ is the inclination angle of the magnetic field to the LoS, $\theta_{B0}$.\\
\noindent
- In panels b3 ($\theta_{B0} = 90^\circ$, $\vartheta_0 = 45^\circ$), $\hat{f}_\alpha(\alpha)$ (center panel) displays the same peak at $\vartheta_0$ as $\hat{f}_\vartheta(\vartheta)$ and $\hat{g}(\vartheta)$ (left panel), and it also exhibits a long tail toward lower values of $\alpha$. This long tail, which results from projection effects on filaments, can be inferred from Eq.~\eqref{eq:vartheta}: with $\theta_{B} = 90^\circ$ and $\vartheta = 45^\circ$, Eq.~\eqref{eq:vartheta} becomes $\sqrt{2}/2 = \sin \theta_{\rm f} \cos \alpha$, which, together with $\theta_{\rm f} \in [45^\circ,135^\circ]$, implies $\lvert\alpha\rvert \in [0^\circ,45^\circ]$.

Figure~\ref{fig:projectionfx_3by3fig}c shows the results obtained when repeating the above three simulations in the more realistic situation where the magnetic field possesses a turbulent component in addition to its uniform component. More specifically, $\boldvec{B} = \boldvec{B}_0 + \delta \boldvec{B}$, with $\boldvec{B}_0$ uniform, $\delta \boldvec{B}$ isotropic, and $\delta B$ following a gaussian distribution peaking at $\delta B = 0$, with a standard deviation $\sigma_{\delta B}\,=\,B_0$. 
\\
\noindent
- In panels c1 ($\theta_{B0} = 90^\circ$, $\vartheta_0 = 90^\circ$), the sharp peaks at $90^\circ$ appearing in panels b1 are smeared out by the inclusion of $\delta \boldvec{B}$, which makes it possible for filaments to have any orientation relative to $\boldvec{B}$, both in 3D and in projection. In this configuration where $\boldvec{B}_0$ lies in the PoS, $\hat{f}_\alpha(\alpha)$ (center panel) is almost identical to $\hat{g}(\vartheta)$ (left panel), which means that the distribution of relative orientations is almost unaffected by projection. \\
\noindent
- In panels c2 ($\theta_{B0} = 45^\circ$, $\vartheta_0 = 90^\circ$), $\hat{f}_\alpha(\alpha)$ is significantly flatter than $\hat{g}(\vartheta)$. As in panels b2, this is due to projection effects. \\
\noindent
- In panels c3 ($\theta_{B0} = 90^\circ$, $\vartheta_0 = 45^\circ$), all three HROs peak at $\vartheta_0$, as in panels b3, but the inclusion of $\delta \boldvec{B}$ gives rise to strong tails on both sides of $\vartheta_0$. The tail of $\hat{f}_\alpha(\alpha)$ toward larger angles ("more perpendicular") is similar to that of $\hat{g}(\vartheta)$, but the tail of $\hat{f}_\alpha(\alpha)$ toward smaller angles ("more parallel") remains much higher than that of $\hat{g}(\vartheta)$.
In other words, projection onto the PoS statistically tends to make filaments look "more parallel" to the magnetic field than in 3D (see again comment below Eq.~\eqref{eq:fraction_parallel}). \\
\noindent
When comparing panels of Fig.~\ref{fig:projectionfx_3by3fig}c to their respective counterparts in Fig.~\ref{fig:projectionfx_3by3fig}b, we find that the addition of a turbulent magnetic field component increases the dispersion of both $\vartheta$ and $\alpha$, and that projection effects tend to amplify this dispersion, thereby weakening the existing trends, regardless of the preferred relative orientations.

Here, we summarize our conclusions from both analytical (Sect.~\ref{sec:projection_analytical}) and numerical (Sect.~\ref{sec:projection_model}) approaches.
\begin{itemize}
    \item If there are no trends in 3D, there will be no trends in 2D. However, the absence of trends in 2D does not imply an absence of trends in 3D.
    \item The fact that PoS projection alone does not create trends if there are none in 3D can be explained by two opposing effects: on the one hand, in 3D, any two vectors are more likely to be closer to perpendicular than parallel; on the other hand, PoS projection is more likely to make any two vectors look "more parallel" than "more perpendicular."
    \item PoS projection tends to broaden HROs, leading to greater dispersion in $\alpha$ than in $\vartheta$, regardless of the preferred relative orientations, in agreement with \cite{PlanckXXXII2016}.
    \item As a general rule, projection effects depend on the distribution of 3D relative orientation angles, $f_\vartheta(\vartheta)$, and are stronger when the considered vectors lie closer to the LoS. For a given GCC field, the observed trends are representative of the real 3D trends if one vector, or preferentially both vectors are close to the PoS.
    \item A statistical study is the best approach to recover the true general 3D trends.
\end{itemize}
Hence, for our HRO analysis, it is advantageous to examine configurations where filaments and/or the magnetic field are reasonably close to the PoS. This was already predicted by numerical simulations (see, e.g., \citealp{Seifried2020,Girichidis2021}). The filaments that we observe are statistically unlikely to be very far from the PoS, otherwise their true lengths would be unrealistically large, so we can expect rather weak projection effects on the orientations of filaments. However, the same cannot be said for the magnetic field. Estimating the magnetic field inclination angle to the LoS, $\theta_{B}$, could help improve analyses of the relative orientations between filaments and magnetic fields.


\section{Discussion}
\label{sec:discussion}

Following previous studies \citep{Malinen2016,Carriere2022a,Carriere2022b}, we combined \textit{Planck} dust polarization maps at 7$^{\prime}$ resolution with $N_{\rm{H_2}}$ maps at 36$^{\prime\prime}$ resolution inferred from \textit{Herschel} dust emission observations of star-forming regions. Our aim in this paper was to compare filament orientations to \textbf{\textit{B}$_{\rm{PoS}}$} orientations at different scales (0.03\,pc to a few 
\jonLEt{parsecs}). We used criteria based on a variety of physical parameters inferred from our data to divide our sample of filaments into different subsamples. This separation enabled us to study the effects of each parameter on the HROs, and to better investigate possible links between filaments and magnetic fields. We both examined individual GCC fields (Sect.~\ref{sec:results_individual-fields}) and performed a statistical analysis of the entire set of GCC fields (Sects.~\ref{sec:results2_FilamentHRO} and~\ref{sec:results3_SourceHRO}). We now summarize our main results (Sect.~\ref{sec:discussion_summary_results}) and discuss some physical interpretation (Sect.~\ref{sec:discussion_physical_interpretation}).

\subsection{Summary of our main results}
\label{sec:discussion_summary_results}

The results we obtained for the relative orientations between filaments and \textbf{\textit{B}$_{\rm{PoS}}$} in the GCC fields extend the general trends observed in previous studies, while also revealing a variety of behaviors. This is expected from the complexity of the ISM and the variety of GCC fields in different Galactic environments with different magnetic field properties.

Statistically, we find that low-$N_{\rm{H_2}}$ filaments tend to be roughly parallel to \textbf{\textit{B}$_{\rm{PoS}}$}, whereas high-$N_{\rm{H_2}}$ filaments tend to be roughly perpendicular (top panel of Fig.~\ref{fig:results_filament2DHRO_all}), and the transition occurs at $(N_{\rm{H_2}})_{\rm{t}}\,\approx\,10^{21.7}\,\rm{cm^{-2}}$. Individually, this transition is clearly identified in 40$\%$ of the GCC fields, and the corresponding $(N_{\rm{H_2}})_{\rm{t}}$ values lie in the range $[0.8$, $8]\,\times\,10^{21}\,\rm{cm^{-2}}$, a range that is consistent with results from previous studies\footnote{Values are given with the assumption that $N_{\rm{H}} = 2\,N_{\rm{H_2}}$.}: $(N_{\rm{H_2}})_{\rm{t}}\,\approx\,[2.5, 25]\,\times\,10^{21}\,\rm{cm^{-2}}$ \citep{PlanckXXXV2016}, $(N_{\rm{H_2}})_{\rm{t}}\,\approx 0.8\,\times\,10^{21}\,\rm{cm^{-2}}$ \citep{Malinen2016}, $(N_{\rm{H_2}})_{\rm{t}}\,\approx 0.6\,\times\,10^{21}\,\rm{cm^{-2}}$ \citep{Alina2019}, $(N_{\rm{H_2}})_{\rm{t}}\,\simeq\,[1.1, 1.4]\,\times\,10^{21}\,\rm{cm^{-2}}$ \citep{Carriere2022b}, $(N_{\rm{H_2}})_{\rm{t}}\,\approx\,[0.45, 2.1]\,\times\,10^{21}\,\rm{cm^{-2}}$ \citep{Jiao2024}. However, exact values found for $(N_{\rm{H_2}})_{\rm{t}}$ also depend on the method used to estimate column densities (molecular lines versus dust emission), the adopted dust model, and different dilution factors due to different angular resolutions or distances.

A novel aspect of our analysis is the impact of the most significant filament bar width, $W_{\rm{b}}^{\star}$, on the above trends. Low-$N_{\rm{H_2}}$ filaments remain mostly parallel to \textbf{\textit{B}$_{\rm{PoS}}$} regardless of their $W_{\rm{b}}^{\star}$. In contrast, high-$N_{\rm{H_2}}$ filaments show different behaviors in the HROs depending on their $W_{\rm{b}}^{\star}$: small-$W_{\rm{b}}^{\star}$ filaments have flat, slightly parallel, or slightly perpendicular HROs, whereas large-$W_{\rm{b}}^{\star}$ filaments have flat, bimodal parallel-perpendicular, or mostly perpendicular HROs. However, these general trends are not systematically observed in all samples, with cases where filaments tend to be more parallel to \textbf{\textit{B}$_{\rm{PoS}}$} regardless of their $N_{\rm{H_2}}$ or $W_{\rm{b}}^{\star}$.

Additional criteria for sampling filaments, such as the physical parameters of the GCC fields introduced in Sect.~\ref{sec:properties_field}, can help us better characterize each observed GCC field, its Galactic environment, its evolutionary stage, and their effects on filament relative orientations. We find correlations between some of the parameters. For example, high-latitude fields tend to have shorter distances and lower ${\langle}N_{\rm{H_2}}{\rangle}$, distant fields tend to have higher ${\langle}N_{\rm{H_2}}{\rangle}$ and lower latitudes, high-${\langle}N_{\rm{H_2}}{\rangle}$ fields tend to have lower ${\langle}p{\rangle}$ (linked to confusion along the LoS), etc. Combining the different criteria can help us better characterize the impact of each parameter and find new trends that were previously harder to identify.

We find that GCC fields with higher ${\langle}N_{\rm{H_2}}{\rangle}$ are more likely to have high-$N_{\rm{H_2}}$ filaments, which tend to be more perpendicular to \textbf{\textit{B}$_{\rm{PoS}}$}, while GCC fields with lower ${\langle}N_{\rm{H_2}}{\rangle}$ tend to have low-$N_{\rm{H_2}}$ filaments, which tend to be more parallel to \textbf{\textit{B}$_{\rm{PoS}}$}. GCC fields at higher latitudes, which tend to have lower ${\langle}N_{\rm{H_2}}{\rangle}$, tend to have filaments mostly parallel to \textbf{\textit{B}$_{\rm{PoS}}$}. We see similar results for nearby fields, which have a higher fraction of high-latitude and low-${\langle}N_{\rm{H_2}}{\rangle}$ fields. GCC fields at shorter distances or higher latitudes, where we can expect less confusion along the LoS, tend to have more pronounced trends in the HROs.

Likewise, GCC fields with higher ${\langle}p{\rangle}$ tend to have more pronounced trends in the HROs regardless of the considered sample.
The polarization fraction, $p$, depends on a number of factors, including the properties of dust grains, the grain alignment efficiency (weaker in high-{\NHH} regions; \citealp{Matthews2001,Whittet2008}), fluctuations in the orientation of \textbf{\textit{B}$_{\rm{PoS}}$} (both across the beam and along the LoS; \citealp{PlanckXIX2015,PlanckXX2015,Clark2018}), and most importantly the magnetic field inclination angle, $\theta_{B}$, averaged along the LoS \citep{PlanckXX2015}. Our results suggest that high ${\langle}p{\rangle}$ values are indicative of a magnetic field close to the PoS, while low ${\langle}p{\rangle}$ values could correspond to a magnetic field close to the LoS (see also \citealp{King2018,Fissel2019}). Properly characterizing the different depolarizing effects could help improve estimations of $\theta_{B}$ based on ${\langle}p{\rangle}$.

Aside from our statistical analysis, our individual analysis presented in Sect.~\ref{sec:results_individual-fields} led us to uncover similarities between the observed regions. We identified various structures in molecular clouds (see Fig.~\ref{fig:results_indiv_stat} and ID cards on Zenodo), such as isolated narrow filaments, medium-sized filaments which seem to connect to wider structures, wide filaments, as well as narrow, high-$N_{\rm{H_2}}$ structures inside wide filaments (crests or clumps), and how they are intertwined. The smallest-$W_{\rm{b}}^{\star}$ ($\approx\,0.03\,\rm{pc}$) and lowest-$N_{\rm{H_2}}$ ($\approx\,0.5\,\times\,10^{21}\,\rm{cm^{-2}}$) filaments are either isolated or connected on one end to wider filaments and may line up in striation patterns (see, e.g., G82.65-2.00 or G163.82-8.44). These filaments tend to be roughly perpendicular to neighboring large-$W_{\rm{b}}^{\star}$ filaments with $N_{\rm{H_2}}\gtrsim\,0.8\,\times\,10^{21}\,\rm{cm^{-2}}$. In most GCC fields, the striation patterns are roughly parallel to \textbf{\textit{B}$_{\rm{PoS}}$} \citep[also identified in the Taurus molecular cloud;][]{Goldsmith2008,Narayanan2008,Palmeirim2013}.

However, our individual analysis of the GCC fields also highlights the complexity of molecular cloud morphology. In some GCC fields, we find that most filaments are aligned with each other at all scales (see, e.g., G159, center column of Fig.~\ref{fig:results_indiv_stat}). In other GCC fields (see, e.g., G271, right column of Fig.~\ref{fig:results_indiv_stat}), we observe peculiar structures in both \textit{Herschel} and \textit{Planck} data, such as bow-shaped fronts, colliding filaments, sharp changes in filament or \textbf{\textit{B}$_{\rm{PoS}}$} orientations, entangled \textbf{\textit{B}$_{\rm{PoS}}$}, etc. In these GCC fields, we find that filaments are either mostly parallel to \textbf{\textit{B}$_{\rm{PoS}}$} (see, e.g., G159, center column of Fig.~\ref{fig:results_indiv_stat}; G202.02+2.85; \citealp{Alina2022,Carriere2022b}) or do not have any preferred orientations relative to \textbf{\textit{B}$_{\rm{PoS}}$} (see, e.g., G271, right column of Fig.~\ref{fig:results_indiv_stat}; G159.23-34.51).

As an additional part of our study, we used the catalog of cold cores detected in the 116 GCC fields \citep{Montillaud2015}. The catalog provides a few physical properties of the cores (e.g., mass, size, evolutionary stage, etc.) and of their surroundings (e.g., background column densities). We used this catalog to focus our analysis on filaments with embedded cores, which we subdivided according to the properties of the GCC fields, the filaments, and the cores themselves. In agreement with \cite{Andre2014}, we find that most of the cores in the GCC catalog are located within filaments, particularly within the densest filaments.

Furthermore, we find a positive correlation between ${\langle}N_{\rm{H_2}}{\rangle}$, the fraction of high-$N_{\rm{H_2}}$ pixels in filaments and the number of cores in a GCC field. More evolved molecular clouds tend to have formed more filaments and high-density structures, and the presence of numerous cores can be linked to later stages of cloud evolution. Hence, we used the filament $N_{\rm{H_2}}$, the presence of cores and their estimated evolutionary stages provided in the GCC catalog as different proxies to study the evolution of the relative orientations between filaments and \textbf{\textit{B}$_{\rm{PoS}}$} across multiple stages of the star-forming process.

In our study, we do not estimate filament $N_{\rm{H_2}}$ contrasts, but we can argue that high-$N_{\rm{H_2}}$ filaments in low-${\langle}N_{\rm{H_2}}{\rangle}$ GCC fields have a high $N_{\rm{H_2}}$ contrast, while low-$N_{\rm{H_2}}$ filaments in high-${\langle}N_{\rm{H_2}}{\rangle}$ GCC fields have a low $N_{\rm{H_2}}$ contrast. In low-${\langle}N_{\rm{H_2}}{\rangle}$ GCC fields, we find a transition from mostly parallel to mostly perpendicular with increasing $N_{\rm{H_2}}$. We find this transition for both filaments regardless of the presence of a core, although the trends are slightly less pronounced for filaments with embedded cores. Our results for low-$N_{\rm{H_2}}$ filaments are consistent with those of \cite{Alina2019}, who revealed that low-contrast filaments are mostly parallel to \textbf{\textit{B}$_{\rm{PoS}}$}. However, we are able to identify additional trends for high-$N_{\rm{H_2}}$ filaments, which is probably due to our combining \textit{Planck} data with \textit{Herschel} data at higher angular resolution, a larger sample of filaments and cores, and different scales of filaments.

We find that the HROs for filaments with embedded cores are similar to the HROs for high-$N_{\rm{H_2}}$ filaments (i.e., flat, bimodal parallel-perpendicular or more perpendicular), but the trends are somewhat less pronounced, in particular for cores within high-$N_{\rm{H_2}}$, small-$W_{\rm{b}}^{\star}$ filaments. These results suggest a weaker link between filaments with embedded cores and \textbf{\textit{B}$_{\rm{PoS}}$} at \textit{Planck} scales. We also find that our cores tend to have their major axes mostly parallel to their parent filaments. This trend is less pronounced for cores within large-$W_{\rm{b}}^{\star}$ filaments or for reliable protostellar cores. The latter trend suggests that cores are initially aligned with their parent filaments, and may turn or deform as they evolve (see Fig.~9 in~\citealp{Juvela2018}; see also Sect.~\ref{sec:discussion_physical_interpretation} for more details). Lastly, cores slightly tend to have their major axes parallel to \textbf{\textit{B}$_{\rm{PoS}}$}, but the trend is even less pronounced. We do not observe any trend for the sample of protostellar cores. These results point to an increased misalignment between more evolved filaments and cores and \textbf{\textit{B}$_{\rm{PoS}}$} at \textit{Planck} scales, possibly due to other physical processes that are dominant at smaller scales.

Projection effects need to be taken into account in the study of relative orientations. Observations do not give direct access to the 3D structure of filaments and molecular clouds, but only to their projections onto the PoS. We quantified projection effects using two complementary approaches, the first one is analytical, and the second one consists of simple numerical simulations. We find that PoS projection has a few effects that are intertwined. In fact, if there are no trends in 3D, there will be no trends in 2D. However, the absence of trends in 2D does not imply an absence of trends in 3D. This indicates that projection alone does not create new trends. This contributes to the second effect, where any dispersion in 3D relative orientations is increased by projection onto the PoS, leading to broader HROs. These effects are amplified if either the filaments or the magnetic fields are nearly parallel to the LoS. For a given GCC field, the observed trends are only representative of the real 3D trends if one vector (representing either filaments or \textbf{\textit{B}$_{\rm{PoS}}$}), or preferentially both vectors are close to the PoS. Our statistical study is the best approach to recover the true general 3D trends.

Another possible impact on the study of relative orientations comes from foreground/background contamination, or more generally from confusion along the LoS due to multiple cloud components and/or magnetic field orientations (see, e.g., \citealp{Pelgrims2024,Angarita2025}). This issue was addressed in previous studies, using different approaches to separate magnetic field components along the LoS (see, e.g., \citealp{PlanckXXXV2016,Alina2019}). More recently, \cite{ferriere&mc_2025} presented a new method to decompose the polarized dust emission onto the contributions from several clouds located at different distances along the LoS. They applied their method to one of the GCC fields, i.e., the G139 field, for which they found that the LoS decomposition is critical to obtain the correct magnetic field orientation in the target dense molecular cloud. This case in point shows that LoS contamination may be an important issue in studies of individual molecular clouds, where the foreground or background magnetic field can have an ordered orientation at the cloud scale and induce an apparent preferential relative orientation of filaments.
The impact of LoS contamination is strongly limited in our statistical analysis, which mixes targets selected over a broad range of Galactic locations. A more detailed discussion of this issue is beyond the scope of the present paper.

\subsection{Physical interpretation}
\label{sec:discussion_physical_interpretation}

Our main results on the different relations between $N_{\rm{H_2}}$, $W_{\rm{b}}^{\star}$, and preferred relative orientations between filaments and \textbf{\textit{B}$_{\rm{PoS}}$} support the idea of a hierarchical process in filament formation in which cloud-scale magnetic fields play an important role. Filaments mostly arise from an interplay between turbulence, magnetic fields and self-gravity (for reviews see, e.g, \citealp{Andre2014,Hacar2023,Pattle2023}). 3D MHD simulations show that these different factors dominate at different scales. Cloud-scale motions are dominated by turbulence. Local compressive motions would create cold gas condensations \citep{Padoan2001,Audit2005,Saury2014} and shear would stretch them together with the frozen-in field lines \citep{Hennebelle2013}, giving rise to elongated structures which tend to be parallel to the magnetic field. At later stages, self-gravity takes the upper hand, and causes gas to contract preferentially along field lines. The converging motions result in high-$N_{\rm{H_2}}$ structures that tend to be perpendicular to the magnetic field \citep{Chen2016,Seifried2020}.

The idea of a hierarchical process in filament formation involving magnetic fields arises from previous observational studies: short and narrow, low-$N_{\rm{H_2}}$ filaments tend to be roughly parallel to \textbf{\textit{B}$_{\rm{PoS}}$} and roughly perpendicular to longer and wider filaments \citep{Sugitani2011}, potentially feeding them \citep{Arzoumanian2013,Palmeirim2013,Andre2014,Saajasto2017,Carriere2022b}. This scenario is consistent with predictions of numerical simulations \citep{Balsara2001,Gomez2014}. The different relations between $N_{\rm{H_2}}$, $W_{\rm{b}}^{\star}$, and preferred relative orientations in most of our results support the idea that narrow filaments, which tend to converge onto the local main filaments, could be inflows along the magnetic field.

\cite{Soler2017} showed that a system in ideal-MHD turbulence tends to evolve toward a configuration where density structures are elongated either parallel or perpendicular to the magnetic field. Additionally, it is only in the presence of a relatively strong magnetic field (i.e., magnetic pressure greater than thermal pressure) that they found that compressive motions can produce a transition in the relative orientation between density structures and magnetic fields from mostly parallel at low $N_{\rm{H_2}}$ to mostly perpendicular at high $N_{\rm{H_2}}$. The fact that we observe a transition in many GCC fields suggests that the magnetic field in star-forming regions is generally strong ($B \gtrsim 6-10\,{\mu}\rm{G}$; \citealp{Li2013}). The presence or absence of a transition $N_{\rm{H_2}}$ in our study could be related not only to the strength of the magnetic field, but also to the particular circumstances of the observed region, or to projection effects (as analyzed in Sect.~\ref{sec:projection_effects}; see also \citealp{Seifried2020}).

The GCC fields with peculiar structures for which the relative orientations differ from the usual trends may suggest different 3D geometries or different histories to molecular cloud evolution and filament formation. Since fields in the GCC sample were chosen to be representative of different Galactic environments, they do not always show the trends found in previous studies of local clouds \citep{PlanckXXXV2016, Fissel2019}. First, the deviation from the usual trends can be due to projection effects, as some of the GCC fields might have a magnetic field roughly along the LoS, thereby making trends harder to identify (see Sect.~\ref{sec:projection_effects}). Second, confusion due to multiple clouds and/or magnetic field orientations along the LoS may have more or less impact depending on the considered GCC field.
Third, some targets might be weakly-magnetized molecular clouds, in which turbulence dominates over the magnetic field, \textbf{\textit{B}$_{\rm{PoS}}$} follows $N_{\rm{H_2}}$ structures, and HROs show preferentially parallel trends regardless of the $N_{\rm{H_2}}$ and $W_{\rm{b}}^{\star}$ of filaments \citep{Padoan2001,Soler2013,Soler2017}. Additionally, external events might drive shear and compression and dictate the shape of molecular clouds, such as expanding shells from supernova remnants \citep{Kun2008}, interstellar radiation fields and stellar winds \citep{Bally2008}, and colliding filaments \citep{Montillaud2019,Montillaud2019b}. Subsequent studies of cloud kinematics using molecular spectral data to have information on the LoS velocity could provide new constraints on the dynamics of structures (e.g., distinguishing between converging and diverging motions) to analyze HROs, and better understand molecular cloud and filament evolution.

Observations suggest that the formation of dense cores occurs in two steps. The first step is the formation of filaments, which can already be observed in non-star-forming clouds (e.g., Polaris; \citealp{Menshchikov2010,Andre2014}). The second step is the fragmentation into dense cores of high-density filaments which, through continuous accretion of matter, become gravitationally unstable \citep{Andre2010}. This scenario is consistent with results of numerical simulations (e.g., \citealp{Inutsuka1997}). The dense cores appear elongated, with their major axes roughly parallel to the parent filaments (Sect.~\ref{sec:results3_SourceHRO_s}; see also \citealp{Loren1989a,Loren1989b,Myers1991}). However, in our study, we find that this parallel alignment is less prominent for more evolved cores. Additionally, we do not find any clear correlation between the orientations of dense cores and the orientation of \textbf{\textit{B}$_{\rm{PoS}}$} at \textit{Planck} scales, consistent with results from \cite{WardThompson2023}. Besides, both observations \citep{Perry2025} and numerical simulations \citep{Chen2018} show that magnetic fields at core scales are preferentially perpendicular to the core major axes. These results, combined with our own findings, lead us to propose the following evolutionary scenario: as they further evolve under gravitational collapse, the self-gravitating cores possibly rotate (collapsing cores tend to spin up) and become increasingly misaligned with their parent filaments and the cloud-scale magnetic field. The rotation drags the core-scale magnetic field, which remains roughly perpendicular to the major axes of the rotating cores. These results suggest that magnetic fields are distorted at smaller scales during more advanced stages of star formation.

Numerical simulations of MHD turbulence show that only a very strong magnetic field can preserve its initial orientation across scales, from molecular cloud to protostellar disk scale \citep{Hull2017}. Observations show that cloud-scale magnetic field orientations are not necessarily inherited from the large-scale Galactic magnetic field (e.g., Orion superbubble; \citealp{Soler2018,Soler2019}). Depending on the magnetic field strength, \textbf{\textit{B}$_{\rm{PoS}}$} orientations at core scales may or may not be directly inferred from the cloud-scale \textbf{\textit{B}$_{\rm{PoS}}$}. Hence, cloud-scale magnetic fields may play a lesser role at smaller scales, or other physical processes such as gravity may dominate star formation at core scales, which affects the orientations relative to \textbf{\textit{B}$_{\rm{PoS}}$} (see, e.g., \citealp{Ibanez2022}). Further statistical studies probing magnetic fields at higher angular resolution are necessary to complete this picture.


\section{Conclusion}
\label{sec:conclusion}

In this paper, we applied our filament extraction method {\tt FilDReaMS} to the 116 fields of the \textit{Herschel}-GCC program, which was selected to ensure broad coverage of physical properties and environments. We examined the differences between the filament orientation angles derived with {\tt FilDReaMS} from \textit{Herschel} maps at a resolution of 36$^{\prime\prime}$ and the local PoS magnetic field orientation angles inferred from \textit{Planck} polarization maps at 7$^{\prime}$ resolution, and we performed both an individual analysis of each GCC field and a statistical analysis of the 116 GCC fields together. The main goal of our statistical analysis was to investigate the relative orientations between filaments and magnetic fields in star-forming regions and their possible dependence on the evolutionary stage and on the Galactic environment. To this end, we considered samples of filaments according to different physical criteria and analyzed the relative orientations for each sample.

We first summarize our key findings from our individual analysis. In the individual GCC fields, we found general trends where low-column density ($N_{\rm{H_2}}$) filaments tend to be roughly parallel to \textbf{\textit{B}$_{\rm{PoS}}$}, whereas high-$N_{\rm{H_2}}$ filaments tend to be roughly perpendicular. The transition from parallel to perpendicular relative orientations can be clearly identified in roughly 40$\%$ of the GCC fields. The changes in preferred orientations occur at transition column densities, $(N_{\rm{H_2}})_{\rm{t}}$, in the range $[0.8, 8]\,\times\,10^{21}\,\rm{cm}^{-2}$. However, about 40$\%$ of the GCC fields do not show this usual trend. There is a variety of behaviors, such as a transition in preferred orientations from perpendicular to parallel, filaments remaining parallel at all $N_{\rm{H_2}}$, and more complex $N_{\rm{H_2}}$ or \textbf{\textit{B}$_{\rm{PoS}}$} morphologies. These may be due to different factors, which include projection effects possibly related to the magnetic field inclination to the LoS, confusion along the LoS, fluctuations in the magnetic field orientation, and different strengths of the local magnetic field.

Here, we summarize our key findings in our statistical analysis.
\begin{itemize}
    \item Low-$N_{\rm{H_2}}$ ($N_{\rm{H_2}}\,\leq\,10^{21.5}\,{\rm{cm^{-2}}}$) filaments tend to be roughly parallel to \textbf{\textit{B}$_{\rm{PoS}}$}, whereas high-$N_{\rm{H_2}}$ ($N_{\rm{H_2}}\,>\,10^{21.5}\,{\rm{cm^{-2}}}$) filaments tend to have flat, bimodal parallel-perpendicular, or mostly perpendicular HROs. The bimodal trend can be explained either by the superposition of many GCC fields having different trends or by peculiar fields with both parallel and perpendicular trends.
    \item Low-$N_{\rm{H_2}}$ filaments remain parallel at all scales.
    The trends for high-$N_{\rm{H_2}}$ filaments depend on the most significant width of the model bar ($W_{\rm{b}}^{\star}$). High-$N_{\rm{H_2}}$, small-$W_{\rm{b}}^{\star}$ filaments have flat, slightly parallel, or slightly perpendicular HROs, whereas high-$N_{\rm{H_2}}$, large-$W_{\rm{b}}^{\star}$ filaments have flat, bimodal, or mostly perpendicular HROs, depending on the considered sample.
    \item In most samples, we find that a transition from mostly parallel to mostly perpendicular orientations occurs at $(N_{\rm{H_2}})_{\rm{t}}\,\approx\,10^{21.7}\,\rm{cm^{-2}}$.
    \item Over 93$\%$ of cores from the GCC catalog are located within the highest-$N_{\rm{H_2}}$ filaments. For filaments with embedded cores, we retrieved the same trends as for high-$N_{\rm{H_2}}$ filaments, but the trends are relatively less pronounced. The cores tend to have their major axes mostly parallel to their parent filaments, but this trend is relatively less pronounced for cores within large-$W_{\rm{b}}^{\star}$ filaments or for protostellar cores.
    \item In high-${\langle}p{\rangle}$ GCC fields, we find more pronounced trends, where HROs show a clearer preference toward either parallel or perpendicular geometries. Our results suggest that high ${\langle}p{\rangle}$ values are indicative of a magnetic field closer to the PoS.
    \item The projection of 3D structures onto the PoS broadens the observed HROs. Nonetheless, regardless of projection effects, trends in 2D HROs (preferentially parallel or perpendicular) are, statistically, indicative of true preferences in the 3D relative orientations.
\end{itemize}

Some of the properties used in this statistical study are subject to limitations due to uncertainties in the estimated distances, the models used, and hypotheses regarding the geometry of filaments, cores, and magnetic fields. However, our results consistently show that the transition in relative orientations occurs in denser environments. We emphasize the importance of studying the relative orientations at multiple scales, from wide filaments down to narrow filaments and cores.

To improve our statistical analysis, we could more reliably estimate the different parameters used in this study, include other important physical parameters such as the magnetic field strength and inclination angle, and apply the same method to other observations or numerical simulations. A similar statistical analysis, this time studying magnetic fields at higher resolution, should provide a more complete picture of the impact of magnetic fields at different evolutionary stages and scales. Finally, we could combine this method with molecular spectral data to gain information on the LoS component. For this purpose, we plan to apply the method developed by \cite{ferriere&mc_2025} to all the GCC fields. In this manner, we could better understand the interplay between the different physical processes (such as turbulence, magnetic fields, and gravity) in 3D, which would contribute to a more complete theory of star formation.

\section*{Data availability}

\jonLEt{The results from our statistical analysis are made available in electronic format on Zenodo} \href{https://doi.org/10.5281/zenodo.20119755}{https://doi.org/10.5281/zenodo.20119755}.


\begin{acknowledgements}
    We extend our thanks to our referee for their careful reading of our paper and for their constructive comments and suggestions.
    We acknowledge useful discussions with Douglas J. Marshall, Yasuo Doi. This work was supported by the Thematic Action "Physique et Chimie du Milieu Interstellaire" (PCMI) of INSU Programme National "Astro", with contributions from CNRS Physique $\&$ CNRS Chimie, CEA, and CNES. M.J. acknowledges the support of the Research Council of Finland Grant No. 348342. D.A. acknowledges support of the Faculty Development Competitive Research Grant Program of Nazarbayev University No. 201223FD8821. \textit{Herschel} is an ESA space observatory with science instruments provided by European-led Principal Investigator consortia and with important participation from NASA. \textit{Planck} (\href{http://www.esa.int/Planck}{http://www.esa.int/Planck}) is an ESA science mission with instruments and contributions directly funded by ESA Member States, NASA, and Canada.
\end{acknowledgements}


\bibliographystyle{aa}
\bibliography{references}


\begin{appendix}

\section{Preferred alignment ratios}
\label{sec:appendix_D}

\begin{table*}
    \caption{Values of the ratios $r_{\parallel}$, $r_{\rm int}$, and $r_{\perp}$ (Eq.~\eqref{eq:perpara_ratios}), and $\mathfrak{r}_{\parallel}$, $\mathfrak{r}_{\rm int}$, and $\mathfrak{r}_{\perp}$ (Eq.~\eqref{eq:perpara_ratios_bis}), together with their uncertainties (derived as explained in Sect.~\ref{sec:method_ratios}), for the samples presented in Sects.~\ref{sec:results2_FilamentHRO} and~\ref{sec:results3_SourceHRO}.}
    \vspace{-10pt}
    \centering
    \setlength\extrarowheight{3pt}
    \begin{tabularx}{0.88\textwidth}{l c c c | c c c}
        \label{tab:perpara_ratios_values}
        \centering
        ~ \vspace{3pt} & $r_{\parallel}\,[\%]$ & $r_{{\rm int}}\,[\%]$ & $r_{\perp}\,[\%]$ & $\mathfrak{r}_{\parallel}\,[\%]$ & $\mathfrak{r}_{{\rm int}}\,[\%]$ & $\mathfrak{r}_{\perp}\,[\%]$ \\ \hline \hline
        \rule{0pt}{15pt} Figure~\ref{fig:results_filamentHRO_all} \vspace{3pt} & ~ & ~ & ~ & ~ & ~ & ~ \\ \hline
        all filaments & \colorbox{Gold1}{42.2} ${\scriptstyle \pm\,0.11}$ & 30.8 ${\scriptstyle \pm\,0.11}$ & 27 ${\scriptstyle \pm\,0.10}$ & \colorbox{Gold1}{41.9} ${\scriptstyle \pm\,0.07}$ & 31.7 ${\scriptstyle \pm\,0.08}$ & 26.4 ${\scriptstyle \pm\,0.07}$ \\ \hline
        $N_{\rm H_2} \leq 10^{21.5}\,{\rm cm^{-2}}$ ($N_{\rm H_2}\leq$) & \colorbox{Gold1}{44.2} ${\scriptstyle \pm\,0.11}$ & 30.8 ${\scriptstyle \pm\,0.10}$ & 25 ${\scriptstyle \pm\,0.09}$ & \colorbox{Gold1}{43.9} ${\scriptstyle \pm\,0.09}$ & 31.9 ${\scriptstyle \pm\,0.09}$ & 24.${\scriptstyle \pm\,0.08}$ \\ \hline
        $N_{\rm H_2} > 10^{21.5}\,{\rm cm^{-2}}$ ($N_{\rm H_2}>$) & 35.3 ${\scriptstyle \pm\,0.06}$ & 31 ${\scriptstyle \pm\,0.06}$ & 33.7 ${\scriptstyle \pm\,0.06}$ & 35.6 ${\scriptstyle \pm\,0.15}$ & 31.3 ${\scriptstyle \pm\,0.16}$ & 33.1 ${\scriptstyle \pm\,0.15}$ \\ \hline
        $W_{\rm{b}}^{\star} \leq 0.2\,{\rm pc}$ ($W_{\rm{b}}^{\star}\leq$) & \colorbox{Gold1}{42.1} ${\scriptstyle \pm\,0.07}$ & 32 ${\scriptstyle \pm\,0.06}$ & 25.9 ${\scriptstyle \pm\,0.06}$ & \colorbox{Gold1}{42.2} ${\scriptstyle \pm\,0.14}$ & 32.6 ${\scriptstyle \pm\,0.15}$ & 25.2 ${\scriptstyle \pm\,0.13}$ \\ \hline
        $0.2\,{\rm pc} < W_{\rm{b}}^{\star} \leq 0.5\,{\rm pc}$ ($<W_{\rm{b}}^{\star}\leq$) & \colorbox{Gold1}{42.3} ${\scriptstyle \pm\,0.07}$ & 31.6 ${\scriptstyle \pm\,0.07}$ & 26.1 ${\scriptstyle \pm\,0.06}$ & \colorbox{Gold1}{42.4} ${\scriptstyle \pm\,0.13}$ & 32.0 ${\scriptstyle \pm\,0.14}$ & 25.6 ${\scriptstyle \pm\,0.12}$ \\ \hline
        $W_{\rm{b}}^{\star} > 0.5\,{\rm pc}$ ($W_{\rm{b}}^{\star}>$) & \colorbox{Gold1}{42.1} ${\scriptstyle \pm\,0.08}$ & 29.4 ${\scriptstyle \pm\,0.08}$ & 28.5 ${\scriptstyle \pm\,0.07}$ & \colorbox{Gold1}{41.3} ${\scriptstyle \pm\,0.11}$ & 31.0 ${\scriptstyle \pm\,0.12}$ & 27.7 ${\scriptstyle \pm\,0.10}$ \\ \hline
        $N_{\rm H_2}\leq$ , $W_{\rm{b}}^{\star}\leq$ & \colorbox{Gold1}{43.4} ${\scriptstyle \pm\,0.06}$ & 31.6 ${\scriptstyle \pm\,0.06}$ & 25.0 ${\scriptstyle \pm\,0.05}$ & \colorbox{Gold1}{43.3} ${\scriptstyle \pm\,0.15}$ & 32.4 ${\scriptstyle \pm\,0.16}$ & 24.3 ${\scriptstyle \pm\,0.14}$ \\ \hline
        $N_{\rm H_2}\leq$ , $<W_{\rm{b}}^{\star}\leq$ & \colorbox{Gold1}{43.6} ${\scriptstyle \pm\,0.06}$ & 31 ${\scriptstyle \pm\,0.06}$ & 25.4 ${\scriptstyle \pm\,0.05}$ & \colorbox{Gold1}{43.6} ${\scriptstyle \pm\,0.15}$ & 31.7 ${\scriptstyle \pm\,0.16}$ & 24.7 ${\scriptstyle \pm\,0.14}$ \\ \hline
        $N_{\rm H_2}\leq$ , $W_{\rm{b}}^{\star}>$ & \colorbox{Orange1}{45.5} ${\scriptstyle \pm\,0.07}$ & 29.8 ${\scriptstyle \pm\,0.06}$ & 24.7 ${\scriptstyle \pm\,0.06}$ & \colorbox{Gold1}{44.6} ${\scriptstyle \pm\,0.14}$ & 31.6 ${\scriptstyle \pm\,0.15}$ & 23.8 ${\scriptstyle \pm\,0.12}$ \\ \hline
        $N_{\rm H_2}>$ , $W_{\rm{b}}^{\star}\leq$ & 33 ${\scriptstyle \pm\,0.02}$ & 34.6 ${\scriptstyle \pm\,0.02}$ & 32.4 ${\scriptstyle \pm\,0.02}$ & 34.3 ${\scriptstyle \pm\,0.41}$ & 34.0 ${\scriptstyle \pm\,0.45}$ & 31.7 ${\scriptstyle \pm\,0.40}$ \\ \hline
        $N_{\rm H_2}>$ , $<W_{\rm{b}}^{\star}\leq$ & \colorbox{PeachPuff1}{37.6} ${\scriptstyle \pm\,0.03}$ & 33.4 ${\scriptstyle \pm\,0.03}$ & 29 ${\scriptstyle \pm\,0.03}$ & \colorbox{PeachPuff1}{38.0} ${\scriptstyle \pm\,0.28}$ & 33.3 ${\scriptstyle \pm\,0.30}$ & 28.7 ${\scriptstyle \pm\,0.26}$ \\ \hline
        $N_{\rm H_2}>$ , $W_{\rm{b}}^{\star}>$ & 34.5 ${\scriptstyle \pm\,0.04}$ & 28.7 ${\scriptstyle \pm\,0.04}$ & 36.8 ${\scriptstyle \pm\,0.04}$ & 34.6 ${\scriptstyle \pm\,0.19}$ & 29.5 ${\scriptstyle \pm\,0.21}$ & 35.9 ${\scriptstyle \pm\,0.19}$ \\ \hline
        \rule{0pt}{15pt} Figure~\ref{fig:results_filamentHRO_bothp} \vspace{3pt} & ~ & ~ & ~ & ~ & ~ & ~ \\ \hline
        $N_{\rm H_2}\leq$ , ${\langle}p{\rangle}\leq 5\,\%$ & \colorbox{Gold1}{40.5} ${\scriptstyle \pm\,0.14}$ & 33.7 ${\scriptstyle \pm\,0.13}$ & 25.8 ${\scriptstyle \pm\,0.12}$ & \colorbox{Gold1}{40.5} ${\scriptstyle \pm\,0.13}$ & 34.1 ${\scriptstyle \pm\,0.14}$ & 25.4 ${\scriptstyle \pm\,0.12}$ \\ \hline
        $N_{\rm H_2}\leq$ , ${\langle}p{\rangle}> 5\,\%$ & \colorbox{Orange1}{47.4} ${\scriptstyle \pm\,0.18}$ & 28.4 ${\scriptstyle \pm\,0.16}$ & 24.2 ${\scriptstyle \pm\,0.16}$ & \colorbox{Orange1}{46.5} ${\scriptstyle \pm\,0.11}$ & 30.1 ${\scriptstyle \pm\,0.12}$ & 23.4 ${\scriptstyle \pm\,0.10}$ \\ \hline
        $N_{\rm H_2}>$ , $W_{\rm{b}}^{\star}>$ , ${\langle}p{\rangle}\leq5\,\%$ & 33.9 ${\scriptstyle \pm\,0.08}$ & 30 ${\scriptstyle \pm\,0.07}$ & 36.1 ${\scriptstyle \pm\,0.08}$ & 34.1 ${\scriptstyle \pm\,0.21}$ & 29.9 ${\scriptstyle \pm\,0.23}$ & 36.0 ${\scriptstyle \pm\,0.21}$ \\ \hline
        $N_{\rm H_2}>$ , $W_{\rm{b}}^{\star}>$ , ${\langle}p{\rangle}>5\,\%$ & \colorbox{PeachPuff1}{39.1} ${\scriptstyle \pm\,0.03}$ & 19.8 ${\scriptstyle \pm\,0.03}$ & \colorbox{Gold1}{41.1} ${\scriptstyle \pm\,0.04}$ & \colorbox{PeachPuff1}{37.8} ${\scriptstyle \pm\,0.45}$ & 27.3 ${\scriptstyle \pm\,0.50}$ & 34.9 ${\scriptstyle \pm\,0.48}$ \\ \hline
        \rule{0pt}{15pt} Figure~\ref{fig:results_filamentHRO_d+_NH2+_p+} \vspace{3pt} & ~ & ~ & ~ & ~ & ~ & ~ \\ \hline
        all filaments & \colorbox{Gold1}{41.6} ${\scriptstyle \pm\,0.46}$ & 30 ${\scriptstyle \pm\,0.43}$ & 28.4 ${\scriptstyle \pm\,0.42 }$ & \colorbox{Gold1}{40.2} ${\scriptstyle \pm\,0.27}$ & 31.4 ${\scriptstyle \pm\,0.28}$ & 28.4 ${\scriptstyle \pm\,0.25}$ \\ \hline
        $N_{\rm H_2}\leq$ & \colorbox{Gold1}{44.5} ${\scriptstyle \pm\,0.43}$ & 29.7 ${\scriptstyle \pm\,0.40}$ & 25.8 ${\scriptstyle \pm\,0.38}$ & \colorbox{Gold1}{42.6} ${\scriptstyle \pm\,0.29}$ & 31.2 ${\scriptstyle \pm\,0.30}$ & 26.2 ${\scriptstyle \pm\,0.26}$ \\ \hline
        $N_{\rm H_2}>$ & 21.7 ${\scriptstyle \pm\,0.14}$ & 31.8 ${\scriptstyle \pm\,0.15}$ & \colorbox{Orange1}{46.5} ${\scriptstyle \pm\,0.16}$ & 23.6 ${\scriptstyle \pm\,0.66}$ & 32.6 ${\scriptstyle \pm\,0.80}$ & \colorbox{Gold1}{43.8} ${\scriptstyle \pm\,0.78}$ \\ \hline
        $N_{\rm H_2}>$ , $W_{\rm{b}}^{\star} \leq 0.5\,{\rm pc}$ & 18.1 ${\scriptstyle \pm\,0.07}$ & 32.4 ${\scriptstyle \pm\,0.09}$ & \colorbox{Orange1}{49.5} ${\scriptstyle \pm\,0.09}$ & 19.5 ${\scriptstyle \pm\,1.23}$ & 33.6 ${\scriptstyle \pm\,1.48}$ & \colorbox{Orange1}{46.9} ${\scriptstyle \pm\,1.38}$ \\ \hline
        $N_{\rm H_2}>$ , $W_{\rm{b}}^{\star} > 0.5\,{\rm pc}$ & 23 ${\scriptstyle \pm\,0.12}$ & 31.5 ${\scriptstyle \pm\,0.13}$ & \colorbox{Orange1}{45.5} ${\scriptstyle \pm\,0.14}$ & 25.2 ${\scriptstyle \pm\,0.77}$ & 32.2 ${\scriptstyle \pm\,0.94}$ & \colorbox{Gold1}{42.6} ${\scriptstyle \pm\,0.91}$ \\ \hline
        \rule{0pt}{15pt} Figure~\ref{fig:results_sourceHRO_d+_p+_density}\vspace{3pt}  & ~ & ~ & ~ & ~ & ~ & ~ \\ \hline
        all filaments w/ cores & 31.5 ${\scriptstyle \pm\,1.98}$ & 29.3 ${\scriptstyle \pm\,1.92}$ & \colorbox{PeachPuff1}{39.2} ${\scriptstyle \pm\,2.05}$ & 32.0 ${\scriptstyle \pm\,1.08}$ & 31.0 ${\scriptstyle \pm\,1.24}$ & \colorbox{PeachPuff1}{37.0} ${\scriptstyle \pm\,1.18}$ \\ \hline
        $n_{\rm H,c}\leq 10^{3.5}\,{\rm cm^{-3}}$ & 33.3 ${\scriptstyle \pm\,1.83}$ & 30.1 ${\scriptstyle \pm\,1.75}$ & 36.6 ${\scriptstyle \pm\,1.88}$ & 34.0 ${\scriptstyle \pm\,1.18}$ & 31.9 ${\scriptstyle \pm\,1.34}$ & 34.1 ${\scriptstyle \pm\,1.26}$ \\ \hline
        $n_{\rm H,c}> 10^{3.5}\,{\rm cm^{-3}}$ & 14.1 ${\scriptstyle \pm\,0.49}$ & 25 ${\scriptstyle \pm\,0.61}$ & \colorbox{Orange3}{60.9} ${\scriptstyle \pm\,0.70}$ & 15.0 ${\scriptstyle \pm\,2.29}$ & 26.4 ${\scriptstyle \pm\,3.31}$ & \colorbox{Orange3}{58.6} ${\scriptstyle \pm\,3.35}$ \\ \hline
        \rule{0pt}{15pt} Figure~\ref{fig:results_sourceHRO_evolstage} \vspace{3pt} & ~ & ~ & ~ & ~ & ~ & ~ \\ \hline
        starless (0) & \colorbox{PeachPuff1}{37.3} ${\scriptstyle \pm\,1.43}$ & 32.6 ${\scriptstyle \pm\,1.41}$ & 30.1 ${\scriptstyle \pm\,1.37}$ & \colorbox{PeachPuff1}{37.7} ${\scriptstyle \pm\,0.83}$ & 33.0 ${\scriptstyle \pm\,0.91}$ & 29.3 ${\scriptstyle \pm\,0.81}$ \\ \hline
        protostellar (0) & 27.7 ${\scriptstyle \pm\,2.31}$ & 32.6 ${\scriptstyle \pm\,2.35}$ & \colorbox{PeachPuff1}{39.7} ${\scriptstyle \pm\,2.55}$ & 28.8 ${\scriptstyle \pm\,1.44}$ & 31.6 ${\scriptstyle \pm\,1.65}$ & \colorbox{PeachPuff1}{39.6} ${\scriptstyle \pm\,1.55}$ \\ \hline
        protostellar (1) & \colorbox{PeachPuff1}{38.6} ${\scriptstyle \pm\,5.47}$ & 30.1 ${\scriptstyle \pm\,5.16}$ & 31.3 ${\scriptstyle \pm\,5.23}$ & \colorbox{PeachPuff1}{39.5} ${\scriptstyle \pm\,3.10}$ & 32.9 ${\scriptstyle \pm\,3.34}$ & 27.6 ${\scriptstyle \pm\,2.90}$ \\ \hline
        \rule{0pt}{15pt} Figure~\ref{fig:results_sourceHRO_sourcePA_f} \vspace{3pt} & ~ & ~ & ~ & ~ & ~ & ~ \\ \hline
        $W_{\rm{b}}^{\star}\leq$ & \colorbox{Orange3}{64.2} ${\scriptstyle \pm\,0.41}$ & 20.5 ${\scriptstyle \pm\,0.34}$ & 15.3 ${\scriptstyle \pm\,0.31}$ & \colorbox{Orange3}{63.2} ${\scriptstyle \pm\,0.84}$ & 22.7 ${\scriptstyle \pm\,0.84}$ & 14.1 ${\scriptstyle \pm\,0.63}$ \\ \hline
        $<W_{\rm{b}}^{\star}\leq$ & \colorbox{Orange3}{55.8} ${\scriptstyle \pm\,0.48}$ & 27.4 ${\scriptstyle \pm\,0.43}$ & 16.8 ${\scriptstyle \pm\,0.36}$ & \colorbox{Orange3}{56.8} ${\scriptstyle \pm\,0.81}$ & 27.5 ${\scriptstyle \pm\,0.83}$ & 15.7 ${\scriptstyle \pm\,0.64}$ \\ \hline
        $W_{\rm{b}}^{\star}>$ & \colorbox{Orange1}{47.7} ${\scriptstyle \pm\,0.55}$ & 29.7 ${\scriptstyle \pm\,0.49}$ & 22.6 ${\scriptstyle \pm\,0.45}$ & \colorbox{Orange1}{48.2} ${\scriptstyle \pm\,0.75}$ & 29.6 ${\scriptstyle \pm\,0.78}$ & 22.2 ${\scriptstyle \pm\,0.65}$ \\ \hline
        \rule{0pt}{15pt} Figure~\ref{fig:results_sourceHRO_sourcePA_b} \vspace{3pt} & ~ & ~ & ~ & ~ & ~ & ~ \\ \hline
        all filaments & 36.0 ${\scriptstyle \pm\,0.76}$ & 33.3 ${\scriptstyle \pm\,0.77}$ & 30.7 ${\scriptstyle \pm\,0.74}$ & 35.9 ${\scriptstyle \pm\,0.46}$ & 32.9 ${\scriptstyle \pm\,0.51}$ & 31.2 ${\scriptstyle \pm\,0.45}$ \\ \hline
        $N_{\rm H_2}\leq$ & \colorbox{PeachPuff1}{37.3} ${\scriptstyle \pm\,0.47}$ & 33.6 ${\scriptstyle \pm\,0.46}$ & 29.1 ${\scriptstyle \pm\,0.44}$ & \colorbox{PeachPuff1}{37.2} ${\scriptstyle \pm\,0.61}$ & 33.0 ${\scriptstyle \pm\,0.66}$ & 29.8 ${\scriptstyle \pm\,0.59}$ \\ \hline
        $N_{\rm H_2}>$ & 34.2 ${\scriptstyle \pm\,0.61}$ & 32.9 ${\scriptstyle \pm\,0.61}$ & 32.9 ${\scriptstyle \pm\,0.59}$ & 33.8 ${\scriptstyle \pm\,0.71}$ & 32.9 ${\scriptstyle \pm\,0.78}$ & 33.3 ${\scriptstyle \pm\,0.71}$ \\
    \end{tabularx}
\end{table*}

\section{ID cards for G219, G159, G271, and G92}
\label{sec:appendix_A}

\begin{figure*}
    \centering
    \includegraphics[height=0.89\textheight]{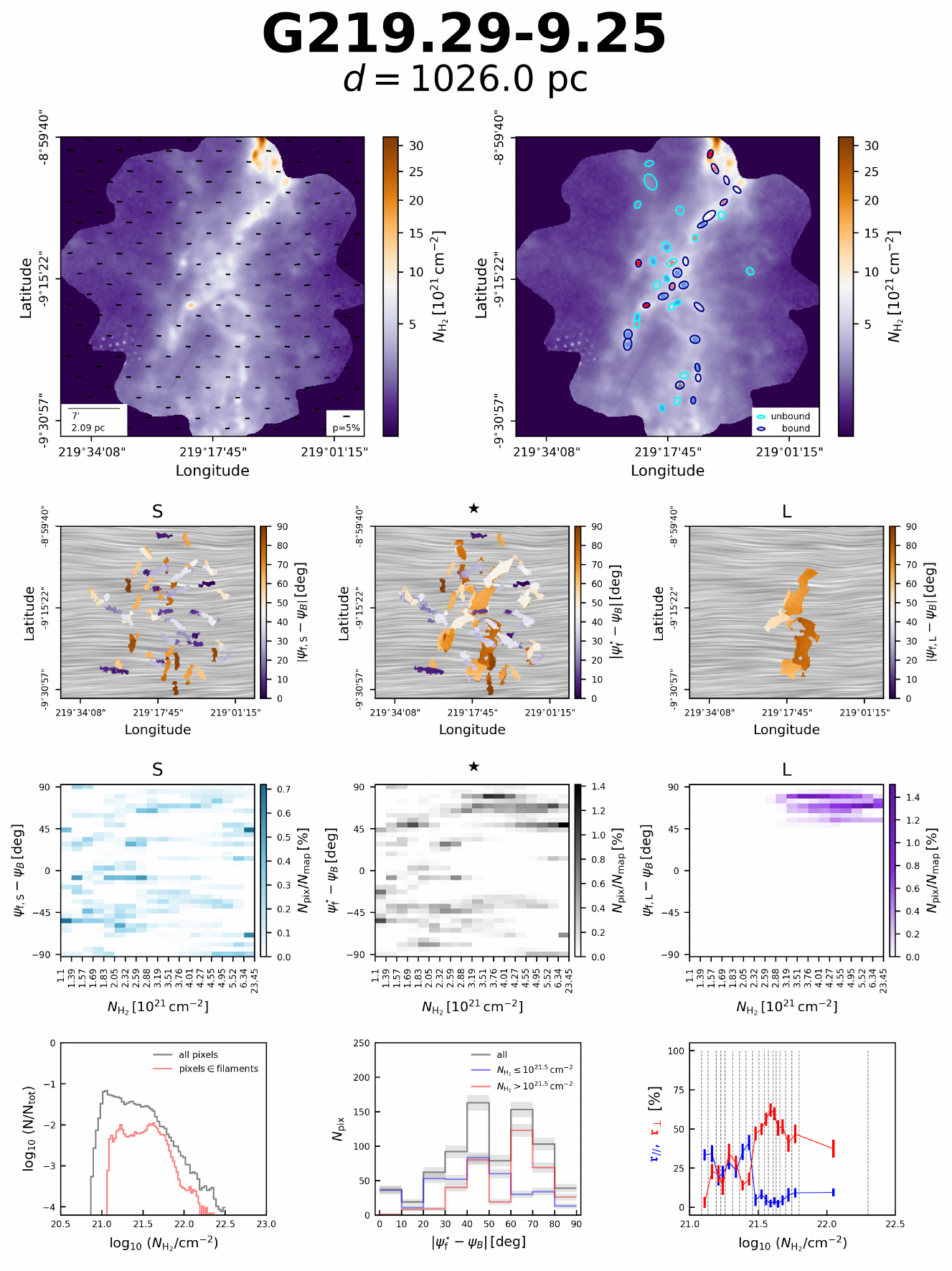}
    \caption{ID card for the \textit{Herschel} G219 field. \textbf{First row: left -} \textit{Herschel} $N_{\rm{H_2}}$ map with \textbf{\textit{B}$_{\rm{PoS}}$} visualized using vectors with lengths proportional to the polarization fraction; \textbf{right -} \textit{Herschel} $N_{\rm{H_2}}$ map with the FWHM ellipses of bound (blue) and unbound (cyan) cores from the GCC catalog. \textbf{Second row:} absolute value of the relative orientations between \textbf{\textit{B}$_{\rm{PoS}}$} and the smallest filaments \textbf{(left)}, the most significant filaments \textbf{(middle)}, or the largest filaments \textbf{(right)} respectively. \textbf{Third row:} Two-dimensional histograms as a function of $N_{\rm{H_2}}$, of the relative orientations between \textbf{\textit{B}$_{\rm{PoS}}$} and the smallest filaments \textbf{(left)}, the most significant filaments \textbf{(middle)}, or the largest filaments \textbf{(right)} respectively. \textbf{Last row: left -} Histogram of $N_{\rm{H_2}}$ of all pixels (black) and pixels belonging to filaments (red); \textbf{middle -} Same as Fig.~\ref{fig:results_filamentHRO_all}a for filaments in G219; \textbf{right -} Same as the bottom panel of Fig.~\ref{fig:results_filament2DHRO_all} for filaments in G219.}
    \label{fig_appendix:G219}
\end{figure*}

\begin{figure*}
    \centering
    \includegraphics[height=0.89\textheight]{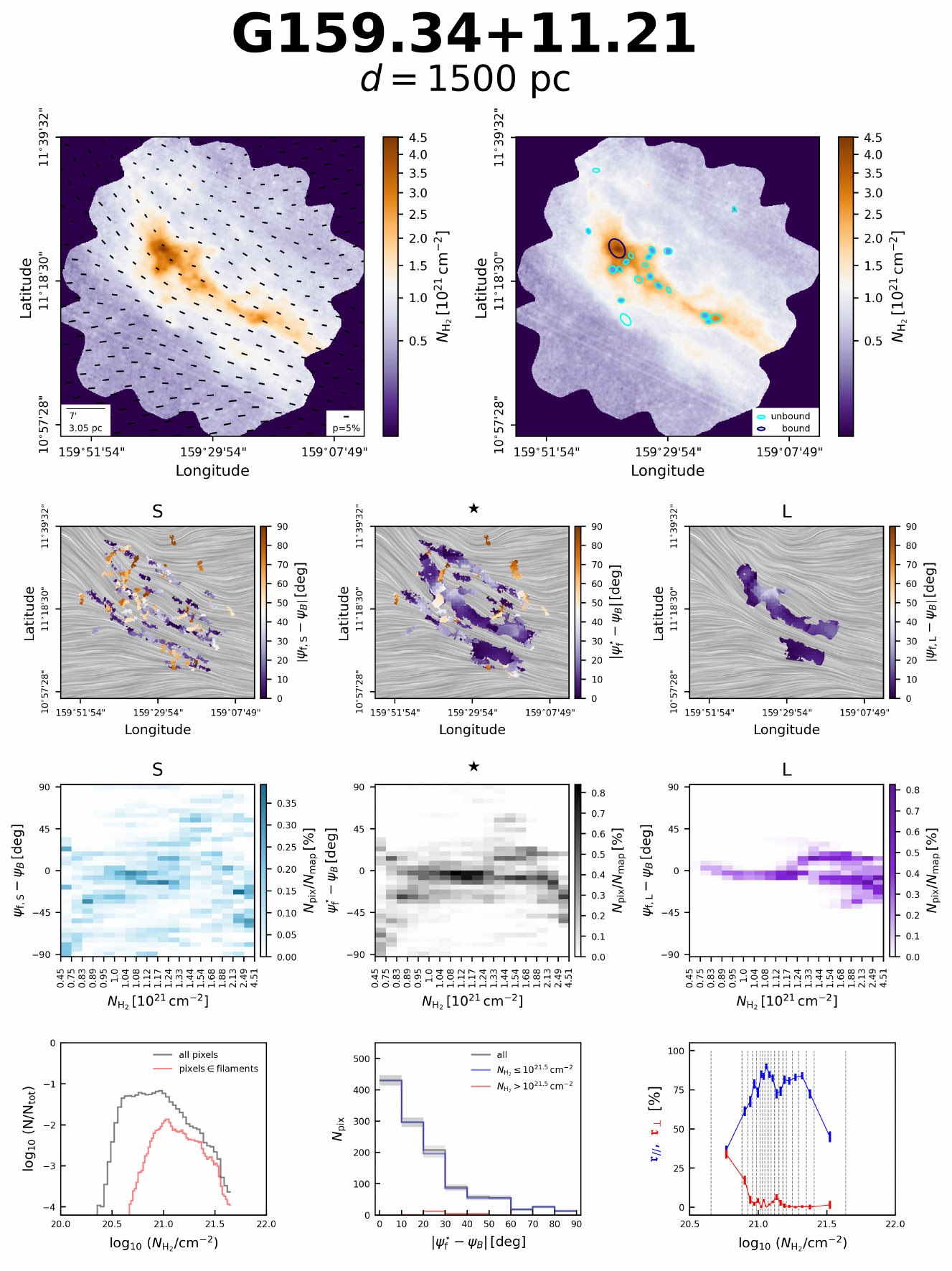}
    \caption{ID card for G159.}
    \label{fig_appendix:G159}
\end{figure*}

\begin{figure*}
    \centering
    \includegraphics[height=0.89\textheight]{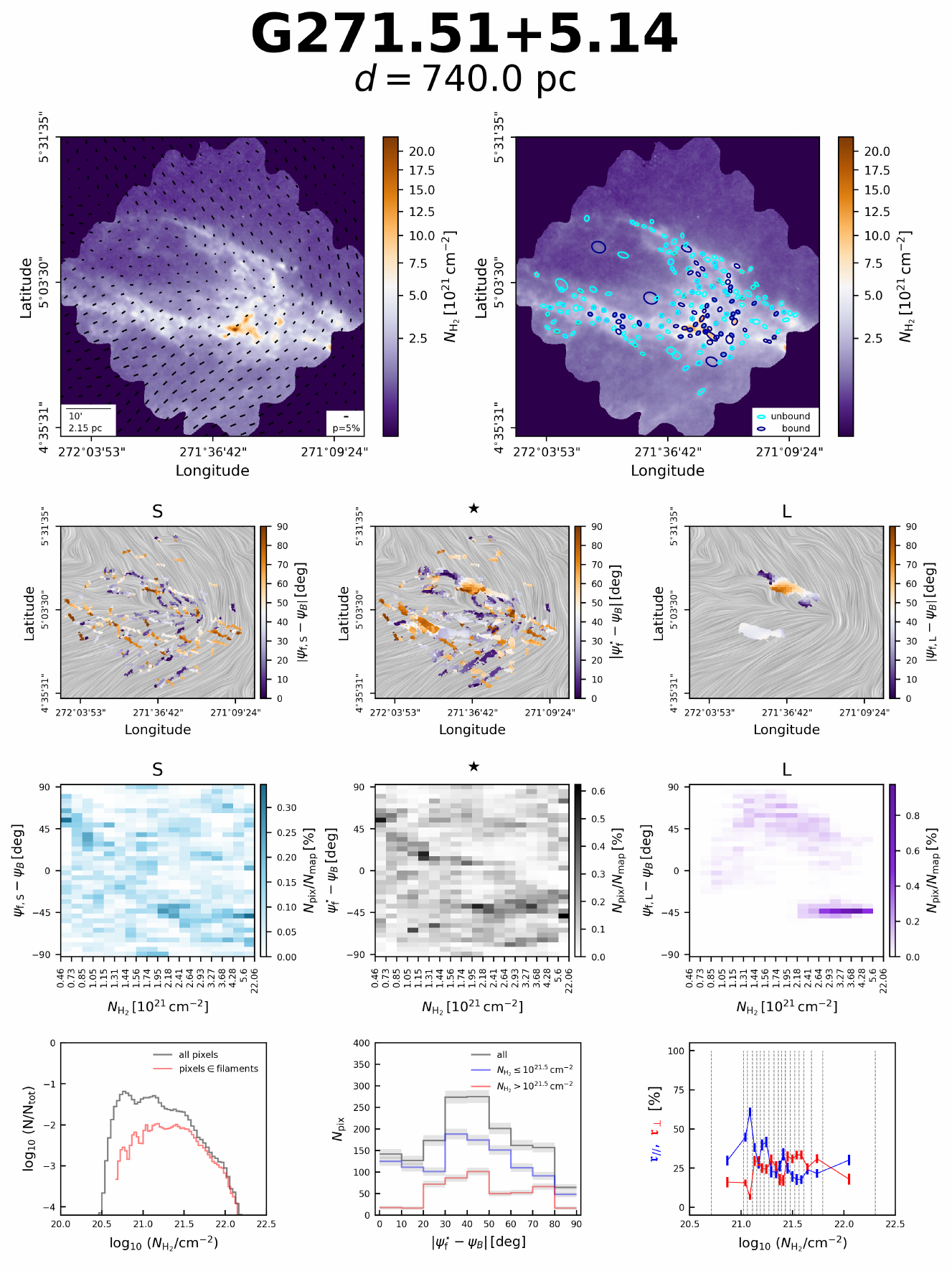}
    \caption{ID card for G271.}
    \label{fig_appendix:G271}
\end{figure*}

\begin{figure*}
    \centering
    \includegraphics[height=0.89\textheight]{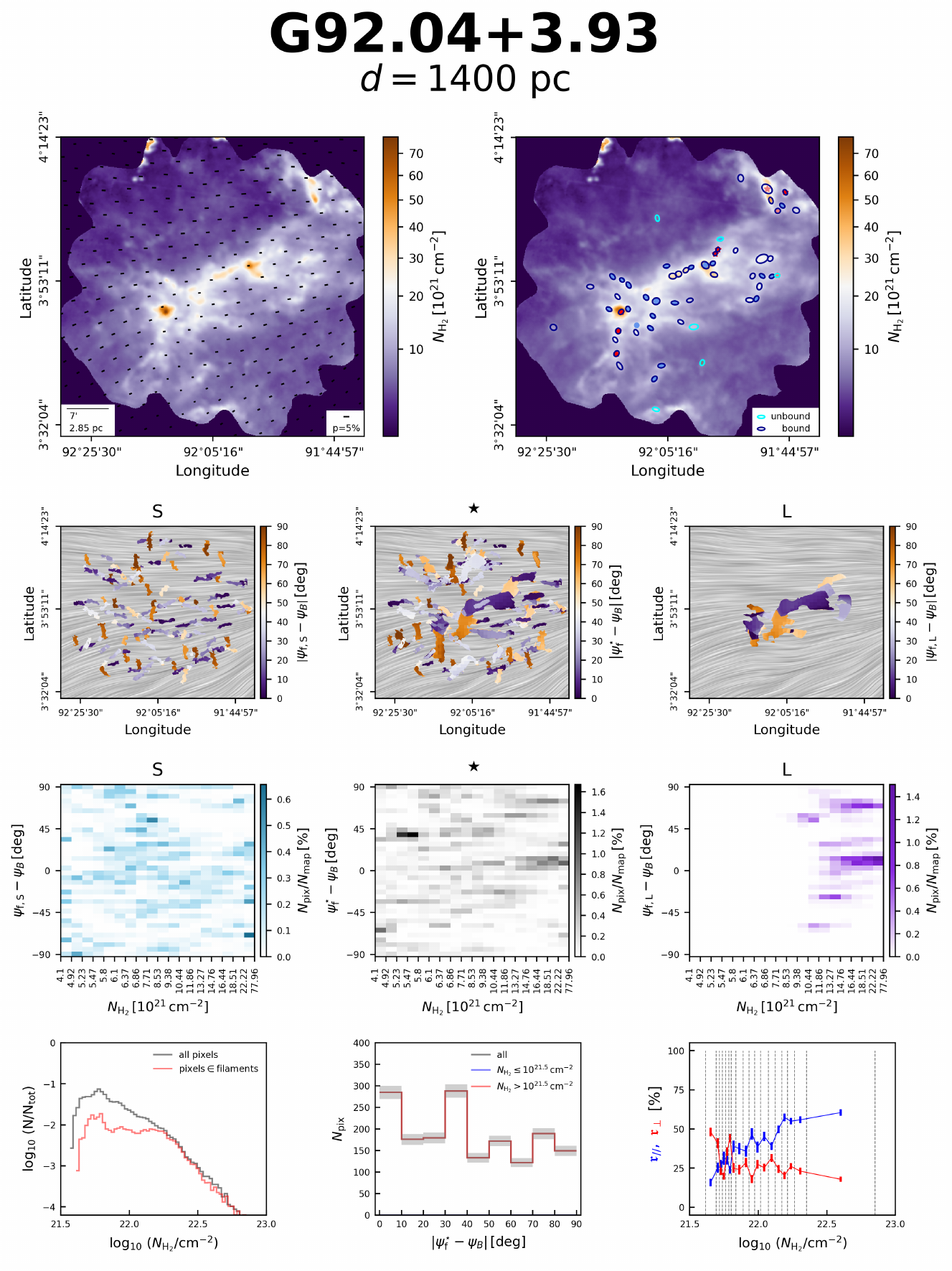}
    \caption{ID card for G92.}
    \label{fig_appendix:G92_04}
\end{figure*}

In Section~\ref{sec:results_3fields}, we showed a summary of the main results obtained when applying {\tt FilDReaMS} to the H$_2$ column density maps of three GCC fields. The main results are presented in the form of ID cards for each GCC field. In this appendix, we show the ID cards for the GCC fields presented in Sect.~\ref{sec:results_3fields} and in Fig.~\ref{fig:results_indiv_stat}, and the ID card for G92, discussed in Sect.~\ref{sec:results_overview-indivfields}. Fig.~\ref{fig_appendix:G219} corresponds to the ID card for G219 (left column of Fig.~\ref{fig:results_indiv_stat}), Fig.~\ref{fig_appendix:G159} corresponds to the ID card for G159 (middle column of Fig.~\ref{fig:results_indiv_stat}), Fig.~\ref{fig_appendix:G271} corresponds to the ID card for G271 (right column of Fig.~\ref{fig:results_indiv_stat}), and Fig.~\ref{fig_appendix:G92_04} corresponds to the ID card for G92. ID cards for the 116 \textit{Herschel} GCC fields can be found \jonLEt{on Zenodo}.

\section{Additional figures for the physical properties}
\label{sec:appendix_B}

We provide extra figures to illustrate the distributions of the physical properties described in Sect.~\ref{sec:properties}.

\begin{figure}
    \centering
    \includegraphics[width=\hsize]{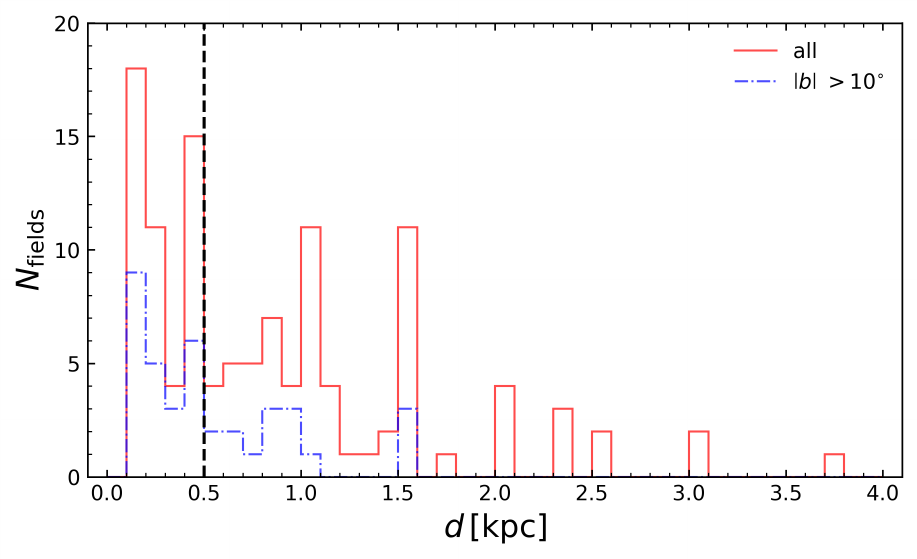}
    \caption{Histograms of distance estimates of the GCC fields. The dashed vertical line represents the threshold at $500\,\rm{pc}$ (see Sect.~\ref{sec:properties_field}).}
    \label{fig_appendix:field_d_b_hist}
\end{figure}

\begin{figure}
    \centering
    \includegraphics[width=\hsize]{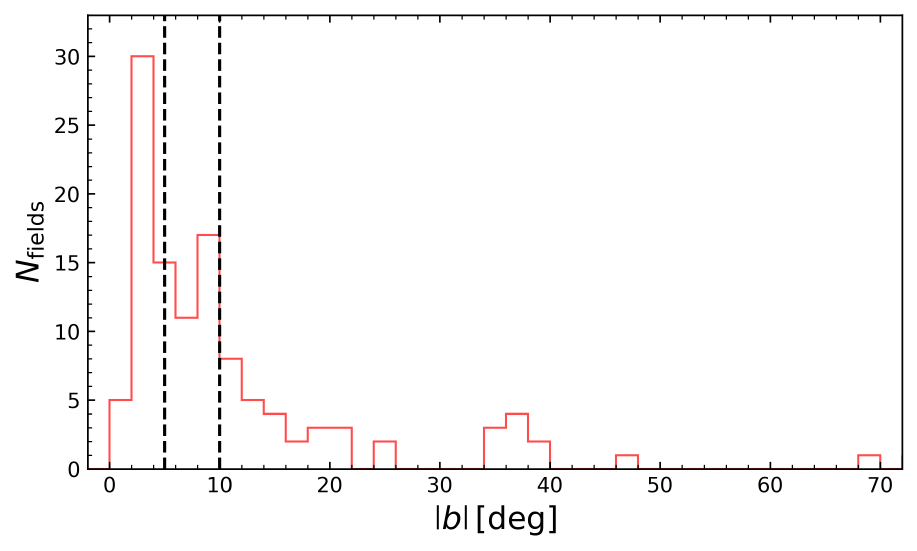}
    \caption{Histogram of absolute latitudes of the GCC fields. The dashed vertical lines represent the thresholds at $5^{\circ}$ and $10^{\circ}$ respectively (see Sect.~\ref{sec:properties_field}).}
    \label{fig_appendix:field_b_hist}
\end{figure}

\begin{figure}
    \centering
    \includegraphics[width=\hsize]{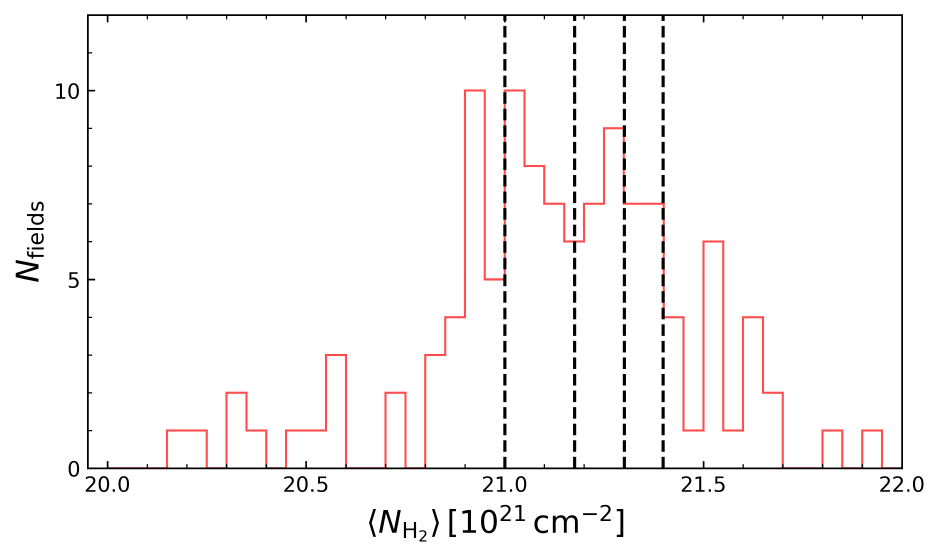}
    \caption{Histogram of ${\langle}N_{\rm{H_2}}{\rangle}$ of the GCC fields. The dashed vertical lines represent the thresholds at $10^{21}$, $1.5\,\times\,10^{21}$, $2\,\times\,10^{21}$ and $2.5\,\times\,10^{21}\,\rm{cm^{-2}}$ respectively (see Sect.~\ref{sec:properties_field}).}
    \label{fig_appendix:field_logNH2avg_hist}
\end{figure}

\begin{figure}
    \centering
    \includegraphics[width=\hsize]{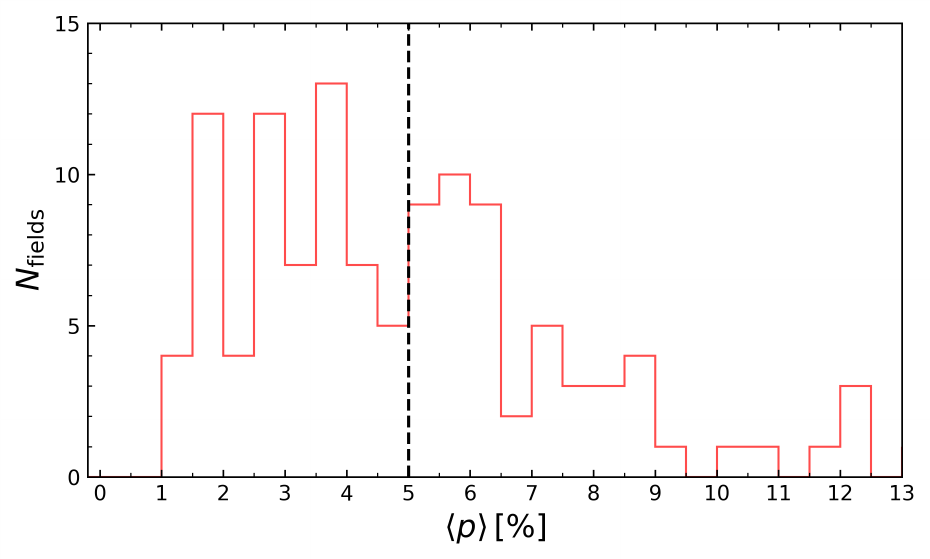}
    \caption{Histogram of ${\langle}p{\rangle}$ of the GCC fields. The dashed vertical line represents the threshold at $5\%$ (see Sect.~\ref{sec:properties_field}).}
    \label{fig_appendix:field_pavg_hist}
\end{figure}

\begin{figure}
    \centering
    \includegraphics[width=\hsize]{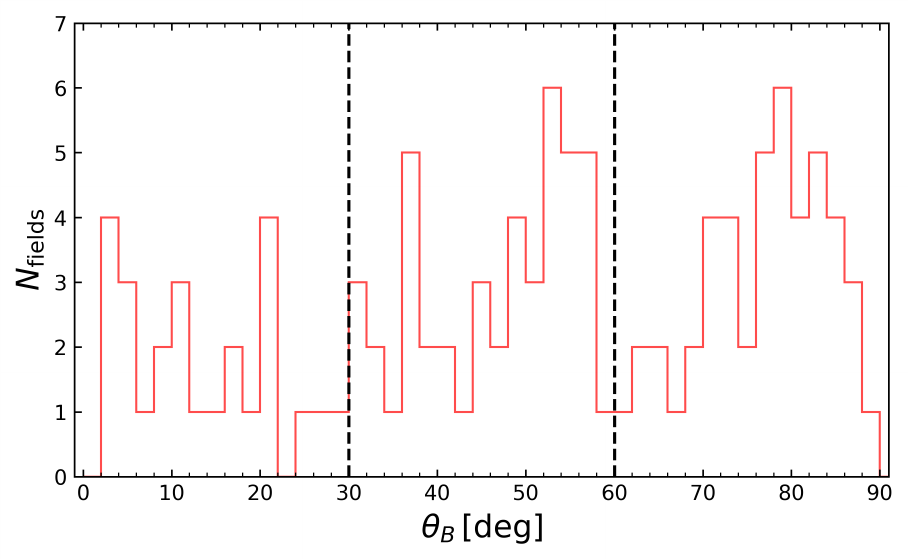}
    \caption{Histogram of $\theta_{B}$ in the GCC fields. The dashed vertical lines represent the thresholds at $30^{\circ}$ and $60^{\circ}$ respectively (see Sect.~\ref{sec:properties_field}).}
    \label{fig_appendix:field_thetab_hist}
\end{figure}

\begin{figure}
    \centering
    \includegraphics[width=\hsize]{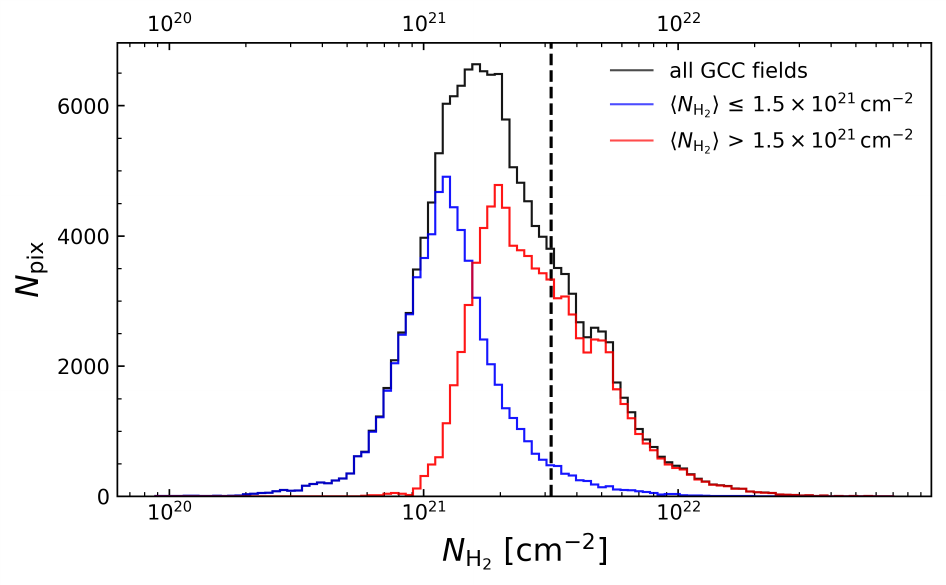}
    \caption{Histograms of $N_{\rm{H_2}}$ of filaments from all the GCC fields (black), filaments from GCC fields with ${\langle}N_{\rm{H_2}}{\rangle}\,\leq\,1.5\,\times\,10^{21}\,\rm{cm^{-2}}$ (blue) and filaments from GCC fields with ${\langle}N_{\rm{H_2}}{\rangle}\,>\,1.5\,\times\,10^{21}\,\rm{cm^{-2}}$ (red). The dashed vertical line represents the threshold at $10^{21.5}\,\rm{cm^{-2}}$}
    \label{fig_appendix:filNH2_NH2avg_hist}
\end{figure}

\begin{figure}
    \centering
    \includegraphics[width=\hsize]{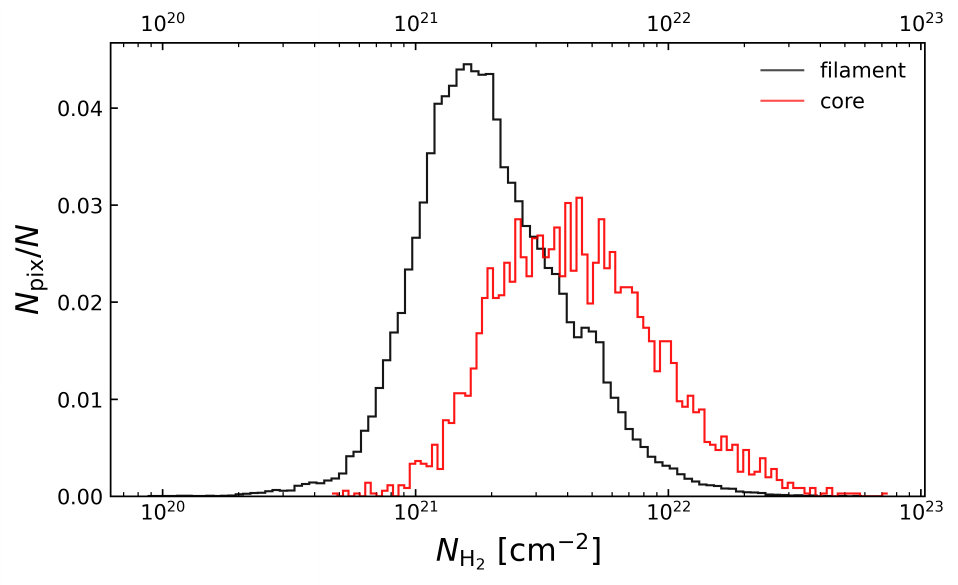}
    \caption{Normalized histograms of $N_{\rm{H_2}}$ of all filaments (red) from the GCC fields and filaments hosting cores (black) from the GCC catalog.}
    \label{fig_appendix:filNH2_sourceNH2_hist}
\end{figure}

\begin{figure}
   \centering
   \includegraphics[width=\hsize]{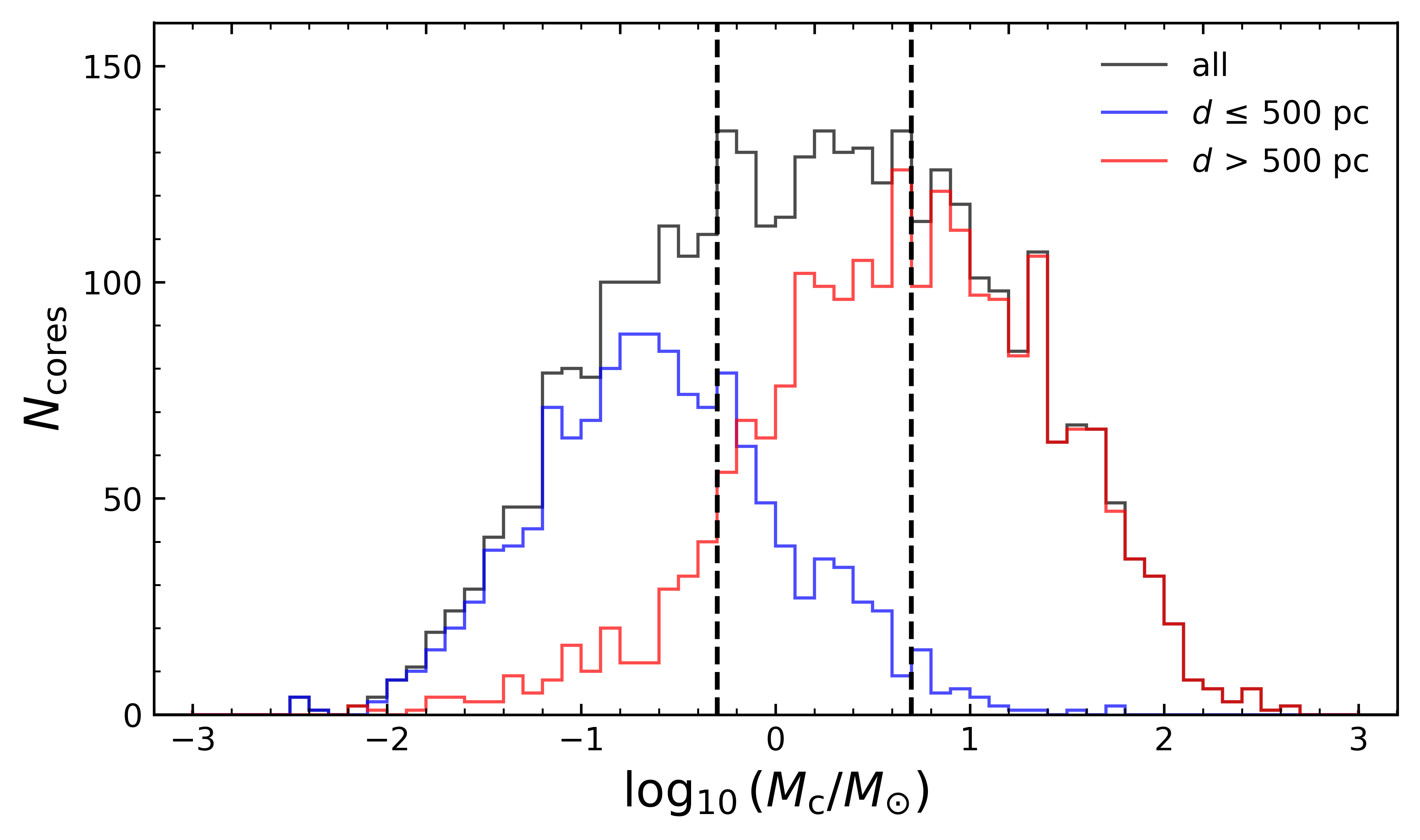}
      \caption{Histograms of core mass for all fields (black), $d \leq 500\,\rm{pc}$ fields (blue) and $d > 500\,\rm{pc}$ fields (red). The dashed vertical lines represent the thresholds at $0.5\,M_{\odot}$ and $5\,M_{\odot}$ (see Sect.~\ref{sec:properties_source}).}
         \label{fig_appendix:results_source_logmass_d_hist}
   \end{figure}

\begin{figure}
    \centering
    \includegraphics[width=\hsize]{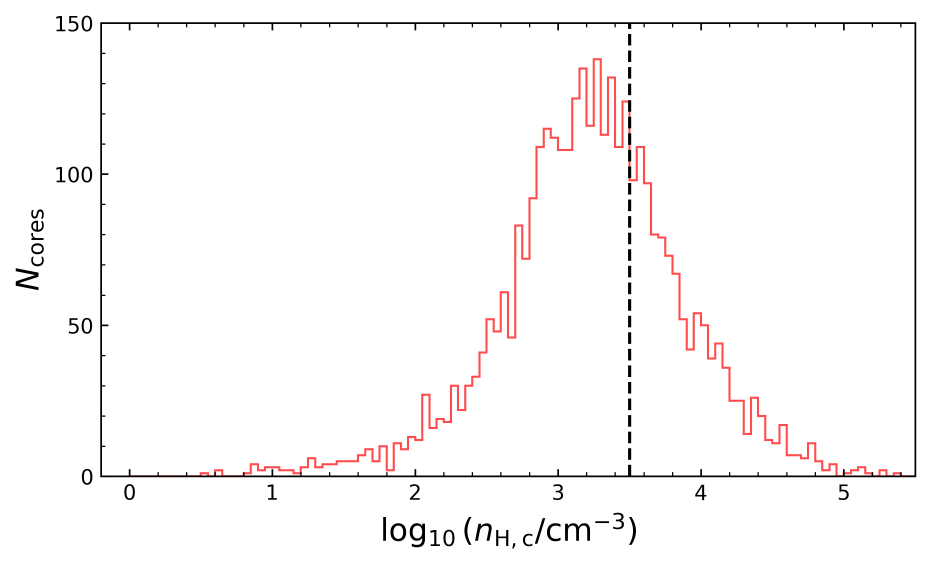}
    \caption{Histogram of $n_{\rm{H}}$ of all cores from the GCC catalog. The dashed vertical line represents the threshold at $10^{3.5}\,\rm{cm^{-3}}$ (see Sect.~\ref{sec:properties_source}).}
    \label{fig_appendix:sourcedensity_hist}
\end{figure}

\section{Additional HROs}
\label{sec:appendix_C}

We provide extra figures to illustrate the results presented in Sect.~\ref{sec:results2_FilamentHRO}.

\begin{figure}
    \centering
    \includegraphics[width=\hsize]{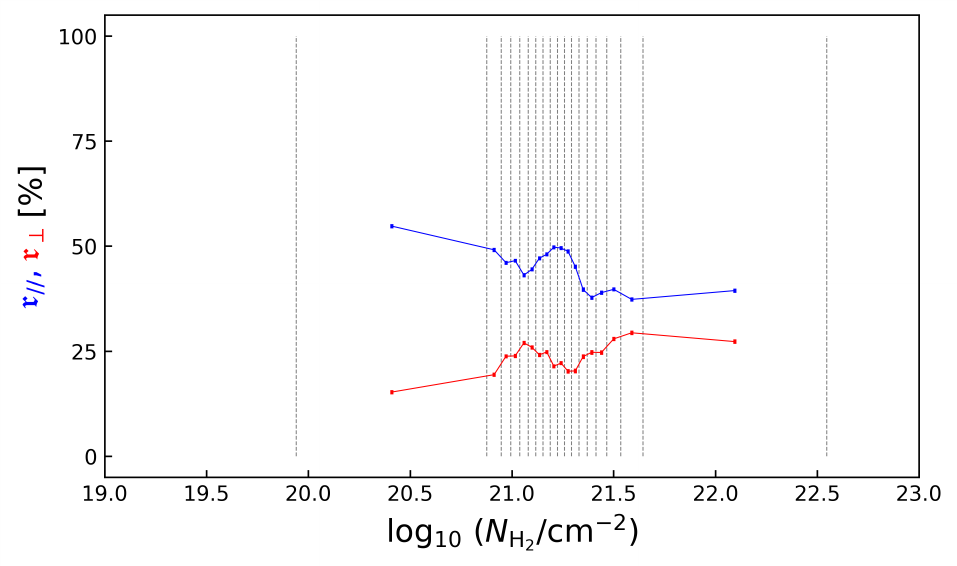}
    \caption{Values of $\mathfrak{r}_{\parallel}$ and $\mathfrak{r}_{\perp}$ (defined in Eq.~\eqref{eq:perpara_ratios_bis}) in 18 $N_{\rm{H_2}}$ bins, for GCC fields with $d \leq 500\,{\rm pc}$.}
    \label{fig_appendix:results_filamentperparaHRO_d-}
\end{figure}

\begin{figure}
    \centering
    \includegraphics[width=\hsize]{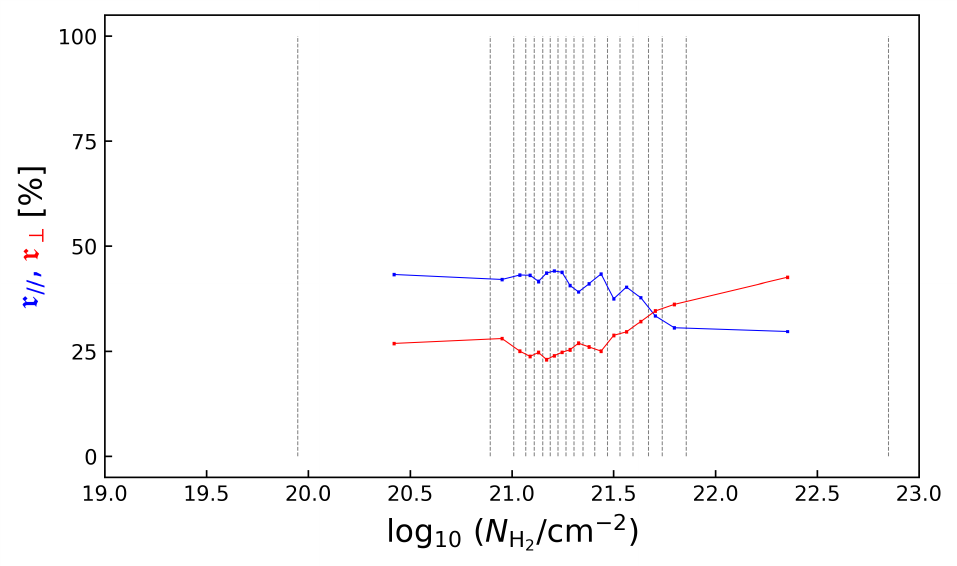}
    \caption{Values of $\mathfrak{r}_{\parallel}$ and $\mathfrak{r}_{\perp}$ (defined in Eq.~\eqref{eq:perpara_ratios_bis}) in 18 $N_{\rm{H_2}}$ bins, for GCC fields with $d > 500\,{\rm pc}$.}
    \label{fig_appendix:results_filamentperparaHRO_d+}
\end{figure}

\begin{figure}
    \centering
    \includegraphics[width=\hsize]{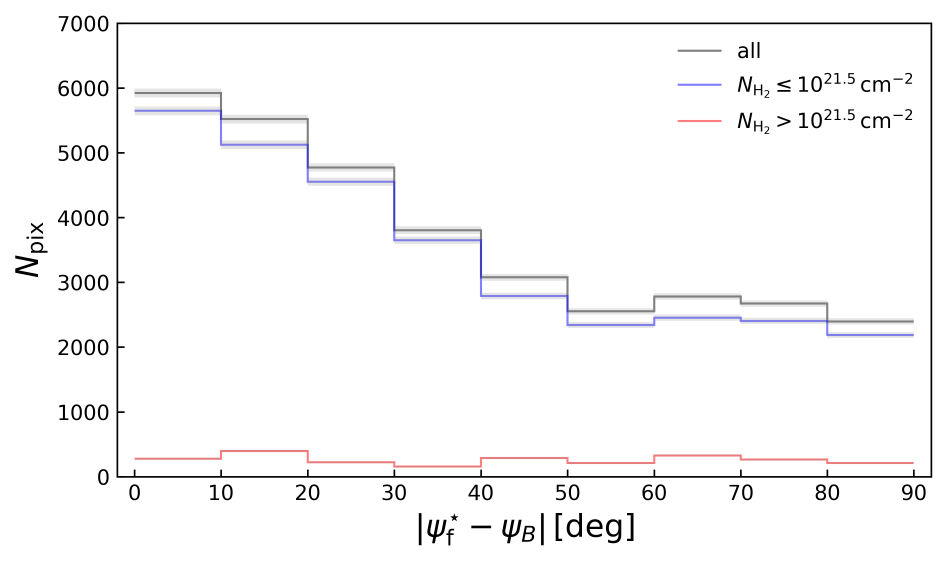}
    \caption{Same HROs as Fig.~\ref{fig:results_filamentHRO_all}a for GCC fields with $\left | b \right | > 10^{\circ}$. The uncertainties are shown in gray shaded areas.}
    \label{fig_appendix:results_filamentHRO_b+}
\end{figure}

\begin{figure}
    \centering
    \includegraphics[width=\hsize]{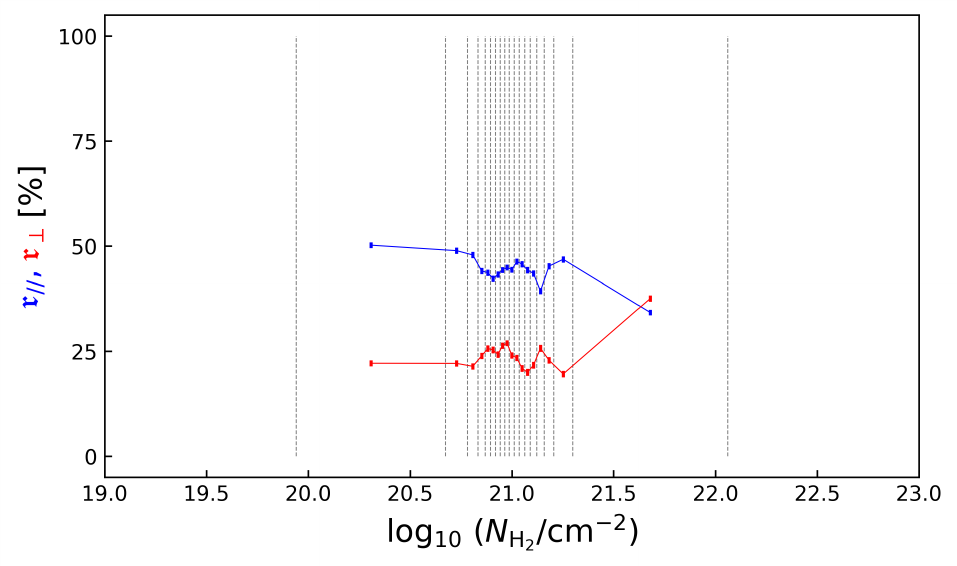}
    \caption{Values of $\mathfrak{r}_{\parallel}$ and $\mathfrak{r}_{\perp}$ (defined in Eq.~\eqref{eq:perpara_ratios_bis}) in 18 $N_{\rm{H_2}}$ bins, for GCC fields with ${\langle}N_{\rm{H_2}}{\rangle} \leq 10^{21}\,{\rm{cm^{-2}}}$.}
    \label{fig_appendix:results_filamentperparaHRO_NH2avg-}
\end{figure}

\begin{figure}
    \centering
    \includegraphics[width=\hsize]{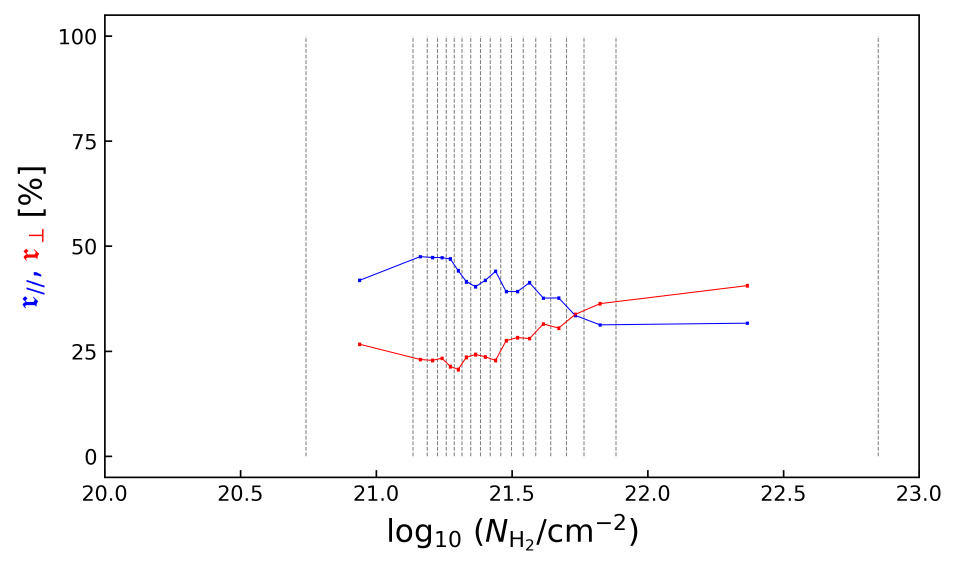}
    \caption{Values of $\mathfrak{r}_{\parallel}$ and $\mathfrak{r}_{\perp}$ (defined in Eq.~\eqref{eq:perpara_ratios_bis}) in 18 $N_{\rm{H_2}}$ bins, for GCC fields with ${\langle}N_{\rm{H_2}}{\rangle} > 1.5 \times 10^{21}\,{\rm{cm^{-2}}}$.}
    \label{fig_appendix:results_filamentperparaHRO_NH2avg+}
\end{figure}

\begin{figure}
    \centering
    \includegraphics[width=\hsize]{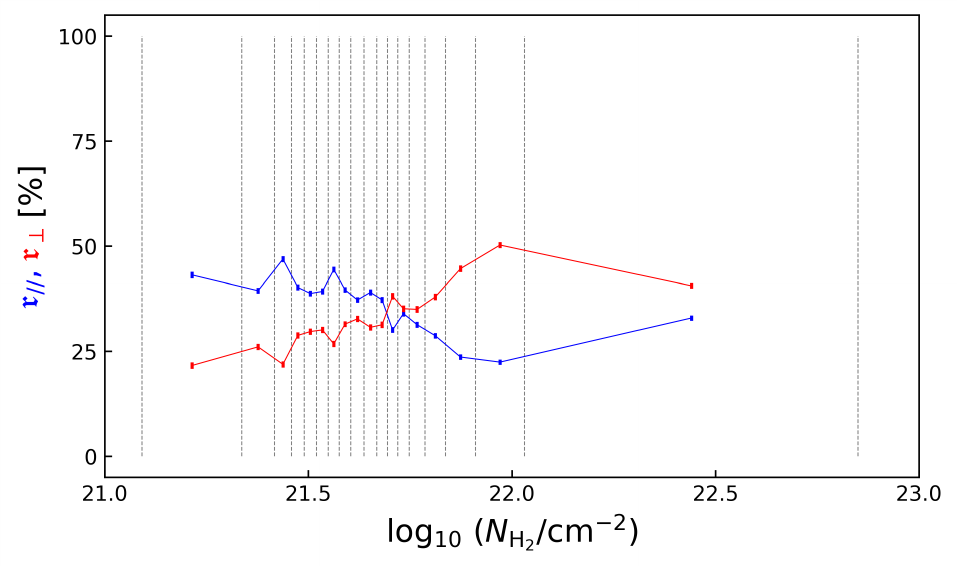}
    \caption{Values of $\mathfrak{r}_{\parallel}$ and $\mathfrak{r}_{\perp}$ (defined in Eq.~\eqref{eq:perpara_ratios_bis}) in 18 $N_{\rm{H_2}}$ bins, for GCC fields with ${\langle}N_{\rm{H_2}}{\rangle} > 2.5 \times 10^{21}\,{\rm{cm^{-2}}}$.}
    \label{fig_appendix:results_filamentperparaHRO_NH2avg++}
\end{figure}

\end{appendix}

\end{document}